\documentclass[10pt]{article} \usepackage{aaspp4}

\usepackage{lscape}

\newcommand{\CI}{C$\;${\small\rm I}\relax}
\newcommand{\NI}{N$\;${\small\rm I}\relax}
\newcommand{\OI}{O$\;${\small\rm I}\relax}
\newcommand{\MgII}{Mg$\;${\small\rm II}\relax}
\newcommand{\SiII}{Si$\;${\small\rm II}\relax}
\newcommand{\SII}{S$\;${\small\rm II}\relax}
\newcommand{\MnII}{Mn$\;${\small\rm II}\relax}
\newcommand{\FeII}{Fe$\;${\small\rm II}\relax}
\newcommand{\ZnII}{Zn$\;${\small\rm II}\relax}

\newcommand{\HI}{H$\;${\small\rm I}\relax}
\newcommand{\HII}{H$\;${\small\rm II}\relax}
\newcommand{\lya}{Ly$\alpha$}
\newcommand{\NHI}{$N({\rm H \; \mbox{\small\rm I}})$}
\newcommand{\Nav}{$N_a (v)$\relax}
\newcommand{\wave}[1]{$\lambda$#1\relax}
\newcommand{\twowave}[1]{$\lambda \lambda$#1\relax}

\newcommand{\kms}{km~s$^{-1}$\relax}
\newcommand{\etal}{{\em et al.}\relax}

\newcommand{\mvlsr}{v_{\rm LSR}\relax}
\newcommand{\percc}{cm$^{-3}$\relax}

\newcommand{\e}[1]{10^{#1}}

\newcommand{\mucol}{$\mu$~Col\relax}
\newcommand{\zoph}{$\zeta$~Oph\relax}
\newcommand{\cloudy}{CLOUDY\relax}
\newcommand{\copernicus}{{\em Copernicus}\relax}

\newcommand{\wham}{WHAM}
\newcommand{\hst}{{\em HST}}
\newcommand{\ghrs}{GHRS}
\newcommand{\costar}{COSTAR}
\newcommand{\fpn}{FPN}
\newcommand{\nav}{$N_a (v)$}

\begin{document}

\title{Abundances and Physical Conditions in the Warm Neutral Medium
Towards $\mu$~Columbae\altaffilmark{1}}

\altaffiltext{1}{Based on observations made with the NASA/ESA Hubble Space
Telescope, obtained from the data archive at the Space Telescope
Science Institute. STScI is operated bythe Association of Universities
for Research in Astronomy, Inc. under the NASA contract NAS 5-26555. }

\author{J. Christopher Howk\altaffilmark{2}, Blair D. Savage, 
	\& Dirk Fabian}
\affil{Department of Astronomy, University of Wisconsin-Madison, 
	Madison, WI 53706 \\ Electronic mail:
     howk@astro.wisc.edu, savage@astro.wisc.edu, dirk@astro.wisc.edu}

\altaffiltext{2}{Current Address: Deptartment of Physics and
Astronomy, The Johns Hopkins University, Baltimore, MD, 21218}

\authoremail{howk@stout.astro.wisc.edu}

\begin{abstract}
	
We present ultraviolet interstellar absorption line measurements for
the sightline towards the O9.5 V star $\mu$ Columbae ($l=237\fdg3$,
$b=-27\fdg1$; $d \approx 400$ pc, $z \approx 180$ pc; $\langle n_{\rm
H I} \rangle \approx 0.06$ \percc) obtained with the Goddard High
Resolution Spectrograph (GHRS) on board the {\em Hubble Space
Telescope}.  These archival data represent the most complete GHRS
interstellar absorption line measurements for any line of sight
towards an early-type star.  The 3.5 \kms\ resolution of the
instrument allow us to accurately derive the gas-phase column
densities of many important ionic species in the diffuse warm neutral
medium, including accounting for saturation effects in the data and
for contamination from ionized gas along this sightline.  We find that
the effects of an \HII\ region around \mucol\ itself do not
significantly affect our derivation of gas-phase abundances.  For the
low velocity material ($-20 \lesssim v_{\rm LSR} \lesssim +15$ \kms)
we use the apparent column density method to derive column densities.
For the individual absorbing components at $v_{\rm LSR}\approx -28.8$,
$+20.1$, $+31.0$, and $+41.2$ \kms, we apply component fitting
techniques to derive column densities and $b$-values.  We have also
used observations of interstellar \lya\ absorption taken with the
\ghrs\ intermediate resolution gratings to accurately derive the \HI\
column density along this sightline.  The resulting interstellar
column density $\log N(\mbox{\ion{H}{1}}) = 19.86\pm0.015$ is in
agreement with other determinations but is significantly more precise.

The low-velocity material shows gas-phase abundance patterns similar
to the warm cloud (cloud A) towards the disk star $\zeta$ Ophiuchi ,
while the component at $v_{\rm LSR}\approx +20.1$
\kms\ shows gas-phase abundances similar to those found in warm halo
clouds.  {\em We find the velocity-integrated gas-phase abundances of
Zn, P, and S relative to H along this sightline are indistinguishable
from solar system abundances.}  We discuss the implications of our
gas-phase abundance measurements for the composition of interstellar
dust grains.  We find a dust-phase abundance $\left[{\rm (Fe+Mg)}/{\rm
Si}\right] _d = 2.7 - 3.3$ in the low-velocity gas; therefore the dust
cannot be composed solely of common silicate grains, but must also
include oxides or pure iron grains.  The low velocity material along
this sightline is characterized by $T\approx 6,000 - 7,000$ K with
$n_e \approx 0.3$ \percc, derived from the ionization equilibrium of
Mg and Ca.

The relative ionic column density ratios of the intermediate velocity
components at $v_{\rm LSR} = +31.0$ and $+41.2$ \kms\ show the imprint
both of elemental incorporation into grains and (photo)ionization.
These clouds have low total hydrogen column densities ($\log N({\rm
H}) \sim 17.4-17.7$), and our component fitting $b$-values constrain
the temperature in the highest velocity component to be $T =
4,000\pm700$ K.  The electron density of this cloud is $n_e
\approx 0.6$ \percc, derived from the  $^2P_{1/2}$ to
$^2P_{3/2}$ fine structure excitation of \ion{C}{2}.  The components
at $v_{\rm LSR} \approx -30$ and $-48$ \kms\ along this sightline
likely trace shocked gas with very low hydrogen column densities.  The
$v_{\rm LSR} \approx -30$ \kms\ component is detected in a few strong
low-ionization lines, while both are easily detected in \ion{Si}{3}.
The relative column densities of the $-30$ \kms\ suggest the gas is
collisionally ionized at moderate temperatures ($T\approx25,000$ K).
This is consistent with the measured $b$-values of this component,
though non-thermal motions likely contribute significantly to the
observed breadths.

\end{abstract}

 
\section{INTRODUCTION}

The measurement and analysis of interstellar absorption lines provides
fundamental information on the content and physical conditions of the
Galactic interstellar medium (ISM).  In particular, measurements of
the gas-phase abundances of the ISM have allowed us to infer the
composition of interstellar dust and trace the variations in the dust
make-up in a wide range of environments (e.g., Savage, Cardelli, \&
Sofia 1992; Spitzer \& Fitzpatrick 1993, 1995; Sembach \& Savage 1996;
Lu \etal\ 1998).  Furthermore, atomic absorption lines allow us to
study the chemical evolutionary history of the Universe over 90\% of
its age through the study of gaseous QSO absorption lines (e.g., Lu
\etal\ 1996; Pettini \etal\ 1997, 1999; Prochaska \& Wolfe 1999),
provided we can understand the imprint dust leaves on the
measurements.  The effects of dust are best constrained by our studies
of gas-phase abundances at zero redshift.

The study of Galactic interstellar absorption lines has been greatly
aided by the Goddard High Resolution Spectrograph (\ghrs) on board the
{\em Hubble Space Telescope} (\hst).  The echelle-mode resolution of
this instrument (FWHM$\, \approx 3.5$ \kms) coupled with its ability
to achieve high signal to noise ultraviolet observations of Galactic
early-type stars makes it a very powerful instrument for studying
absorption lines in the Galactic ISM.

The star $\mu$ Columbae is the \ghrs\ high- and
intermediate-resolution radiometric standard.  As such, it was
observed extensively during the Servicing Mission Orbital Verification
(SMOV) stages after the installation of the Corrective Optics Space
Telescope Axial Replacement (COSTAR).  In the post-\costar\ era, more
than 500 echelle-mode observations were made of \mucol\ using the
\ghrs, as well as a similar number of observations with the
lower-resolution first-order gratings.  This makes \mucol\ the most
extensively observed early-type star with the \ghrs.

The design of the \ghrs\ calibration observations during the SMOV
period were such that the extensive \mucol\ echelle-mode dataset is
characterized by extensive wavelength coverage and relatively high
signal-to-noise ratios.  The resulting high-resolution absorption line
dataset is almost ideal for studying abundances along the low-density
sightline to this star.  Though this sightline has been extensively
studied by the \copernicus\ satellite (Shull \& York 1977; hereafter
SY) and with the pre-\costar\ \ghrs\ by Sofia, Savage \& Cardelli
(1993; hereafter SSC), the extremely rich dataset acquired after the
installation of \costar\ represents a significant increase in
resolution over the observations of SY and in S/N over those of SSC.
The extensive wavelength coverage of the dataset includes observations
of a wide assortment of ionic species, with most being observed in
several transitions with a range of $f$-values.

We have reduced and analyzed the extensive archival \ghrs\ ultraviolet
absorption line dataset for the \mucol\ sightline.  Our main
objectives in this work are to very accurately derive the gas-phase
elemental abundances in the low- and intermediate-velocity gas along
this sightline as well as information on the physical conditions of
the gas.  From the gas-phase abundances we infer the dust content of
the ISM in this direction and discuss the implications of these data
for understanding the make-up of dust grains in the diffuse ISM.

Our work is presented as follows.  We discuss the properties of the
\mucol\ sightline and the previous studies of this sightline in \S
\ref{sec:sightline}.  In \S \ref{sec:processing} we describe our
reductions of the \ghrs\ dataset, as well as our extraction and
analysis methods for analyzing the ISM absorption line data.  In \S
\ref{sec:lowvelocity} we discuss the gas-phase abundances of the
low-velocity material along this sightline, including a detailed
analysis of the contributions from an \HII\ region about \mucol\
itself and the derivation of physical conditions for the absorbing
gas.  We present an analysis of the abundances of the intermediate
velocity gas along this sightline in \S \ref{sec:highvelocity}.  A
discussion of the implications of this interstellar absorption line
dataset is included in \S \ref{sec:discussion}, and we summarize our
major conclusions in \S \ref{sec:summary}.

\section{THE $\mu$ COL SIGHTLINE}
	\label{sec:sightline}

The star $\mu$ Columbae (HD 38666) lies in the direction ($l, \ b$) =
(237$^\circ$, $-27^\circ$) along a relatively unreddened sightline
[E($\bv$) = 0.02; Bastiaansen 1992].  Classified as an O9.5~V star by
Walborn (1973), \mucol\ is a runaway star with a radial velocity
relative to the local standard of rest (LSR) of $v_{\rm LSR} = +90$
\kms\ (Gies 1987; Keenan \& Dufton 1983).  Lesh (1968) has placed the 
star as far away as 1000 pc.  The absolute magnitude scale of Vacca,
Garmany, \& Shull (1996) yields a slightly smaller distance of
$\approx790$ pc, given the observed magnitude $V=5.16$.  However,
recent {\em Hipparcos} measurements have constrained its parallax to
be $\pi = 2.52\pm0.55$ milliarcseconds (Perryman \etal\ 1997),
implying a distance of only $400^{+100}_{-70}$ pc.  This result is
more consistent with the distance derived from Str\"{o}mgren
photometry (410 pc) by Keenan \& Dufton (1983).  This discrepancy in
distance scale may be related to problems in assigning luminosity
classes to rapidly rotating spectral standards (Lamers \etal\ 1997).
We will adopt the {\em Hipparcos} results throughout this work, though
the reader should be aware of the continuing gaps in our knowledge of
the stellar distance scale.



The adopted distance, coupled with the observed neutral hydrogen
column density of $\log N(\mbox{\HI}) = 19.86 \pm 0.015$ (see Appendix
\ref{appendix:lya}), implies an average line of sight density of
$\langle n_{\rm H} \rangle
\approx 0.06$ cm$^{-3}$.  Molecular hydrogen makes a negligible 
contribution to the total hydrogen column density with $\log N({\rm
H_2}) = 15.50$ summed over $J=0$ to 4 (Spitzer, Cochran, \& Hirshfeld
1975).

Highly ionized gas along the \mucol\ sightline implies the presence of
both hot ($T >10^5$ K) collisionally ionized material and photoionized
gas near the star.  York (1974) fit the strong interstellar \ion{O}{6}
absorption along this sightline with a column density $\log
N(\mbox{\ion{O}{6}}) = 13.82\pm0.02$ and a Doppler parameter $b_{\rm
O\,VI} = 43.8$ \kms.  Brandt \etal\ (1999) present GHRS observations
of weak
\ion{C}{4}, \ion{Si}{4} and \ion{N}{5} absorption at 3.5 \kms\
resolution.  Figures \ref{fig:norm1}-\ref{fig:excited} show the
normalized interstellar absorption line profiles for the
\mucol\ sightline observed with the \ghrs, including the profiles of
\ion{C}{4} and \ion{Si}{4} (see Figure \ref{fig:highions}).  
The profile widths of the highly ionized atoms increase with
ionization potential.  The breadth of the profile implies $T < 2
\times 10^6$ K, while the width of the \ion{C}{4} line, for example,
implies temperatures $T < 2.3 \times 10^5$ K; much of the velocity
widths may be due to non-thermal motions.  The \ion{O}{6} and
\ion{C}{4} profiles are centered near $\langle v_{\rm LSR} \rangle =
-2$ \kms, while the \ion{Si}{4} is centered at $\langle v_{\rm LSR}
\rangle = +5$ \kms.  The ratio $N(\mbox{\ion{C}{4}}) /
N(\mbox{\ion{O}{6}}) = 0.11\pm0.01$ is consistent with that observed
for other sightlines intercepting Galactic disk gas (Spitzer 1996).
Brandt \etal\ suggest much of the
\ion{O}{6} and some \ion{C}{4} arises in an evolved supernova remnant
along the line of sight, though some of the high-ion absorption likely
arises in the interface between the Local Cloud and the Local Bubble.
They also show that a significant amount of the
\ion{Si}{4} column density is likely produced in a low-density \HII\
region surrounding \mucol.  We will discuss in more detail the
possible contributions of ionized gas to the absorption line
measurements.

The low-ionization material along the sightline to \mucol\ was studied
with the \copernicus\ satellite by SY at a resolution of $\approx 13$
\kms\ and with the echelle gratings of the pre-COSTAR GHRS by SSC with
a resolution of $\approx 3.5$ \kms.  These studies have identified
low-ion absorption in four main absorbing components, the properties
of which are given in Table \ref{table:components}.  This table
includes the number by which we will refer to each of these absorbing
regions, the approximate central velocity and the range of velocity
over which the absorption from each region extends, both in the LSR
frame, as well as the identifications of these regions in the
nomenclature of SY and SSC.  Regions 2, 3, and 4 are relatively well
separated in velocity and may represent individual absorbing
``clouds;'' region 1 is a blend of several clouds that overlap in
velocity.  The region we designate as component 5, which appears at
negative velocities, was described by SY as ``trailing absorption.''
Each of the absorbing regions in Table
\ref{table:components} shows different gas-phase abundances,
suggesting changing patterns of elemental incorporation into dust
grains and ionization.  In general the absorption along this sightline
is consistent with the Routly-Spitzer effect (Routly \& Spitzer 1952),
where the gas-phase abundances of refractory elements increase with
velocity (SY; SSC; Hobbs 1978).  Also given in Table
\ref{table:components} are the temperatures, $T$, and electron
densities, $n_e$, for each component from our analysis below (see \S
\ref{subsec:physicalconditions} and \S
\ref{subsec:IVphysicalconditions}).  

Lockman (1991) has discussed the \HI\ 21-cm emission profile towards
\mucol, which is reproduced in Figure \ref{fig:ground}.  These data,
from Lockman, Hobbs, \& Shull (1986), have a 21\arcmin\ beam and have
been corrected for stray radiation.  The column of \HI\ from the 21-cm
data is more than twice that derived from observations of
Lyman-$\alpha$ along the sightline to \mucol, implying much of the
\HI\ column observed in 21-cm radiation comes from beyond the star.
In the direction of \mucol\ Galactic rotation is expected to carry the
gas to positive velocities, with a maximum velocity of $v_{LSR}
\approx +5.2\pm0.6$ \kms.  The majority of the \HI\ emission and the
UV-absorption (i.e., the region 1 blend) resides at velocities allowed
by Galactic rotation.  A quite narrow component at $\mvlsr \approx 3$
\kms\ shows the peak brightness temperature in the \HI\ emission
profile, though several broad absorbing clouds also contribute to the
blend of region 1.  It is clear that the \HI\ content of components 2,
3, and 4 is significantly less than that of component 1.  Though UV
absorption profiles along this sightline show gas out to $\mvlsr
\approx +42$ \kms, the \HI\ emission profile shows an extended wing to
$\mvlsr \gtrsim +60$ \kms.

Component 1 is centered at $\mvlsr \approx +3$ \kms\ and contains
$\sim 90\%$ of the neutral gas along the sightline, as evidenced by
the relative strengths of the absorbing regions in the undepleted
species \SII.  Ground based observations of the \ion{Na}{1} D2 line at
a resolution of 1 \kms\ show the complex to consist of several (at
least three or four) velocity components (Hobbs 1978; see Figure
\ref{fig:ground}).  Though the absolute velocity of the absorption is
uncertain, SY tentatively associate the observed H$_2$ absorption with
component 1.

The papers of SY and SSC have shown absorbing complex 1 likely has
depletion characteristics similar to that of the warm cloud towards
$\zeta$ Oph (component A of Savage \etal\ 1992).  The low
velocity resolution of the \copernicus\ data make the separation of
absorption from components 1 and 2 difficult; the higher resolution
observations of SSC are superior in this respect.  Those elements with
high condensation temperatures, such as \ion{Fe}{2} and \ion{Cr}{2},
exhibit gas phase abundances that are sub-solar with respect to \SII\
by $\approx 1.3$ dex.  Sofia \etal\ (SSC) suggest that the level of
depletion exhibited by the \mucol\ component 1 and the warm component
along the $\zeta$ Oph sightline may be typical of the low-density,
warm neutral medium (WNM) of the Galactic disk.

The presence of ionized material in the velocity range encompassed by
component 1 is suggested by \ion{N}{2} and \ion{Si}{3} absorption in
the data of SY.  The fine structure lines of \ion{N}{2}$^{**}$, which
trace the densest ionized regions, are centered at $\mvlsr = 0$ to +2
\kms.  Shull \& York argue that the majority of the ionized gas along
the sightline is associated with an \HII\ region surrounding \mucol\
and estimate a density $\langle n_e \rangle \approx 0.2$ cm$^{-3}$.
Recent observations of this sightline with the \wham\ Fabry-Perot
spectrometer (Reynolds \etal\ 1998), which are reproduced in Figure
\ref{fig:ground}, show that the ionized gas is indeed concentrated
near $\mvlsr = 0$ (M. Haffner, 1998, priv. comm.).  The imprint of
ionized regions on the column densities of (primarily) neutral gas
tracers can be one of the largest uncertainties in studying the
gas-phase abundances in the Galactic WNM, such as in component 1.  We
will discuss this contamination of our dataset in \S
\ref{subsec:ionization}.

Component 2 appears distinctly in species that tend to be highly
depleted in component 1.  A quick comparison of the \ion{Ca}{2} and
\ion{Na}{1} profiles of Hobbs (1978) shows that the ratio of these two
species changes significantly as the velocity of the gas increases
(these data are reproduced in Figure \ref{fig:ground}).  This
difference is likely due to the return of elements to the gas phase
due to dust destruction.  Shull \& York (SY) and Sofia \etal\ (SSC)
have shown that the refractory elements have a much higher relative
abundance in this component than in the lower velocity material.  This
cloud is at a velocity that would place it well beyond the star if it
were simply participating in Galactic rotation.

The intermediate-velocity gas along the \mucol\ sightline has been
less-well observed due to the relatively low column of material
present in these clouds.  Centered at $\mvlsr \approx +33$ and +42
\kms, respectively, components 3 and 4 show patterns of abundances
quite different than the low-velocity material.  Both SY and SSC
examine the gas-phase abundances in component 4, though component 3
was only identified by SSC.  In both cases, the effects of ionization
may be significant, though the gas-phase abundances of these clouds
may also be high, suggesting substantial grain destruction (SSC).
Shull \& York also noted the presence of ``trailing'' absorption in
the \ion{Si}{3} profile.  This material, which we identify as
component 5, has a velocity relative to the local standard of rest
$v_{LSR} \approx -30$ \kms, which is inconsistent with Galactic
rotation.  

Some $90\%$ of the gas towards \mucol\ is associated with component 1,
which has velocities roughly consistent with expectations due to
Galactic rotation.  Most of the remaining 10\% of the material has
velocities which are inconsistent with rotation.  Sembach \& Danks
(1994) have found that on average $\sim10\%$ of \ion{Ca}{2} absorption
is at forbidden velocities.  They estimate a cloud-to-cloud velocity
dispersion in this forbidden-velocity gas of $\sigma \approx 22$ \kms.
The forbidden-velocity gas towards \mucol, therefore, seems not to
show highly unusual kinematics compared with the observations of
low-density sightlines by Sembach \& Danks.

\section{DATA PROCESSING AND ANALYSIS}
\label{sec:processing}

Table \ref{table:log} lists basic information of the individual
spectra used in our analysis of the \mucol\ sightline.  The present
set of observations was acquired for the purpose of evaluating the
in-flight performance of and flux-calibrating the high-resolution
modes of the \ghrs\ following the installation of
\costar.\footnote{Details about the \ghrs\ and its in-flight
performance characteristics can be found in Robinson \etal\ (1998) and
Heap \etal\ (1995).}  The data have been collected over a large span
of time, beginning in early 1994 after the installation of \costar.
As such, the observations do not represent a completely homogeneous
data set.  We rely most heavily on measurements made with the
echelle-mode Ech-A and Ech-B gratings, which give a resolution of
$\approx 3.5$ km~s$^{-1}$ (FWHM).  An extensive dataset exists for the
first-order gratings as well, and we have used observations taken with
the G140M and G160M gratings to derive the column densities of \HI\
(see Appendix \ref{appendix:lya}) and \ion{Fe}{3} along the \mucol\
sightline.  Typically the star was observed for 30-120 seconds through
the large science aperture (LSA; 1\farcs74$\times$1\farcs74) with four
substeps per diode.  Appropriate measurements were made of the
inter-order scattered light in all cases (see Cardelli \etal\ 1993),
and the observations employed the comb-addition routine with the
on-board doppler compensator enabled.  We have in general restricted
ourselves to using the post-\costar\ data for this sightline due to
the degradation in resolution of the pre-\costar\ LSA observations.
We have, however, made use of pre-COSTAR small science aperture (SSA;
0\farcs22$\times$0\farcs22) data in our component fitting analysis
(see \S \ref{subsec:compfitting}).

\subsection{Data Processing}

Our calibration and reduction of the data follows procedures similar
to those discussed in Savage \etal\ (1992) and Cardelli \etal\ (1995).
Our determination and propagation of errors follows Sembach \& Savage
(1992) for our measurements of the integrated equivalent widths and
column densities. The basic calibration was performed at the \ghrs\
computing facility at the Goddard Space Flight Center and at the
University of Wisconsin-Madison using the standard {\tt CALHRS}
routine.\footnote{{\tt CALHRS} is part of the standard Space Telescope
Science Institute pipeline and the STSDAS IRAF reduction package.  It
is also distributed via the \ghrs\ Instrument Definition Team for the
IDL package.}  The {\tt CALHRS} processing includes conversion of raw
counts to count rates and corrections for particle radiation
contamination, dark counts, known diode nonuniformities, paired pulse
events and scattered light.  The wavelength calibration was derived
from the standard calibration tables and should be accurate to
approximately $\pm 3.5$ \kms.

The final data reduction was performed using software developed and
tested at the University of Wisconsin-Madison.  This includes the
merging of individual spectra and allowing for additional refinements
to the scattered light correction.  The inter-order scattered light
removal discussed by Cardelli \etal\ (1990, 1993) is based upon
extensive pre-flight and in-orbit analysis of \ghrs\ data and is used
by the {\tt CALHRS} routine; the coefficients derived by these authors
are appropriate for observations made through the SSA.  We find that
many of the LSA observations required an additional correction to
bring the cores of strongly saturated lines to the appropriate zero
level.  The final scattered light coefficients, $d_{\rm c}$, used for
each group of spectra are given in Table \ref{table:log}.  Many of the
observations had no strong lines which would allow us to refine the
values of $d_{\rm c}$; those cases for which we have adopted the
Cardelli \etal\ (1993) values, having no independent measure of
$d_{\rm c}$, are marked with a colon in Table \ref{table:log}.  In
general the signal-to-noise ratio and spacing of the individual
spectra did not warrant solving explicitly for the \fpn\ spectrum.
Those regions for which we have derived the noise spectrum and removed
it (following the algorithm of Cardelli \& Ebbets 1994) are identified
in Table \ref{table:log}.

To bring all of the species into a common velocity reference we have
applied a ``bootstrap'' technique similar to that discussed in
Cardelli \etal\ (1995).  We have aligned in velocity space lines of
the same species found in different observations.  We have further
attempted to align several ions with similar ionization and depletion
characteristics, or similar velocity structure.  For example, the
strong lines of \ion{Mg}{1} and {\small II}, \ion{Fe}{2}, \ion{Al}{2},
and \SiII\ have been shifted in velocity space to be brought into
alignment with the \SiII\ \wave{1304} line.  Component 4 is well
separated from lower-velocity absorption for the stronger lines of
these species, making the alignment relatively straight-forward.  The
\SiII\ \wave{1808}\ line was brought into this reference frame using
the wings of \SiII\ \wave{1304}, and the weak lines of \ion{Fe}{2}
have been aligned with the wings and the distinct component 2 of the
stronger \ion{Fe}{2} lines.  The \OI\ \wave{1302} line appears in the
same exposures as \SiII\ \wave{1304} and was used to tie \NI\ and
\ion{C}{2} into the velocity frame of the more depleted species.  The
lines of the heavily depleted species \ion{Ni}{2} \wave{1370} and
\ion{Cr}{2} \twowave{2056, 2062} were aligned to the weaker
\twowave{2249 and 2261} lines of \ion{Fe}{2}; the other lines of these
species lacked sufficient signal-to-noise to enable any correction to
their velocity zero-points.  \ion{Cr}{2} \wave{2062} was then used to
align \ZnII\ \wave{2062}, which appears in the same observations.  The
latter line was then used to bring \ZnII\ \wave{2026}, \ion{P}{2}
\wave{1152}, and \SII\ \wave{1250} into the common velocity reference.
The two stronger lines of \SII\ were aligned to the \wave{1250} line.
Lastly, the lines \MgII\, \twowave{1239 and 1240}, and \ion{Mn}{2}
\twowave{2577 and 2594} were aligned as well as possible with \SiII\
\wave{1808}.  The \MgII\ to \SiII\ alignment should be reasonably
secure given the somewhat similar component structure of these
transitions, but the \MnII\ lines show different component structure,
particularly for component 2 centered near $\mvlsr \approx +21$ \kms,
which seems to be indicative of the different depletion
characteristics of these elements.  The alignment of these last lines
is more uncertain than most.

To determine an absolute velocity frame we have measured the
heliocentric velocity of the \ion{O}{1}$^*$ \wave{1304} and
\ion{O}{1}$^{**}$ \wave{1306} telluric absorption lines, which are
present in the same specrum as the \SiII \wave{1304} line.  We have
determined the central velocities of each of these telluric lines and
compared those with the velocity of the spacecraft at the time of the
observation plus the correction for the Earth's motion towards the
star.  A correction of $-0.3$ \kms\ was needed to bring the \SiII\
\wave{1304} line observed through the SSA into the heliocentric
rest-frame.  An additional correction was then made to convert
heliocentric velocities to the LSR frame.  Assuming a solar
neighborhood speed of +16.5 \kms\ in the direction
$(l,b)=(53^\circ,25^\circ)$ (Mihalas \& Binney 1981) implies $v_{\rm
LSR} - v_{helio} = -16.5$ \kms\ for the sightline to \mucol.  We have,
however, applied a shift of $-19.9$ \kms\ to all heliocentric
velocities in order to be consistent with previous studies of this
sightline which have assumed a solar motion of $+20$ \kms\ in the
direction $(l,b)=(56^\circ,22^\circ)$ [$(\alpha,\delta)_{1900} =
(18^h,+30^\circ)$; see York \& Rogerson 1976; also adopted by York
1974, SY, and SSC].

\subsection{Absorption Profiles and Measurements}
	\label{subsec:measurements} 

\begin{figure}
\epsscale{1.1}
\plotone{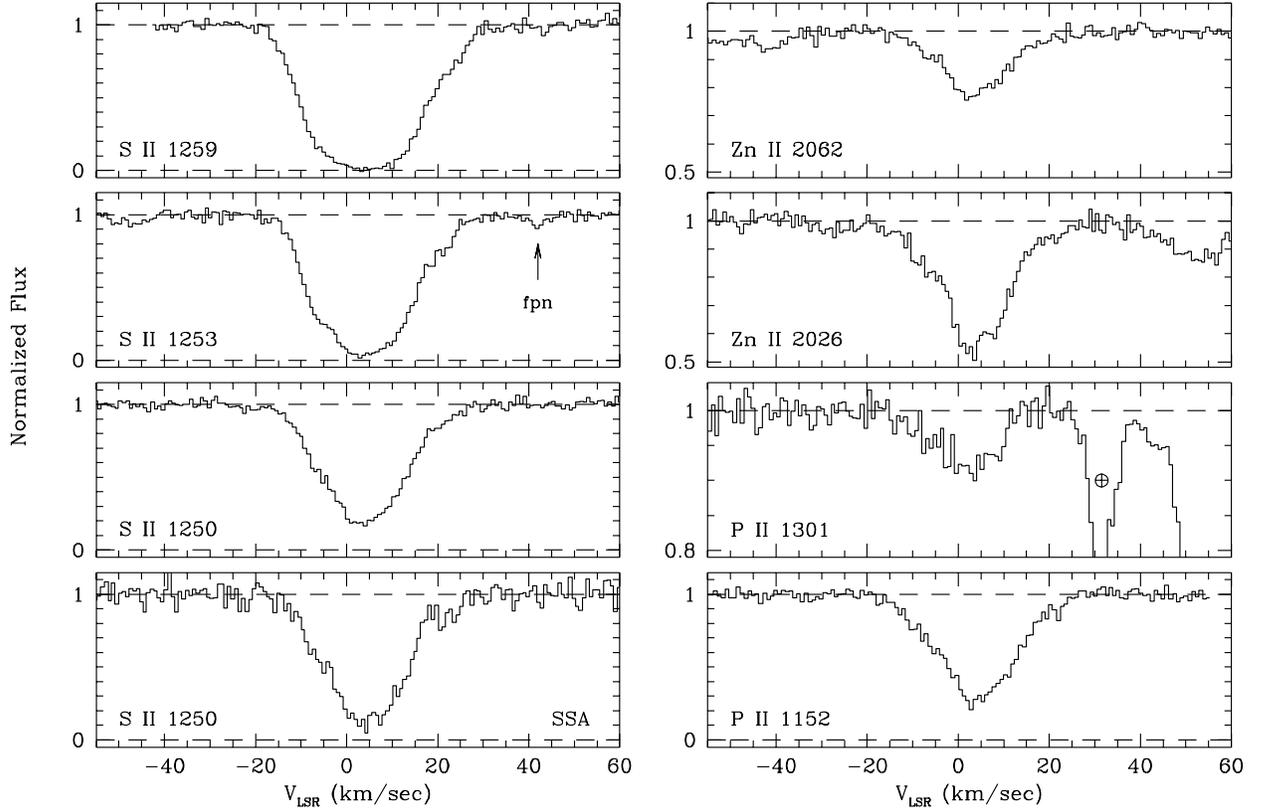}

\figcaption{Continuum-normalized intensity vs. LSR velocity absorption
profiles for the species \protect\ion{S}{2}, \protect\ion{P}{2}, and
\protect\ZnII.  These species generally exhibit very low levels of
depletion in the WNM.  The stellar continuum for each line was
approximated by a low-order polynomial ($\leq 5$) fitted to regions on
either side of the line.  For $\mu$ Col, $v_{\rm LSR} = v_{\rm helio}
- 19.9$ km s$^{-1}$.  A telluric absorption line of \protect\ion{O}{1}
in the \protect\ion{P}{2} \protect\wave{1301} profile is marked with
the $\oplus$ symbol.  The \protect\ion{S}{2} \protect\wave{1253}
observation contains a fixed pattern noise feature.  We have marked
observations made with the SSA.  The absorption features extending
from $\sim40$ to 60 km s$^{-1}$ in the \protect\ion{Zn}{2}
\protect\wave{2026} and \protect\ion{P}{2} \protect\wave{1301} panels
are \protect\ion{Mg}{1} \protect\wave{2026} and \protect\ion{O}{1}
\protect\wave{1302}, respectively.
\label{fig:norm1}}
\end{figure}

\begin{figure}
\epsscale{1.1}
\plotone{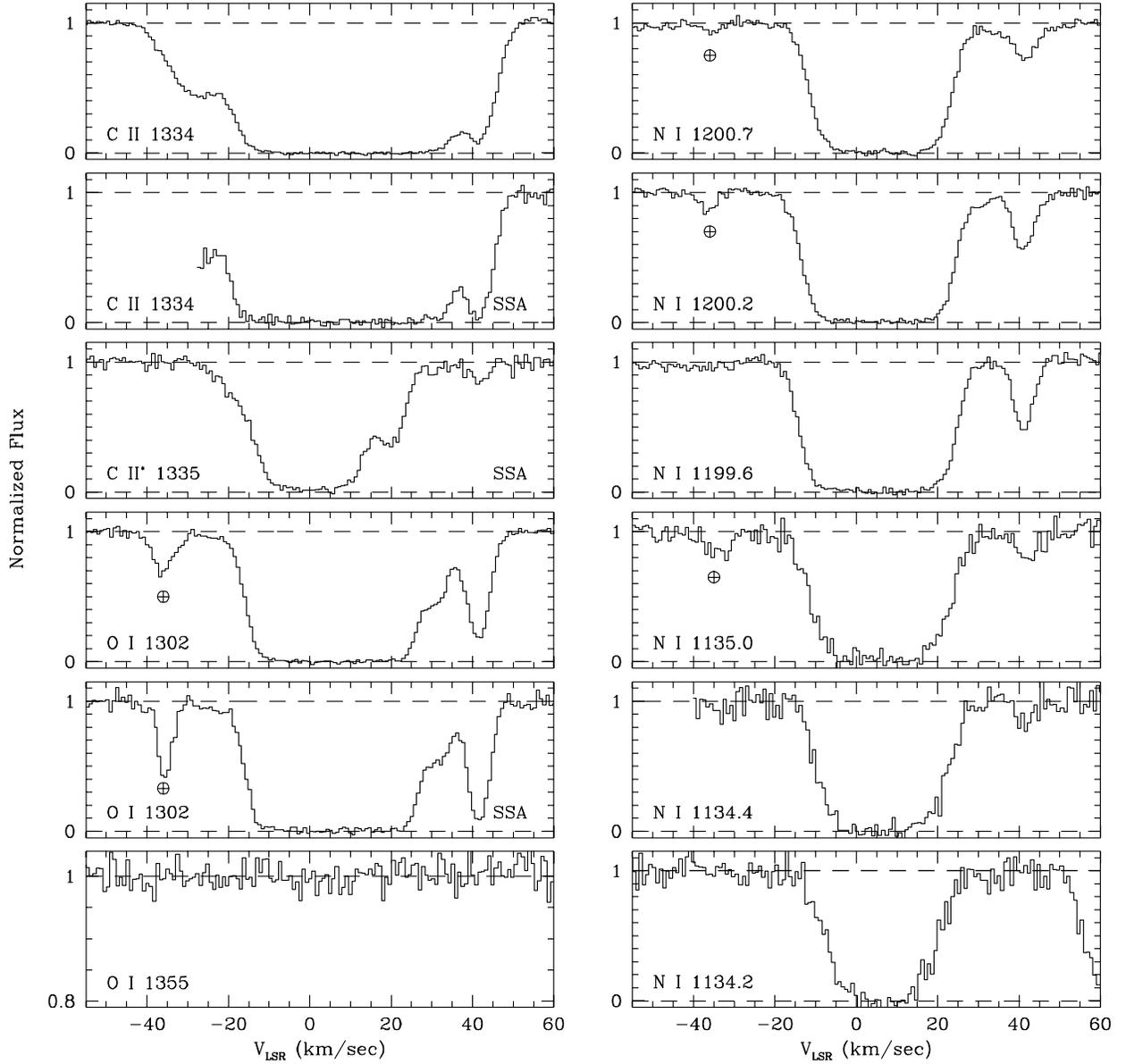}

\figcaption{As Figure \ref{fig:norm1}, but for the relatively lightly
depleted species \protect\ion{C}{2}, \protect\ion{O}{1}, and
\protect\ion{N}{1}.  Telluric absorption lines of \protect\ion{O}{1}
and \protect\ion{N}{1} are marked with the $\oplus$
symbol. \label{fig:norm2}}
\end{figure}

\begin{figure}
\epsscale{1.1}
\plotone{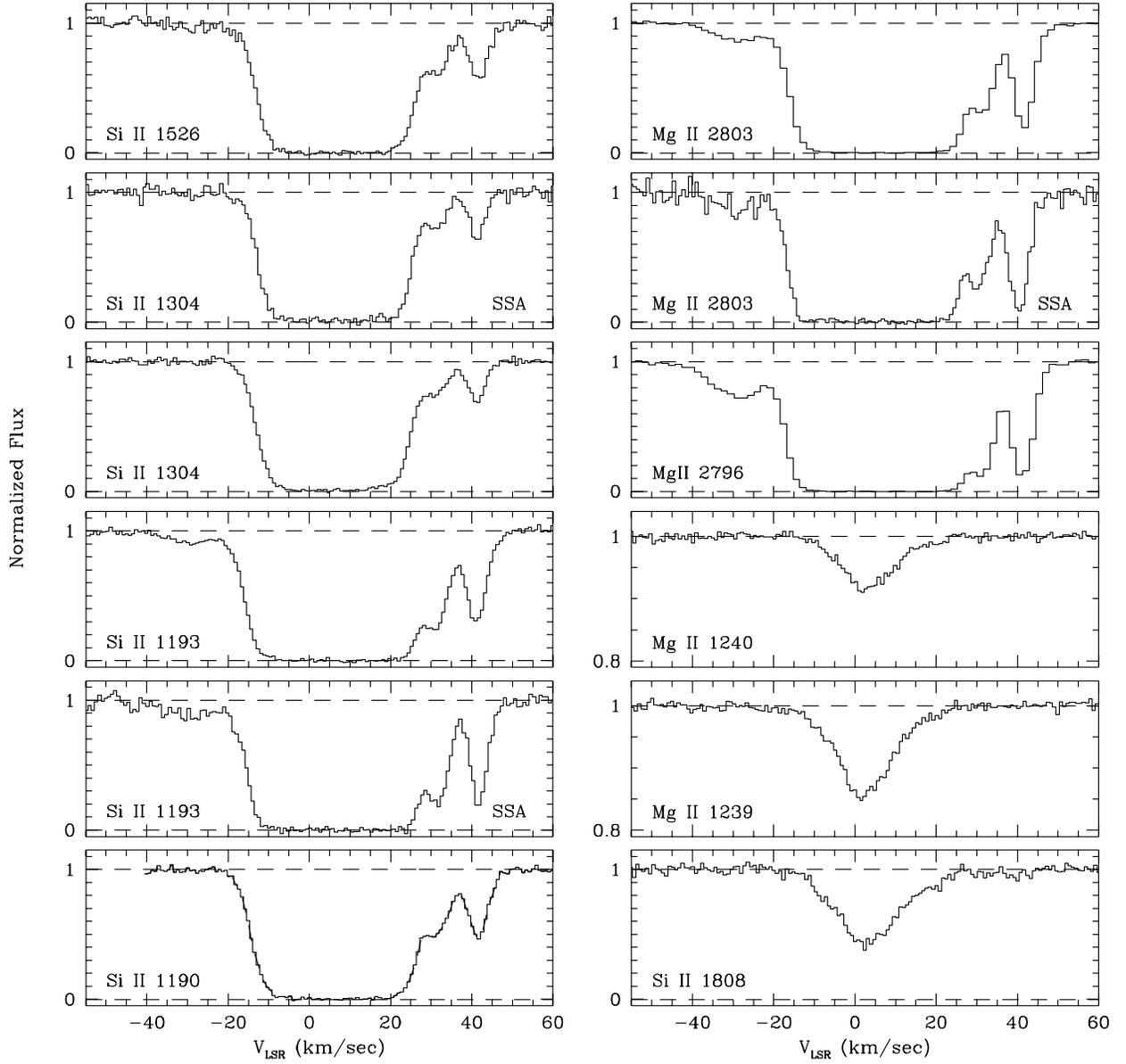}
\figcaption{As Figure \ref{fig:norm1}, but for the moderately depleted
species \protect\ion{Mg}{2} and \protect\ion{Si}{2}.\label{fig:norm3}}
\end{figure}

\begin{figure}
\epsscale{1.1}
\plotone{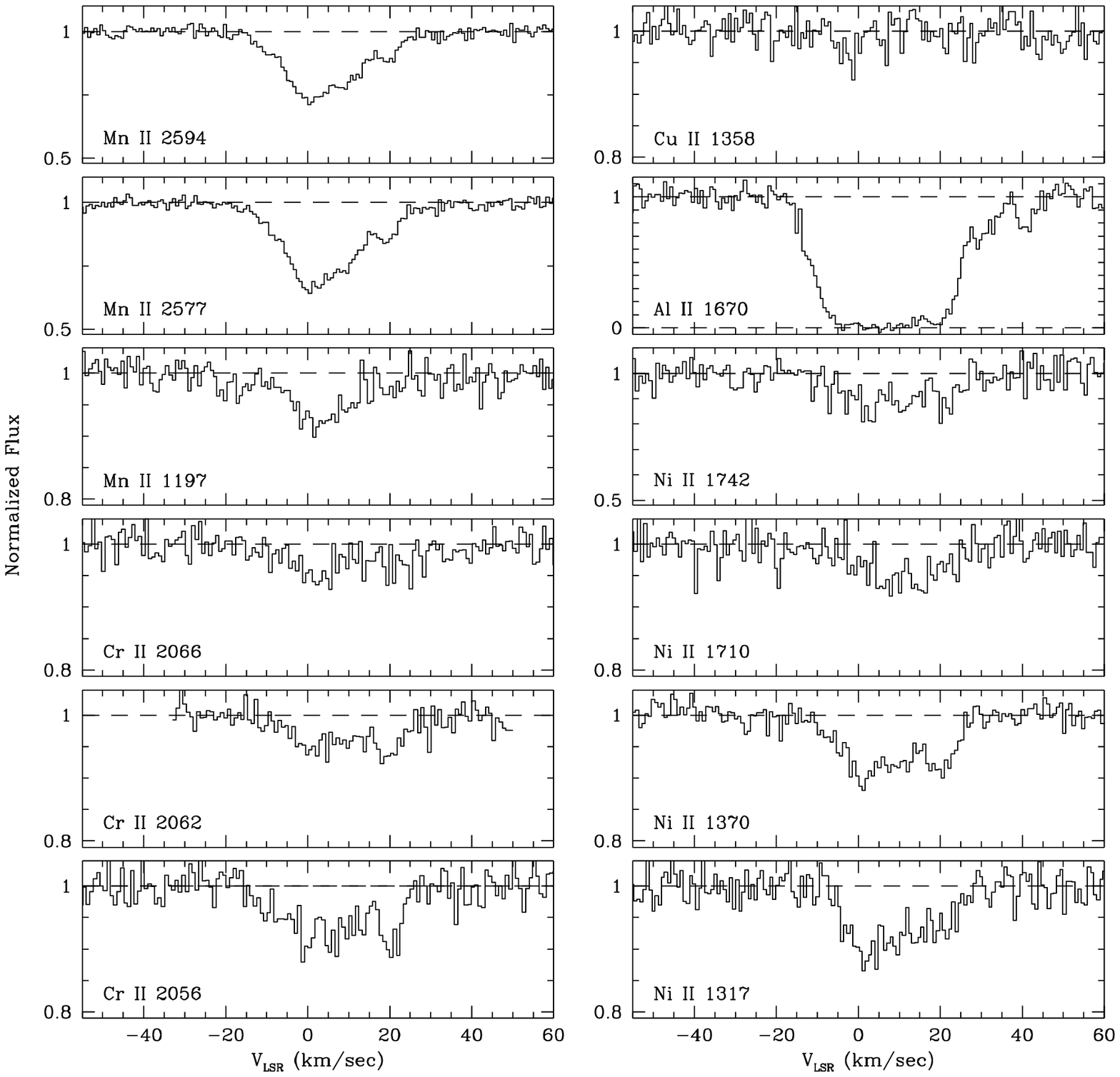}

\figcaption{As Figure \ref{fig:norm1}, but for the moderately to
heavily depleted species \protect\ion{Mn}{2}, \protect\ion{Cr}{2},
\protect\ion{Cu}{2}, \protect\ion{Al}{2}, and
\protect\ion{Ni}{2}. \label{fig:norm4}}
\end{figure}

\begin{figure}
\epsscale{1.1}
\plotone{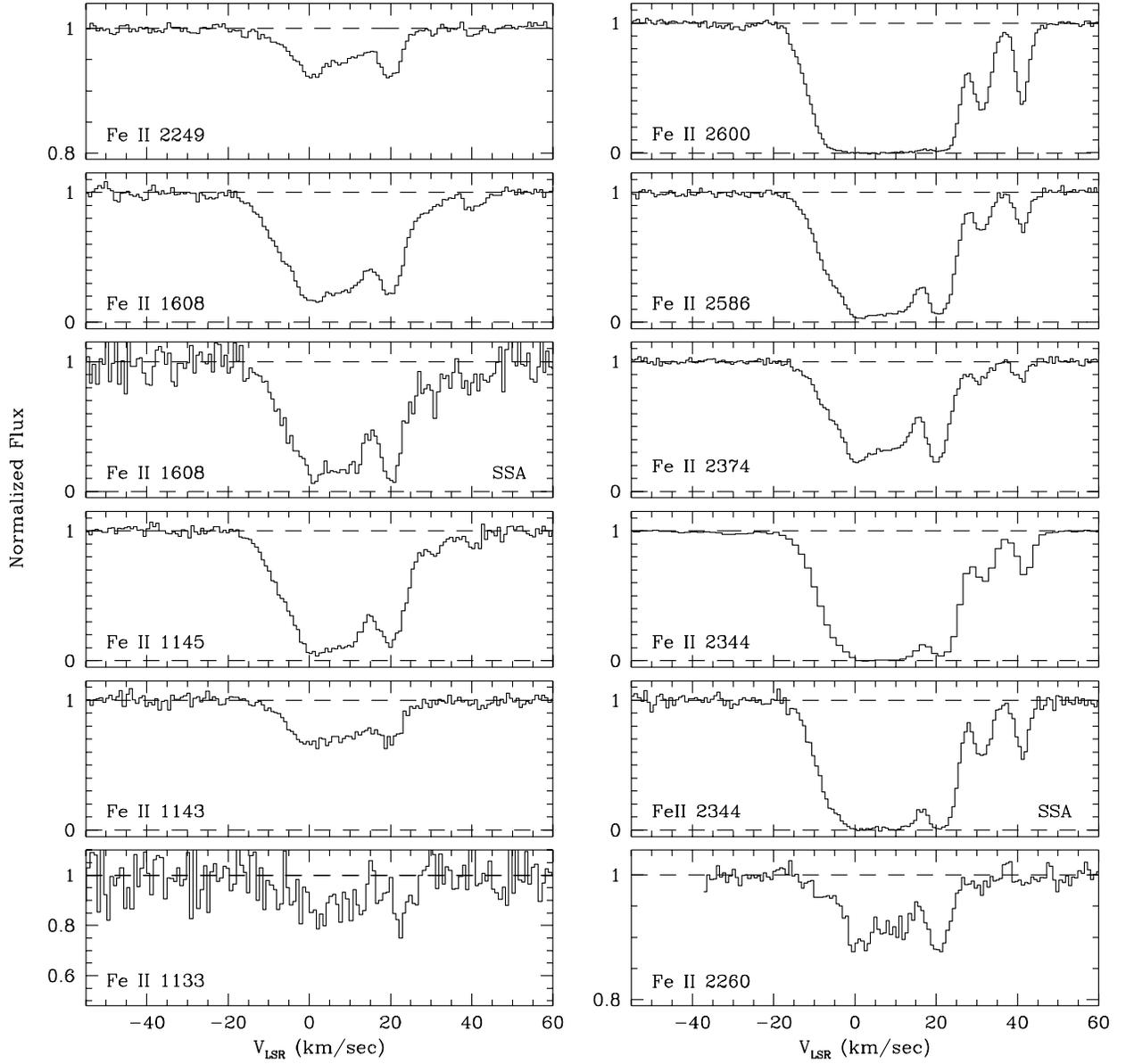}
\figcaption{As Figure \ref{fig:norm1}, but for the heavily depleted
species \protect\ion{Fe}{2}. \label{fig:norm5}}
\end{figure}

\begin{figure}
\epsscale{1.1}
\plotone{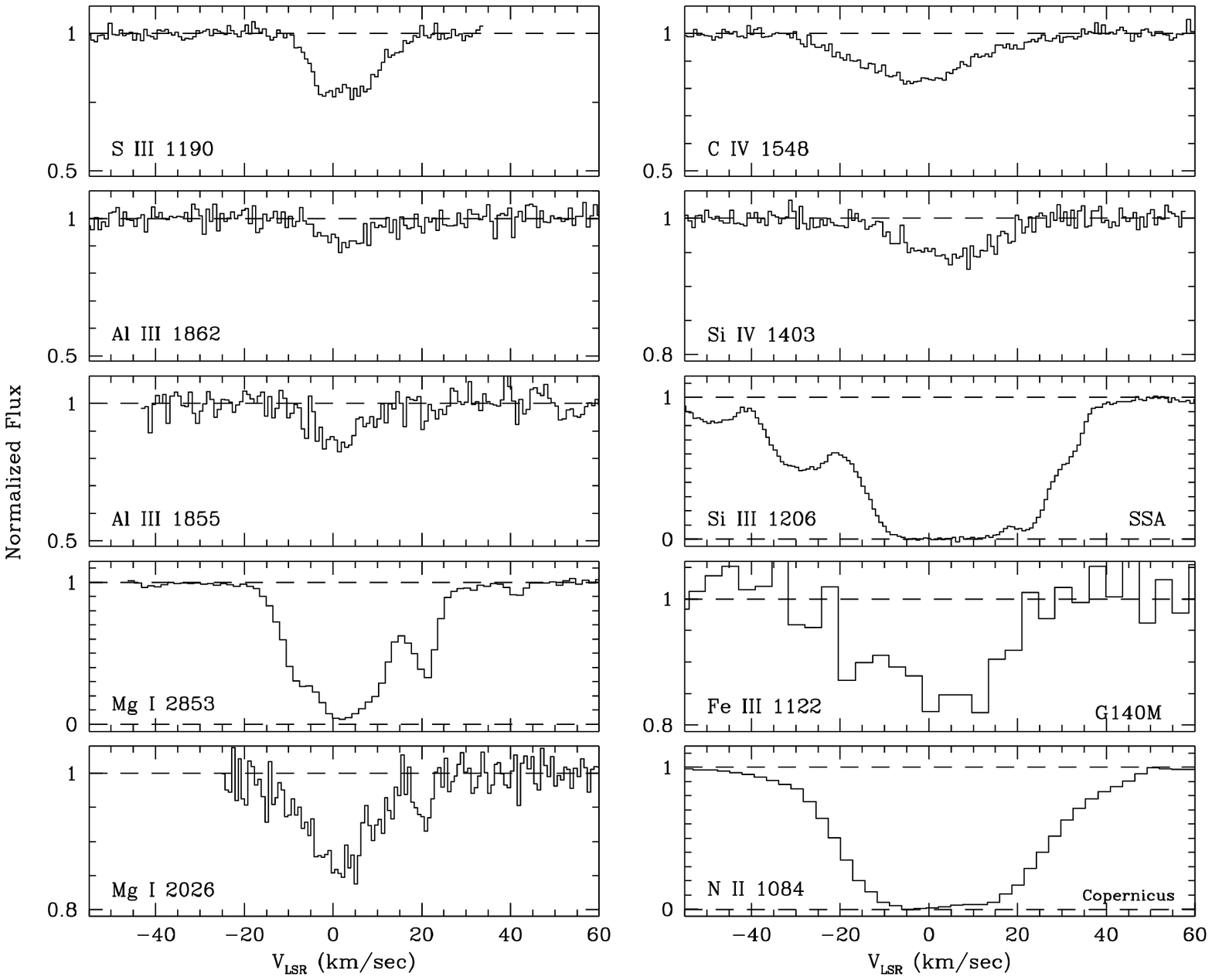}
\figcaption{As Figure \ref{fig:norm1}, but for the non-dominant
species \protect\ion{Mg}{1}, \protect\ion{C}{4}, \protect\ion{Si}{2},
\protect\ion{Si}{3}, \protect\ion{Si}{4}, \protect\ion{Al}{3},
\protect\ion{S}{3}, \protect\ion{N}{2} and \protect\ion{Fe}{3}.  The
\protect\ion{C}{4} and \protect\ion{Si}{4} profiles are from Brandt
\etal\ (1998).  The \protect\ion{Fe}{3} observations were made with
the G140M grating.  The \protect\ion{N}{2} are archival {\em
Copernicus} data and were analyzed by Shull \& York (1977).
\label{fig:highions}}
\end{figure}

\begin{figure}
\epsscale{1.1} 
\plotone{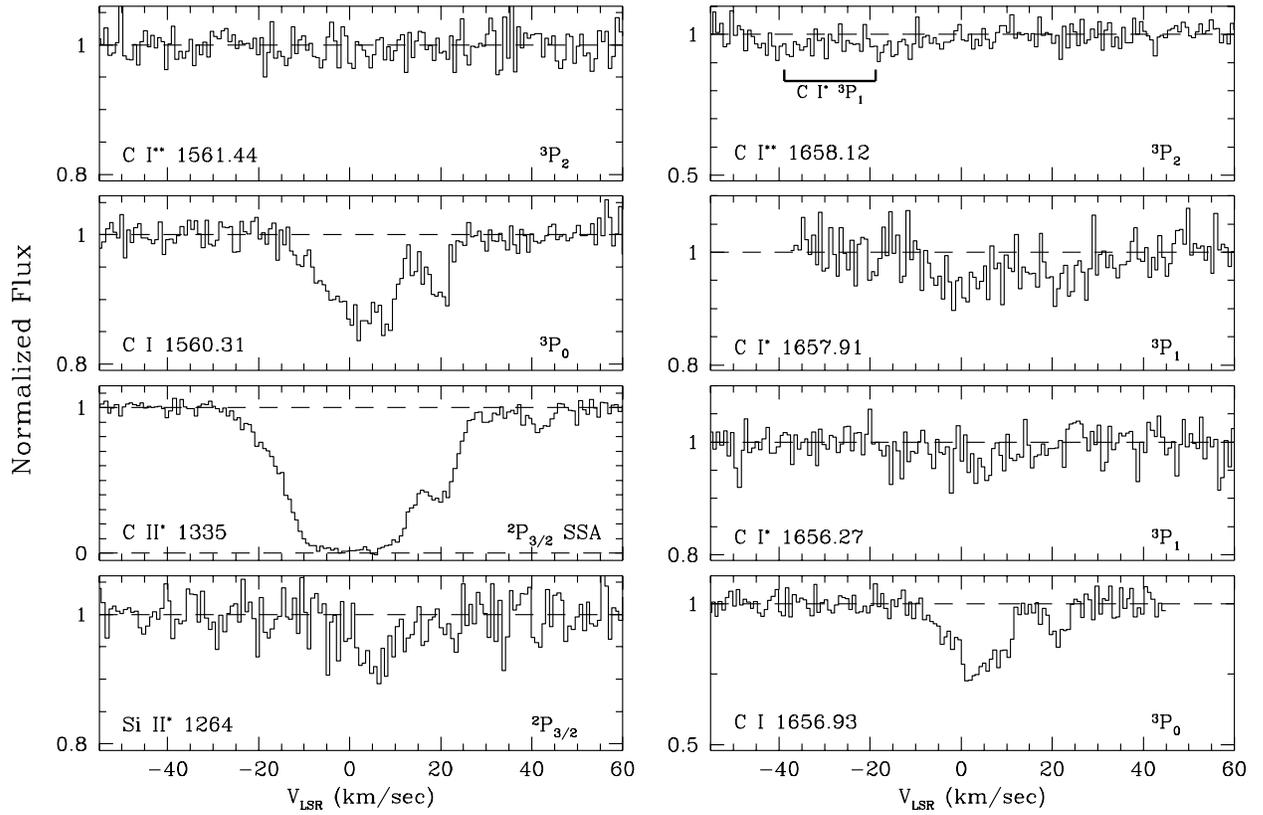}
\figcaption{As Figure \ref{fig:norm1}, but showing transitions from
excited states of \protect\ion{Si}{2}$^*$ \protect\wave{1264},
\protect\ion{C}{2}$^*$ \protect\wave{1335}, and several transitions of
\protect\ion{C}{1}, \protect\ion{C}{1}$^*$, and
\protect\ion{C}{1}$^{**}$.  The term symbols for the lower level of
the transition are given for each species.
\label{fig:excited}}
\end{figure}

\begin{figure}
\epsscale{1.1}
\plotone{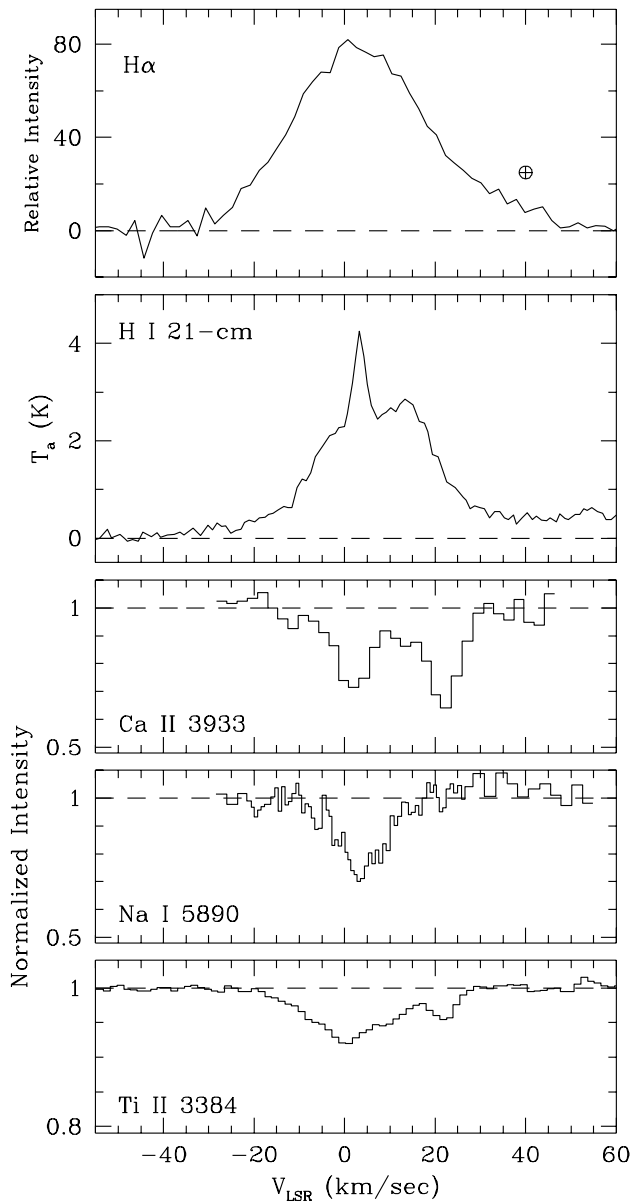}
\figcaption{Ground-based data for the $\mu$ Col sightline.  The top
panel is the \wham\ spectrum of H$\alpha$ towards $\mu$ Col taken with
a 1$^\circ$ beam and 10 km s$^{-1}$ resolution.  An atmospheric OH
line which coincides with component 4 is present at $\mvlsr \approx
+40$ km s$^{-1}$.  The next panel shows the \HI\ profile of Lockman,
Hobbs, and Shull (1986), taken with a 21\arcmin\ beam and 2 km
s$^{-1}$\ resolution.  The next three panels from top to bottom are
the continuum-normalized profiles of \protect\ion{Ca}{2} and
\protect\ion{Na}{1} from Hobbs (1978) and \protect\ion{Ti}{2} from
Welsh \etal\ (1997).  These data have velocity resolutions (FWHM) of
$\Delta v \approx 4.5$, 1.0, and 4.5 km s$^{-1}$, respectively.  The
data for \protect\ion{Na}{1} include contamination from telluric
absorption lines.  Those most likely to be present in the region of
interest for this paper are near $\mvlsr \approx -8$, 5, and 14 km
s$^{-1}$\ (see Hobbs 1978).
\label{fig:ground}}
\end{figure}

Continuum normalized interstellar line profiles for all species
treated in this work are shown in Figures \ref{fig:norm1}$ -
$\ref{fig:excited}.  Each profile was normalized by fitting low-order
($<5$) Legendre polynomials to the local stellar continuum in regions
free from interstellar absorption (Sembach \& Savage 1992).  In
general the continuum of the star, which has a radial velocity $v_{\rm
LSR} = +90$ km~s$^{-1}$ (Keenan \& Dufton 1983) and projected
rotational velocity $v \sin i \approx 111$ km~s$^{-1}$ (Penny 1996),
was well behaved, making the fit to the stellar continuum relatively
certain.  In some cases, however, the interstellar absorption
coincides with stellar lines in a way that makes the continuum
placement more ambiguous.  Examples of such occurences include the
lines \SiII\ \wave{1193}, \ion{Si}{3} \wave{1206}, and \NI\
\wave{1199}.  We have marked lines with less than certain continuum
placement in the tables of data presented herein.  For comparison with
the low-ionization GHRS data presented here, we also include in Figure
\ref{fig:ground} ground based absorption profiles of \ion{Ti}{2} from
Welsh \etal\ (1997) and the \ion{Ca}{2} and \ion{Na}{1} profiles from
Hobbs (1978) as well as the \HI\ emission profile from Lockman, Hobbs,
\& Shull (1986).  Also shown in Figure \ref{fig:ground} is the \wham\
spectrum of H$\alpha$ emission along this sightline (M. Haffner, 1998,
priv. comm.).  The GHRS data for \ion{Si}{4} and \ion{C}{4} from
Brandt \etal\ (1999) are shown in Figure \ref{fig:highions}.

The integrated equivalent widths, $W_\lambda$, are given in Table
\ref{table:eqwidths} for each species, along with the 1$\sigma$ error
estimates (see \S \ref{subsec:aodmethod}).  The range over which the
equivalent width and apparent column density integrations extend for
each absorbing region are given in Table \ref{table:components}.  Also
listed in Table \ref{table:eqwidths} are the ionization potentials of
the measured ionic species and the next lower ionization state of the
element, vacuum wavelengths, adopted values of the oscillator
strengths for each transition and the studies from which we have drawn
these values, and the empirically estimated signal-to-noise ratios.
Oscillator strengths used in our analysis are generally taken from the
compilation of Morton (1991), using the recommended updates listed in
Table 2 of Savage \& Sembach (1996a) with a few updates for new
determinations detailed below.

For the \ion{Ni}{2} and the weak \MgII\ transitions recent
determinations suggest oscillator strength revisions by factors of
$\sim2$.  We choose to adopt the recent empirical determination of the
oscillator strengths for \MgII\ \twowave{1239 and 1240} by Fitzpatrick
(1997).  These $f$-values are determined from a comparison of the
strong \MgII\ transitions near \wave{2800} with those near \wave{1240}
in high signal to noise \ghrs\ observations.  Fitzpatrick's
recommended $f$-values are a factor of $\sim2.4$ larger than the
Hibbert \etal\ (1983) theoretical calculation and a factor of $\sim2$
smaller than the emperical determination of Sofia, Cardelli, \& Savage
(1994).

The \ion{Ni}{2} oscillator strengths are derived from a combination of
the Fedchak \& Lawler (1999) and Zsarg\'{o} \& Federman (1998)
results.  Zsarg\'{o} \& Federman (1998) have placed many of the
\ion{Ni}{2} $f$-values on a consistent relative scale using \ghrs\
observations of several stars.  Their compilation includes all of the
transitions we observe with the exception of the \wave{1317} line.  We
find no significant evidence in our data that the ratios of the
\twowave{1317 and 1370} $f$-values from Morton (1991) should be
modified, though the signal-to-noise ratio for the latter line is less
than ideal for this type of study.  Fadchek \& Lawler (1999) have very
recently provided absolute laboratory measurements of the oscillator
strengths of a number of vacuum ultraviolet \ion{Ni}{2} transitions,
including the \twowave{1709 and 1741} transitions observed in this
work.  The ratios of the absolute $f$-values derived by these authors
for the \twowave{1709 and 1741} transitions are in excellent agreement
with the results of Zsarg\'{o} \& Federman.  Fedchak \& Lawler suggest
using the $f$-values derived by Zsarg\'{o} \& Federman multiplied by a
scale factor of $0.534\pm0.05$.  We adopt this recommendation in this
work, using values $\log \lambda f$ that are $-0.272$ dex below the
values suggested by Zsarg\'{o} \& Federman.  We note however that
there is a discrepancy between the implied column densities derived
from the \wave{1741} transition and the \twowave{1317 and 1370}
transitions.  The measurements of the \wave{1741} transition in our
dataset are based on a single Ech-B exposure with $S/N \sim 21$.  The
possibility exists that a \fpn\ feature is present in these data;
however, the velocity structure appears similar to the weak
\ion{Cr}{2} and \ion{Fe}{2} lines with comparable signal-to-noise
ratios.  We believe that there may still be uncertainties in the
relative oscillator strengths between the lines longward and shortward
of 1700 \AA\ in our dataset.  The change in oscillator strengths
suggested by Fedchak \& Lawler (1999) and Zsarg\'{o} \& Federman
(1998) not only has implications for the dust content of diffuse
interstellar clouds in the Milky Way but also, and perhaps more
importantly, clouds the interpretation of [Ni/Fe] measurements in
high-redshift damped \lya\ systems (e.g., Lu \etal\ 1996; Prochaska \&
Wolfe 1999; Kulkarni, Fall, \& Truran 1997).  We discuss the
implications of the new \ion{Ni}{2} oscillator strengths in more
detail in \S \ref{sec:discussion}.

Our estimation of the errors inherent to our measurements of the
integrated equivalent widths and apparent column densities (see below)
includes contributions from photon statistics, continuum placement
uncertainties, and zero-level uncertainties (Appendix A of Sembach \&
Savage 1992).  We have adopted a 2\% zero-level uncertainty
throughout.  Though this may overestimate the errors in regions near
heavily saturated lines, we feel it is appropriate given the uncertain
scattered-light properties of the LSA.  Continuum placement
uncertainties were estimated based upon the effects of adjusting the
continuum level by $\pm0.4$ times the rms noise about the fit.  These
sources of error are independent and have been added in quadrature to
produce the final error estimate quoted with our measurements.

The sources of error discussed above make no allowance for the
existence of \fpn\ features in our data.  The strength of \fpn\
features is reduced significantly by co-adding spectra that have been
shifted along the diode array from one another, so that features
constant in diode-space are shifted in wavelength-space.  Even given
this improvement, \fpn\ features are found in our reduced data, often
mimicking weak interstellar absorption lines.  For example a \fpn\
feature is present at $v_{\rm LSR} \approx +41$ km~s$^{-1}$ in our
data for the \SII\ \wave{1253} line (see Figure \ref{fig:norm1}),
almost exactly coincident with the expected absorption from component
4.  In this case, however, we can identify it as a \fpn\ feature
because it appears in only one of the two co-added observations of
this wavelength region.  For those wavelength regions covered by only
one observation, our ability to discriminate between true interstellar
absorption and weak \fpn\ features becomes less robust.  In these
cases we are aided by the excellent wavelength coverage of the current
\mucol\ dataset: we are often able to check the reality of absorption
features in many different transitions of the same ionic species.

The nominal short-wavelength limit of the GHRS is 1150 \AA, given the
inefficiency of the magnesium fluoride coatings of the \hst\ optics at
wavelengths less than 1150 \AA.  However, the short wavelength Digicon
detector has a LiF window, and observations shortward of this are
possible.  We include in our data measurements of the \ion{Fe}{3}
\wave{1122.526} transition as well as the \NI\ triplet at \wave{1134},
and transitions of \FeII\ at \twowave{1133, 1143, and 1145}.
Unfortunately, two of the ground-state transitions of neutral carbon
are nearly coincident with the \ion{Fe}{3} transition, at
\twowave{1122.518 and 1122.438}, which lie at $-2.1$ and $-23.5$
km~s$^{-1}$ relative to the \ion{Fe}{3} velocity zero point,
respectively.  We do not believe the \ion{C}{1} contamination of
\ion{Fe}{3} is a significant problem for the following reason.  We
detect the much stronger \CI\ line at \wave{1560.309} with an
integrated equivalent width of $W_\lambda = 14.7\pm0.9$ m\AA.  This is
equivalent to a combined equivalent width from the \CI\ transitions at
\twowave{1122.518 and 1122.438} of $W_\lambda = 0.66\pm0.04$ m\AA,
implying the \CI\ transitions make a negligible contribution to the
\ion{Fe}{3} measurement ($W_\lambda = 19\pm4$ m\AA).

	\subsection{Analysis Methods}

In theory one can separate an observed interstellar absorption profile
into individual absorbing clouds along the line of sight, if the data
fully resolve these individual entities.  By fitting models for the
absorption from each cloud, or component, one can determine the column
densities, central velocities, and Doppler parameters for each
absorbing cloud along a given line of sight.  An excellent example of
applying this approach to study the abundances in the diffuse ISM is
the work of Spitzer \& Fitzpatrick (1993), who used the \ghrs\ to
study the abundances and physical conditions towards HD 93521.  In
practice one can run into significant uncertainties with this
approach, particularly for clouds closely-spaced in velocity.  An
example of a region where the component fitting techniques become
difficult, leading to a lack of uniqueness, is the principal absorbing
region along the line of sight towards \mucol\ (component 1; $-17.0
\lesssim v_{\rm LSR} \lesssim 15.5$ km~s$^{-1}$).  The \ion{Fe}{2} and
\ion{Mg}{1} profiles suggest that there are multiple blended
components in this velocity range.  However, constraining a fit to the
data for the less depleted species of \SII\ and \SiII\ is more
difficult.

To derive accurate column densities for the components along the line
of sight to \mucol, we will apply the component fitting technique for
the higher velocity gas (components 2--5) but will primarily rely upon
the apparent column density, \nav, method described by Savage \&
Sembach (1991) for dealing with the central low-velocity blend we have
designated component 1.

	\subsubsection{Apparent Column Density Method}
	\label{subsec:aodmethod}

In analyzing the low-velocity components towards \mucol, we will make
use of the so-called apparent column density, or \nav, method, which
gives information on the velocity structure of the absorbing material
that is model-independent (Savage \& Sembach 1991).  In short, a
continuum normalized absorption profile $I(v) \equiv e^{-\tau_a (v)}$,
for a transition having wavelength, $\lambda$, and an oscillator
strength, $f$, is related to the apparent column density per unit
velocity, \nav, by
\begin{equation}
N_a (v) = \frac{m_e c}{\pi e^2} \frac{\tau_a (v)}{f \lambda}
= 3.768 \times 10^{14} \frac{\tau_a (v)}{f \lambda({\rm \AA})},
\end{equation}
in units ${\rm atoms \ cm^{-2} \ (km \ s^{-1})^{-1}}$, where $\lambda$
is given in \AA.  In the absence of {\em unresolved} saturated
structure, which can be identified by comparing \nav-profiles for
different transitions of the same species, Savage \& Sembach (1991)
have shown this method provides a valid, instrumentally blurred
representation of the true column density as a function of velocity,
$N(v)$.

\begin{figure}
\epsscale{1.1} 
\plotone{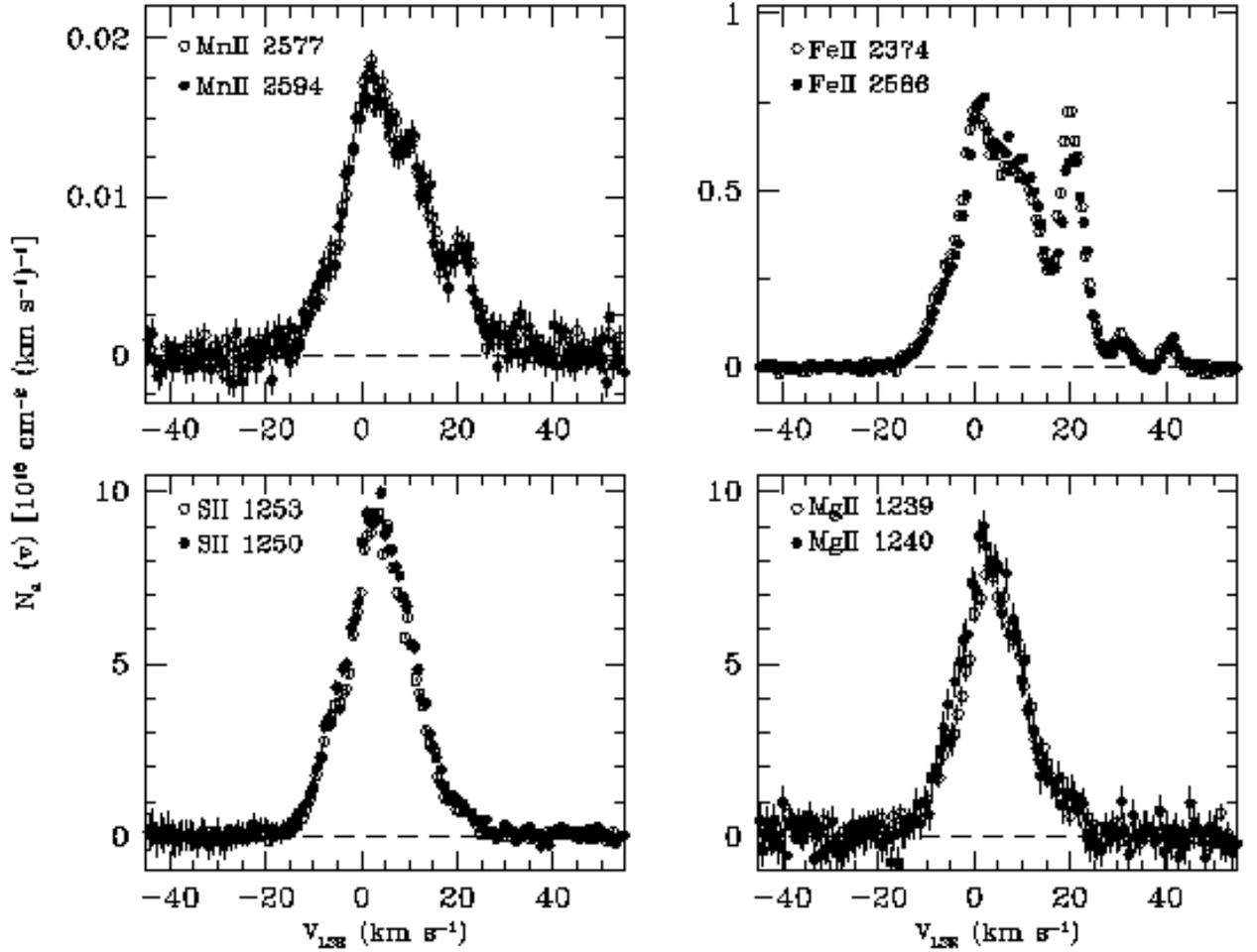}
\figcaption{Representative plots of the $N_a (v)$ profiles for the
ions \protect\ion{Mn}{2}, \protect\ion{Fe}{2}, \protect\ion{S}{2}, and
\protect\ion{Mg}{2}.  For most of the absorbing components, these
profiles exhibit little in the way of unresolved saturated structure.
The $N_a (v)$ profiles of \protect\ion{Fe}{2} are discrepant near
$\mvlsr \approx +20$ km s$^{-1}$\ (component 2), suggesting the
presence of unresolved saturation in these transitions.
\label{fig:navprofiles}}
\end{figure}

Examples of \nav\ profiles for the (presumably) non-depleted species
\SII, the moderately-depleted \MgII\ and \ion{Mn}{2}, and the
highly-depleted \ion{Fe}{2}, are shown in Figure
\ref{fig:navprofiles}.  For each of these ionic species two
transitions with different oscillator strengths are plotted.  One can
see that the examples we have chosen, in general, show good agreement
between the two transitions.  An example where our profiles exhibit
unresolved saturated structure is seen in the \ion{Fe}{2} profiles.
The \nav\ values near $v_{\rm LSR} \approx +21$ km~s$^{-1}$ are lower
in the stronger \wave{2586} line than the weaker \wave{2374} line.
This is evidence for unresolved saturated structure in component 2
within these \ion{Fe}{2} lines.  For
\ion{Fe}{2}, however, it is still possible to accurately derive $N(v)$
with transitions weaker than the \wave{2586} transition.

Table \ref{table:coldens} contains the velocity-integrated apparent
column densities (Savage \& Sembach 1991) and estimated errors for the
transitions studied in this work.  The sources of errors for these
integrations were taken to be the same as those described above for
deriving the integrated equivalent widths.  We have not included the
uncertainties in the $f$-values in this error budget.  In the absence
of unresolved saturated structure, these column densities are
equivalent to the true column densities in the velocity ranges
outlined in Table \ref{table:components}.  The column densities are
representative of the individual components considered here if there
exists no significant blending between the components.

\subsubsection{Component Fitting}
	\label{subsec:compfitting}

The column densities derived through the \nav\ method above may be
subject to large uncertainties in cases where the individual absorbing
regions overlap significantly in velocity.  In particular component 2
may be heavily contaminated by overlap with absorption due to
component 1 in lightly depleted species (cf., Figure \ref{fig:norm1}).
To more cleanly separate the higher velocity gas seen in components 2,
3, 4, and 5 we use component fitting techniques to derive the
interstellar column densities.  

The component fitting approach begins with a model of the interstellar
absorption spectrum, which consists of $k$ individual components
described by their central velocities, $v_k$, Doppler spread
parameters, $b_k$, and column densities, $N_k$.  The individual model
components are assumed to be well approximated by a Voigt profile.
This model is then convolved with an appropriate instrumental line
spread function (LSF), and the value of $\chi^2$ minimized between
this blurred model and the observed line profile to determine the
best-fit parameters.  Our component fitting analysis makes use of
software kindly provided by E. Fitzpatrick (1998, priv. comm.) and
described in Spitzer \& Fitzpatrick (1993) and Fitzpatrick \& Spitzer
(1997).

Where they exist, we have used observations taken through the SSA of
the GHRS in our component fitting analysis.  While these data often
have lower signal to noise, the LSF of the SSA has been carefully
studied by Spitzer \& Fitzpatrick (1993) and has a slightly better
resolution than the LSF of the LSA.  However, most of the data
presented here were taken through the GHRS LSA.  The LSF for the GHRS
LSA has not been well characterized.  Robinson \etal\ (1998) present a
LSF for the post-COSTAR LSA; however, we found this LSF inadequately
matched the results derived from our fitting of SSA profiles and was
not able to fit narrow deep lines correctly (e.g., component 4).  In
Appendix \ref{appendix:lsf} we derive a new LSF for the post-COSTAR
GHRS LSA.  The new LSF is a sum of a strong narrow Gaussian and a weak
broad Gaussian.  The narrow component has a FWHM of 1.09 diodes, while
the weak component has a FWHM of $\sim4.21$ diodes with a peak
approximately 4.5\% that of the narrow component at \wave{2800}.
Therefore the broad weak component contains 15.1\% of the spread
function area.  This fraction is a function of wavelength, however.
We discuss this LSF in more detail in Appendix \ref{appendix:lsf}.

The best fit $b$-values and column densities from our component
fitting analysis are given in Table \ref{table:compfitting}.  The best
fit central velocities are given in Table \ref{table:compvelocities}.
We fit the blend making up component 1 with three components having
approximate central velocities $\langle v_{\rm LSR} \rangle \approx
-8$, 1, and 7 \kms, though there is some variation in the best fit
values.  Due to the lack of uniqueness in this central blend, we do
not report here the results for these individual components, but only
the sum of their column densities in Table \ref{table:compfitting}.
Our purpose in fitting component 1 was not to disentangle this blend,
but to approximately account for the overlap of this region with the
more distinct component 2.

Where several lines exist for a given ionic species, we have fit all
of the profiles simultaneously.  For several species it was necessary
to adopt $b$-values derived from fits to other ions.  This
approximation was necessary when either the signal-to-noise ratio of
the spectra were not high enough for the fitting to be reliable (e.g.,
\ion{Cr}{2} or \ion{Ni}{2}) or when the profiles of the components
were not distinct enough to provide appropriate information to
constrain the fit (e.g,. for component 2 in \ion{P}{2} and
\ion{Zn}{2}).  In these cases we have adopted relatively
well-constrained $b$-values from lines of similar atomic mass.  To
assess the error contribution of the adopted $b$-values to our derived
column densities, we have also calculated the best fit models for
$b\pm1 \sigma$.  The differences between the $b\pm1 \sigma$ results
and the best fit $b$-value results were added in quadrature to the
formal fitting error.  For the \ion{S}{2} profile we have fixed the
central velocity of component 2 to that derived for \ion{Si}{2}.  This
was done because the unconstrained fit yielded velocity structure and
relative column densities in the central blend that were significantly
different than that of any other ion, and these differences impacted
the fitted component 2 parameters.  We found, however, that by holding
the velocity of this component we were able to obtain a fit in good
agreement with our results for the other ions.

In general the component fitting results for component 1 agree with
the \Nav\ integrations, suggesting little in the way of unresolved
saturated structure or confusion from component overlap.  We find the
column densities derived for component 2 in lightly depleted species
are typically $\sim 0.2$ dex lower than the results derived from
integrating the \Nav\ profiles.  This is a result of the overlap from
wings of components in the low-velocity blend that are included in the
\Nav\ integration.  We see that component 4 may be significantly
saturated in a number of profiles by comparing the component fitting
results with the column densities derived from a straight integration
of the \Nav\ profiles (e.g., \ion{Mg}{2} and \ion{O}{1}).  The
$b$-values derived from our component fitting analysis are $b \leq
2.4$ \kms\ (FWHM$\, \la 4.0$ \kms).  Thus it is important to use the
component fitting results for this cloud.

\subsection{Adopted Column Densities}

To derive the best column densities for component 1, we take the
weighted average of all the transitions for a given species that show
no evidence for unresolved saturated structure in their \nav\
profiles.  We present our adopted final column densities in Table
\ref{table:finalcoldens}, noting where we have chosen not to use an
observed transition in deriving these column densities.  We are
relatively certain that the column densities in Table
\ref{table:finalcoldens} for the low-velocity absorbing components
toward \mucol\ are not significantly affected by saturation effects.
The individual clouds making up the blend of component 1 are
relatively broad, and the absorption profile seems to be fully
resolved by the echelle-mode resolution of 3.5 km~s$^{-1}$.

In a few cases we have chosen to use the component fitting results for
component 1.  These cases have been marked in Table
\ref{table:finalcoldens} and include \ion{Ni}{2}, \FeII, and \ZnII.
The \ZnII\ observations show a systematic offset of 0.08 dex between
the integrated column densities of the two transitions.  This cannot
be due to saturation effects since the stronger of the two lines gives
a higher apparent column density, exactly the opposite of the expected
behavior in the presence of saturation.  This behavior may be caused
by uncertainties in the oscillator strengths of the transitions.  We
have chosen to adopt the component fitting results for \ZnII\ because
we believe the column densities are a better compromise between the
two profiles than the \Nav\ values and the formal errors are more
representative of the true errors.

The final adopted column densities for components $2-5$ are from our
component fitting results.  Most of the values given in Table
\ref{table:finalcoldens} are the result of simultaneously fitting all
the available transitions of a given species, with exceptions noted in
the table.  Table \ref{table:finalcoldens} also includes our
derivation of the \HI\ column density along this sightline, which is
described in Appendix \ref{appendix:lya}.

\section{ABUNDANCES IN THE LOW VELOCITY GAS}
\label{sec:lowvelocity}

In this section we discuss the observed abundances in the low-velocity
absorbing regions (components 1 and 2) along the line of sight to
\mucol\ and the implications of these abundances for interstellar
dust.  This velocity range (from $\mvlsr \approx -17$ to $+29$ km~s$^{-1}$)
not only contains the vast majority of the warm neutral absorbing
column but also material associated with ionized gas along the path to
\mucol.  It is necessary to examine the degree to which material in
primarily ionized gas may affect the derivation of relative abundances
in the neutral material along this sightline.  We discuss this
contamination and our assessment of it in \S \ref{subsec:ionization}.

Throughout this paper we will be discussing the normalized gas-phase
abundance of elemental species.  We define the normalized gas-phase
elemental abundance of a species $X$ relative to $Y$ as a function of
velocity to be
\begin{equation}
  [X/Y]_v  \equiv \log \{N_a(v)_X/N_a(v)_Y \} 
	- \log \{X/Y\}_\odot, 
\label{eqn:abundance}
\end{equation}
%
where \Nav$_X$ and \Nav$_Y$ are the apparent column density per unit
velocity for the elements $X$ and $Y$, respectively, and the quantity
$\{X/Y\}_\odot$ is the solar or cosmic reference abundance ratio of
the two species $X$ and $Y$ (e.g., the meteoritic abundances from
Anders \& Grevesse 1989).  We will use $[X/Y]$ to denote the
equivalent of Equation (\ref{eqn:abundance}) when one uses
velocity-integrated total column densities in place of \Nav$_X$ and
\Nav$_Y$.  This nomenclature is equivalent to others' definition of
the logarithmic depletion, $D(X)$ (e.g., SSC; Spitzer \& Fitzpatrick
1993).  When deriving gas-phase abundances it is typically assumed
$[X^+/Y^+] \approx [X/Y]$, if $X^+$ and $Y^+$ are the dominant stages
of ionization in the warm neutral medium.  For the most part this
assumption is justified, though the effects of ionized gas along the
line of sight may modify the ions in a different manner, causing
this assumption to break down.  We will show in \S
\ref{subsec:ionization} that this is not a significant effect for
component 1.  However, for components $3-5$ this assumption will not
be appropriate (see \S \ref{sec:highvelocity}).

\subsection{Ionization Effects}
	\label{subsec:ionization}

The presence of absorption due to the ions \ion{S}{3}, \ion{Si}{3},
\ion{Si}{4}, \ion{Al}{3}, and \ion{Fe}{3} in our \ghrs\ spectra and
the \copernicus\ observations of \ion{N}{2} strongly suggest the
presence of ionized hydrogen (H$^+$) along the line of sight to
\mucol.  The \wham\ observations of this region imply the presence of
ionized gas in this direction at velocities compatable with those of
the ionized tracers observed by \ghrs, though much of the emission may
come from beyond the star.  For regions primarily containing H$^{\rm
o}$, an element $X$ whose first ionization potential falls below that
of hydrogen is predominantly found in its singly ionized form, $X^+$.
Thus measurements of $X^+/{\rm H^o}$ or $X^+ / Y^+$ are generally good
indicators of the gas-phase abundance of the element $X$ in the
neutral material.  However, the inferred presence of H$^+$ along the
sightline to \mucol\ complicates this simple picture since the
relative contributions of $X^+$ and $X^{+2}$ may be different for each
element in the H$^+$-containing region and will be dependent upon the
ionization structure of the region.  Therefore, it is important to
investigate the effects of H$^+$-containing regions along the line of
sight on our derived gas-phase abundances.

\mucol\ is an \ion{O9.5}{5} star with $T_{eff} \approx 33,000$ K
(Howarth \& Prinja 1989)\footnote{There are varying determinations of
the stellar effective temperature.  Keenan \& Dufton (1983) have
published $T_{eff} \approx 31,400$ K, while the temperature scale of
Vacca \etal\ (1996) suggests $T_{eff} \approx 34,600$ K is
appropriate.  We will adopt the intermediate Howarth \& Prinja (1989)
value.  Changes over this range of temperatures do not significantly
affect the results of this section.}  and is therefore hot enough to
ionize its immediate surroundings.  A SIMBAD search of the area within
5$^\circ$ of \mucol\ reveals two \ion{B2.5}{4} stars along the
sightline to \mucol: $\gamma$ Col and HD 41534.  These stars, with
$T_{eff} \sim 20,000$ K, lie $\ga$20 pc from the sightline to the star
and should contribute very little to the ionized column along the
sightline compared with \mucol\ itself.  Within 10$^\circ$ of the star
there are a total of seven B stars of types B2.5 or later.  Given the
lack of any O-type stars within 300 pc of the \mucol\ sightline (SY)
and the late spectral types of the B-type stars found near the
sightline, we will continue under the assumption that the majority of
the warm ionized gas that occurs along this sightline is contained in
a photoionized nebula around \mucol.  The column of \HII\ along the
line of sight can be predicted using
\begin{eqnarray}
N({\rm H}^+) & = & N({\rm H^o}+{\rm H}^+) - N({\rm H^o}) \\
             & \approx &
 N({\rm S}^+) \cdot \{{\rm H/S}\}_\odot - N({\rm H^o}). \nonumber
\label{eqn:calchii}
\end{eqnarray}
Using this approach SY predict log $N({\rm H}^+) = 19.94$.  Using our
accurate determination of \NHI\ towards \mucol\ (Appendix
\ref{appendix:lya}), coupled with our high-quality S$^+$ measurements,
we predict $\log N({\rm H}^+) = 19.17\pm0.14$.  This estimate relies
on the key assumption that the total sulfur to hydrogen abundance
along this sightline is solar.  The \wham\ spectra and the
\ghrs\ observations of \ion{S}{3}, \ion{Al}{3} and \ion{Si}{2}$^*$
(see below) suggest that ionized gas contamination is most significant
for component 1.

To more reliably assess the contribution of material associated with
an ionized nebula about \mucol\ to our absorption line measurements,
we have used the photoionization equilibrium code \cloudy\ (v90.04;
Ferland 1996, Ferland \etal\ 1998) to model the ionization and
temperature structure of such an \HII\ region.  To estimate the
stellar spectrum from \mucol\ for use in our photoionization models,
we adopt an ATLAS line-blanketed model atmosphere (Kurucz 1991) with
the stellar parameters $T_{eff} = 33,000$, $\log (g) = 4.0$ and $L_* =
2.6\times10^4 \ L_\odot \ = 1 \times 10^{38}$ ergs s$^{-1}$, close to
the values estimated by Howarth \& Prinja (1989) for
\mucol.\footnote{As noted above, some of the fundamental properties of this
star have a range of published values.  We continue to adopt the
intermediate Howarth \& Prinja (1989) results, but see also the
results of Keenan \& Dufton (1983) and the relationships with spectral
types given in Vacca \etal\ (1996).}  For a range of densities and
filling factors of the ambient ISM, we have calculated the ionization
and temperature structure of model nebulae from a distance 0.03 pc
from the star to the point where the electron density, $n_e$, falls to
$5\%$ of the total hydrogen density, $n_{\rm H}$.  We have used solar
abundances throughout.\footnote{We have run models with Orion nebula
abundances and abundances appropriate for warm disk gas to assess the
effects of the different abundances on the temperature and ionization
structure of the nebula.  We have also included opacity due to dust
grains to test the robustness of our results.  The derived results
from these models are completely consistent with our approach.}  The
ambient densities used in the models presented here are $n_{\rm H}=
0.02, 0.05$, 0.2, and 0.5 cm$^{-3}$.  The densities 0.05 and 0.2
cm$^{-3}$ are approximately the average line of sight density, and the
estimated density of the \HII\ region gas from the excited states of
\ion{N}{2} (SY) and \SiII\ (see
\S \ref{subsec:physicalconditions}). The highest value was used to
study regions more dense than the limit of $n_e \gtrsim 0.2$
cm$^{-3}$, while the lowest is used to show the relative constancy of
the results.  The first three models are those presented in Brandt
\etal\ (1999) to discuss the source of the \ion{Si}{4} absorption
along the line of sight to \mucol.  Similar models have also been
presented by Howk \& Savage (1999) to derive the gas-phase abundance
of [Al/S] in the ionized medium of the Galaxy.  

In order to match the observational constraints, we assume all of the
column density of \ion{S}{3} arises in the photoionized region about
the star.  For the densities considered, we vary the volume filling
factor of the material until a match to the observed \ion{S}{3} column
density of $\log N(\mbox{\ion{S}{3}}) = 13.82$ is obtained.  Table
\ref{table:cloudy} contains the physical parameters for each of our
models and the predicted column densities of important ionic species.
The predicted column densities of H$^+$ for the models given in Table
\ref{table:cloudy} are in the range $\log N(\mbox{\HII}) = 19.01 -
19.10$, which is in rough agreement with the \HII\ column density
predicted by scaling our \SII\ column density according to
Equation (\ref{eqn:calchii}).  The \HII\ column densities from our
\cloudy\ models are mildly sensitive to where the model nebulae are
truncated. 

Among the most important conclusions we have drawn from our modelling
of the \mucol\ \HII\ region is that the ratio of $N({\rm Al}^{+2}) /
N({\rm S}^{+2})$ predicted to occur in an \HII\ region is relatively
insensitive to the model assumptions.  Howk \& Savage (1999) have used
similar models and observed values of $N({\rm Al}^{+2})$ and $N({\rm
S}^{+2})$ to determine the normalized gas-phase abundance of Al to S
in the \HII\ region surrounding \mucol.  Their model-corrected result,
\begin{displaymath}
[{\rm Al/S}]_{{\rm H \, II}}  = -0.78 \pm0.08 ,
\end{displaymath}
is insensitive to the adopted density or filling factor of the nebula,
and the ionizing photon flux of the star.  Further, the sensitivity to
the shape of the ionizing spectrum (i.e., to $T_{eff}$), is also
relatively small.  The error estimate comes from the standard
deviation about the mean predicted value of $N({\rm Al}^{+2}) / N({\rm
S}^{+2})$ for models using input spectra in the range $27,000 \leq
T_{eff} \leq 39,000$ K (see Howk \& Savage 1999).

A similar calculation can be made regarding the gas-phase abundance of
Fe relative to S in the \HII\ region about \mucol.  Howk \& Savage
(1999) have shown that values of $N({\rm Fe}^{+2}) / N({\rm S}^{+2})$
predicted by the models are much more sensitive to uncertainties in
the input effective temperature of the ionizing source.  Using
observations of $N({\rm Fe}^{+2})$ and $N({\rm S}^{+2})$ and \cloudy\
models, they determine
\begin{displaymath}
[{\rm Fe/S}]_{{\rm H \, II}}
	= -0.87 \pm 0.21
\end{displaymath}
in the \mucol\ \HII\ region.

The sub-solar abundances of Al and Fe in the \mucol\ \HII\ region
likely implies the existence of dust in the ionized ISM about this
star.  An Al depletion of 0.8 dex is similar to that found for
refractory elements along warm disk+halo sightlines (Sembach \& Savage
1996).  The similar depletion of Fe is also consistent with warm
disk+halo sightlines.

For the undepleted ions \SII\ and \ion{P}{2}, which are the dominant
ionization stages in the WNM, our models predict absorption from
material in the \HII\ region may account for $\sim10$\% of the total
observed absorbing column density.  Thus the presence of a low-density
\HII\ region surrounding \mucol\ may add a systematic contribution of
$\approx 0.04$ to 0.05 dex to the column densities of undepleted
species in component 1.  This is a relatively small contribution.  We
have chosen not to report the photoionization model results for the
undepleted species \ion{Zn}{2} given the large uncertainties in the
adopted atomic parameters, particularly the recombination coefficients
and ionization cross sections, for this element.

Assessing the impact of the \HII\ region to the measurements of highly
depleted species is more complicated.  We have assumed solar
abundances for our model \HII\ region.  Given our analysis of the
\ion{Al}{3} and \ion{Fe}{3} absorption above, using solar abundance
models will over-estimate the contribution of the \HII\ region to the
total line of sight absorption measurements for depleted species.
This can easily be seen in the results tabulated for \ion{Fe}{2} in
Table \ref{table:cloudy}.  In some cases we predict more \ion{Fe}{2}
absorption than is actually measured for component 1.  Because there
is evidence for sub-solar abundances in the measured ratios $N({\rm
Fe}^{+2}) / N({\rm S}^{+2})$ and $N({\rm Al}^{+2}) / N({\rm S}^{+2})$,
we have confidence that dust exists in the ionized gas about \mucol.

The estimated abundance [Fe/S] in the \HII\ region is roughly
consistent with that of warm disk+halo sightlines.  We can use this
abundance to estimate the corresponding abundances of the elements Mg,
Si, and Mn in the \HII\ region.  The compiled logarithmic depletions
(equivalent to our normalized gas-phase abundances) of these elements
are tabulated in Savage \& Sembach (1996) for such sightlines to be
\begin{displaymath}
	[ {\rm Mg / H}]    \approx -0.31; \ 
	[{\rm Si / H}] \approx -0.25; 
 \ {\rm and} \ [{\rm Mn / H}]  \approx -0.66,
\end{displaymath}
where we have corrected the value of [Mg/H] for our adopted
$f$-values.  If we apply these depletions to our model \HII\ region
results, the relative contribution of \HII\ region gas to the measured
total column densities of these species is approximately
\begin{displaymath}
\mbox{\ion{Mg}{2}}: 0.05 \ {\rm dex}; \ 
\mbox{\ion{Si}{2}}: 0.06 \ {\rm dex}; \ 
\mbox{\ion{Mn}{2}}: 0.09 \ {\rm dex}; \ {\rm and} \ 
\mbox{\ion{Fe}{2}}: 0.06 \ {\rm dex},
\end{displaymath}
where we have assumed [Fe/H]$ = -0.87$.  Thus, material associated
with the \HII\ region may provide a relatively small, though not
insignificant, contribution (uncertainty) to the measured column
densities.  This contribution is not large enough to hinder our major
conclusions regarding the gas-phase abundances of the primarily
neutral medium.  We will assume that the contribution of \HII\ region
gas to the measurements of \ion{Zn}{2} is similar to that of the
undepleted elements \SII\ and \ion{P}{2}.  For \ion{Cr}{2} and
\ion{Ni}{2} we find results similar to those of \ion{Fe}{2} and \MnII,
respectively.  It is important to point out that the expected
contribution of an ionized nebula to the depleted and non-depleted
species are very similar.  Thus while the contributions to the \SII\
and \FeII\ column densities from the ionized nebula are of order 0.05
to 0.06 dex, the ratio $N(\mbox{\FeII})/N(\mbox{\SII})$ is only
changed by $\sim 0.01$ dex.  However, when comparing singly-ionized
species to neutral hydrogen, this contribution is more significant.

The aforementioned uncertainties in the fundamental stellar parameters
(temperature, luminosity, and surface gravity) do not seriously affect
our estimates of contamination from gas in an \HII\ region around
\mucol.  The temperature and surface gravity of the star change the
shape of the spectrum, but over the range of allowable values, we have
found varying these parameters changes little in our models.  Changing
the luminosity does not affect the results at all.  In this case the
important parameter is the ``ionization parameter'' (see Howk
\& Savage 1999), which can be made made constant with varying
luminosity by simply adjusting the ambient density and/or filling
factor.

It is important to constrain the ionization fraction of the primarily
neutral gas along the sightline.  There may be partially ionized gas
in the \HI-containing regions along the sightline that do not
contribute to the \ion{S}{3} used to constrain our \HII\ region
models, but which may contribute (particularly) to the \SII\ column
density along this sightline.  We can roughly estimate the ionization
fraction $x_e \equiv n_e / n_{\rm H}$ of the neutral regions using the
\ion{Ar}{1} measurements of SY with the work of Sofia \& Jenkins
(1998; hereafter SJ).  Sofia \& Jenkins have suggested that the
intrinsic Ar/H abundance in the ISM is close to to the solar system
abundance [following SJ we adopt $12.0 + \log ({\rm Ar/H})_\odot =
6.52$], and that sub-solar \ion{Ar}{1}/\HI\ measurements reflect the
over-ionization of \ion{Ar}{1} relative to \HI.  This over-ionization
of \ion{Ar}{1} is a result of its very large ionization cross section
relative to that of \ion{H}{1}.  Sofia \& Jenkins develop a formalism
to relate the observed discrepancy, $[{\rm Ar/H}]$, between the
observed and expected Ar abundance to the ionization fraction, $x_e$,
along an interstellar sightline (see their Equation 11).  Their
treatment incorporates the relevant ionization and recombination
rates, effects of charge exchange reactions, and photoionization into
high ionization states of Ar through a quantity they denote $P\arcmin
_{\rm Ar}$.  Characteristic values of $P\arcmin _{\rm Ar}$ for various
interstellar conditions and ionizing spectra are tabulated in their
Table 4.  Using the \ion{Ar}{1} measurements of SY, corrected for the
suggested \ion{Ar}{1} oscillator strengths used in SJ, we find $[{\rm
Ar/H}] = -0.24\pm0.10$ towards \mucol\ using our $N(\mbox{\HI})$
measurement.  Thus it would seem that some degree of ionization is
present even in the neutral gas along this sightline, unless the
reference abundance of Ar is incorrect.  Using Equation (11) from SJ,
with an estimate of $P\arcmin _{\rm Ar} = 10\pm5$ from their Table 4,
we find $x_e = 0.09\pm0.07$.  Therefore we expect the contribution of
ionized hydrogen to the total neutral column density due to
partially-ionized gas to be $0.037\pm0.028$ dex.  Thus there could be
an additional correction of $\sim -0.04$ dex needed to account for
partially-ionized gas in the \HI\ regions along the \mucol\ sightline.

We note that the relative abundance [Ar/S] derived with \ion{Ar}{1}
and \SII\ measurements is extremely sensitive to relatively small
amounts of ionized gas along a given sightline since the measured gas
phase \SII\ and \ion{Ar}{1} increase and decrease, respectively, as
the level of ionization increases.  Furthermore, neither of these
elements is expected to be incorporated into dust grains in large
amounts (see SJ), and their nucleosynthetic origins are the same, both
being $\alpha$-process elements primarily produced in high-mass stars.
The usefulness of [Ar/H] measurements for identifying the existence of
partially-ionized gas has been emphasized by SJ; we suggest that the
[Ar/S] ratio, which can be determined as a function of velocity, is an
even more sensitive indicator of the combined effects of
partially-ionized gas in \HI-bearing regions and \HII\ region
contamination.

\subsection{Gas-Phase Abundances}
	\label{subsec:abundances}

Table \ref{table:abundances} contains the values of $[X/{\rm S}]$ in
the low-velocity components 1 and 2 for the elements considered here.
For comparison with other sightlines and QSO absorption line systems
we also give the sightline integrated values of [$X$/H] for each
element $X$.  These results are derived from the adopted column
densities presented in Table \ref{table:finalcoldens}.  Also given in
this table is the adopted solar reference abundance for each element
relative to hydrogen, and the normalized gas-phase abundances for
components 3 and 4, which will be discussed in \S
\ref{sec:highvelocity}.  The quoted errors here contain only those
described in \S \ref{subsec:measurements}.  Our adopted solar
reference abundance system is taken from Savage \& Sembach (1996a),
who rely primarily on the meteoritic abundances of Anders \& Grevesse
(1989) with updates for C, N, and O by Grevesse \& Noels (1993).

We have plotted the data from Table \ref{table:abundances} in Figure
\ref{fig:totalabundances}.  The ordinate of these plots is the
gas-phase abundance for the elements listed along the abscissa,
normalized to S.  The top panel of Figure \ref{fig:totalabundances}
contains only the data from components 1 and 2.  The bottom panel also
shows these data, but we have overlayed the values $[X/{\rm S}]$ for
the warm cloud along the $\zeta$ Oph sightline (cloud A of Savage
\etal\ 1992) as the dashed line, and for the spread of warm halo cloud
abundances from Sembach \& Savage (1996) as the hatched region.  Data
for the gas-phase abundances of interstellar Ti in individual halo
clouds are sparse.  For Ti we plot an upper limit equal to that of Ni
and a lower limit that is equal to the lower limit of Fe in these
clouds. The values of $[X/{\rm S}]$ from Savage \etal\ (1992) and
Sembach \& Savage (1996) have been adjusted to reflect our choice of
oscillator strengths.

The values of $[X/{\rm S}]$ for elements in the component 1 blend are
quite similar to those found in the warm $\zeta$ Oph cloud.  This was
also the conclusion of SSC, though with considerably more uncertainty.
This level of sub-solar abundances is consistent with other ``warm
disk''-like clouds as categorized by Sembach \& Savage (1996).
Similarly, the relative abundance pattern seen for component 2 is
mostly within the small range of ``halo'' cloud values.  This is
somewhat deceiving, however, since the earlier GHRS data for the
\mucol\ sightline (SSC) were used by Sembach \& Sembach (1996) in
defining this spread of values.  The new data presented here for
component 2 support the claim of Sembach \& Savage that the
upper-limit of $[X/{\rm S}]$ for many elements shows relatively little
cloud-to-cloud variation about the mean for each species.  What is
perhaps surprising, given the new distance to \mucol, is that a cloud
within $\approx 400$ pc of the sun ($z \la 180$ pc) has abundance
patterns similar to clouds at relatively large heights from the plane.
If one adopts the distance derived from the Vacca \etal\ (1996)
absolute magnitude scale, namely $d\sim790$ pc, the $z$-height becomes
$z\sim360$ pc from the midplane.  This is more than one \HI\ scale
height above the midplane.

In the gas of component 1, the values [Si/S] and [Mg/S] are slightly
higher than those seen along the $\zeta$ Oph sightline warm cloud.
Using the oscillator strengths suggested by Fitzpatrick (1997), Si and
Mg show quite similar abundance patterns.  Given the similarity of the
\SiII\ \wave{1808} line profile to those of \MgII\ \twowave{1239,
1240}, this is not a surprising result.  The similar abundance
behavior of these elements is seen even more clearly in plots of
$[X/{\rm S}]_v$, the gas-phase abundance of an element $X$ as a
function of velocity.

Figure \ref{fig:velabundances} presents plots of the normalized
gas-phase abundances for several elements as a function of velocity,
using the adopted solar system abundances listed in Table
\ref{table:abundances} (principally the meteoritic abundances of Anders 
\& Grevesse 1989).  We have plotted the ratios of \ion{P}{2},
\ZnII, \MgII, \SiII, \MnII, \ion{Fe}{2}, \ion{Ni}{2}, and \ion{Cr}{2}
to \SII\ in this figure.  For all of the species but \ion{Ni}{2}, the
width of the data bins is half a resolution element (i.e., $\approx
1.75$ km~s$^{-1}$ or two of the original data points for profiles
taken with four substeps per diode).  We have binned the \ion{Ni}{2}
profile to one point per resolution element.  The error bars represent
the sources of errors discussed in \S \ref{subsec:measurements} as
well as a contribution from possible velocity scale offsets. The
latter was derived by shifting one profile relative to the other by
half a resolution element (in both the positive and negative velocity
directions) and adding the resulting errors in quadrature to those of
\S \ref{subsec:measurements}.  These velocity shift error estimates are
the primary cause of the asymmetric error bars in Figure
\ref{fig:velabundances}.  The range over which data points are plotted
in this figure is determined by the significance of each point once we
have added all of the sources of noise.  This presentation assumes
that the $N_a(v) \approx N(v)$ and $[X^+/{\rm S}^+]_v \approx [X/{\rm
S}]_v$.  The former of these assumptions is justified by our
examination of the \Nav\ profiles of each of the ions for which
unresolved saturation might be present.  In no case do we find
evidence for saturation effects in the transitions presented in Figure
\ref{fig:velabundances}.  The latter of these assumptions is likely
valid for material in the warm neutral medium.  In the lower-right of
each panel we have included a bar representing the maximum degree of
uncertainty the \mucol\ \HII\ region is thought to add to the
individual measurements, as discussed in the previous subsection.

There is an increase in the gas-phase abundances of the elements Fe,
Mn, Cr, Ni and possibly Si near $\mvlsr \approx +20$ km~s$^{-1}$,
which is associated with the significant strengthening of component 2
in moderately- and heavily-depleted species.  However, one of the more
striking aspects of Figure \ref{fig:velabundances} is the relative
constancy of the ratios plotted here over the range $-10 \lesssim
\mvlsr \lesssim +14$ km~s$^{-1}$.  The absorbing clouds making up the
component 1 blend have similar gas-phase abundances.  The $[X/{\rm
S}]_v$ profiles for Si and Mg are quite similar, with $[{\rm Si/S}]_v
\approx [{\rm Mg/S}]_v \approx-0.4$ dex.  The Mg abundance profile is
slightly flatter than the Si profile.  Figures
\ref{fig:totalabundances} and \ref{fig:velabundances} show that Mg and
Si trace each other well (see Fitzpatrick 1997).

The abundances $[X/{\rm S}]_v$ for the elements P and Zn are
consistent with solar system abundances, i.e., [P/S]$_v \approx$
[Zn/S]$_v$ $\approx$ 0.  This behavior holds over the whole velocity
range considered in Figure \ref{fig:velabundances}.  This supports our
choice of solar system abundances as a reference for this dataset.  B
stars in the solar neighborhood may have lower intrinsic abundances
than the sun (e.g., Kilian-Montenbruck \etal\ 1994; Gies \& Lambert
1992), suggesting the proper ``cosmic'' reference abundance may be
sub-solar by $\sim 0.2$ dex.  In the case of the sightline towards
\mucol, we observe gas with solar system abundance ratios of P and Zn
to S, which we believe argues for using solar system reference
abundances for these clouds.  The choice of solar relative abundances
for these clouds is particularly strong given the sources of metal
production for S, an $\alpha$-element, are thought to be different
stars than those producing Zn, which traces Fe-peak elements.  Given
these elemental abundance ratios, we will rely primarily upon the
solar system abundances of Anders \& Grevesse (1989) as a cosmic
reference (as compiled in Savage \& Sembach 1996a).

	\subsection{Implications for Interstellar Dust}
	\label{subsec:depletion}

\begin{figure}
\epsscale{0.85}
\plotone{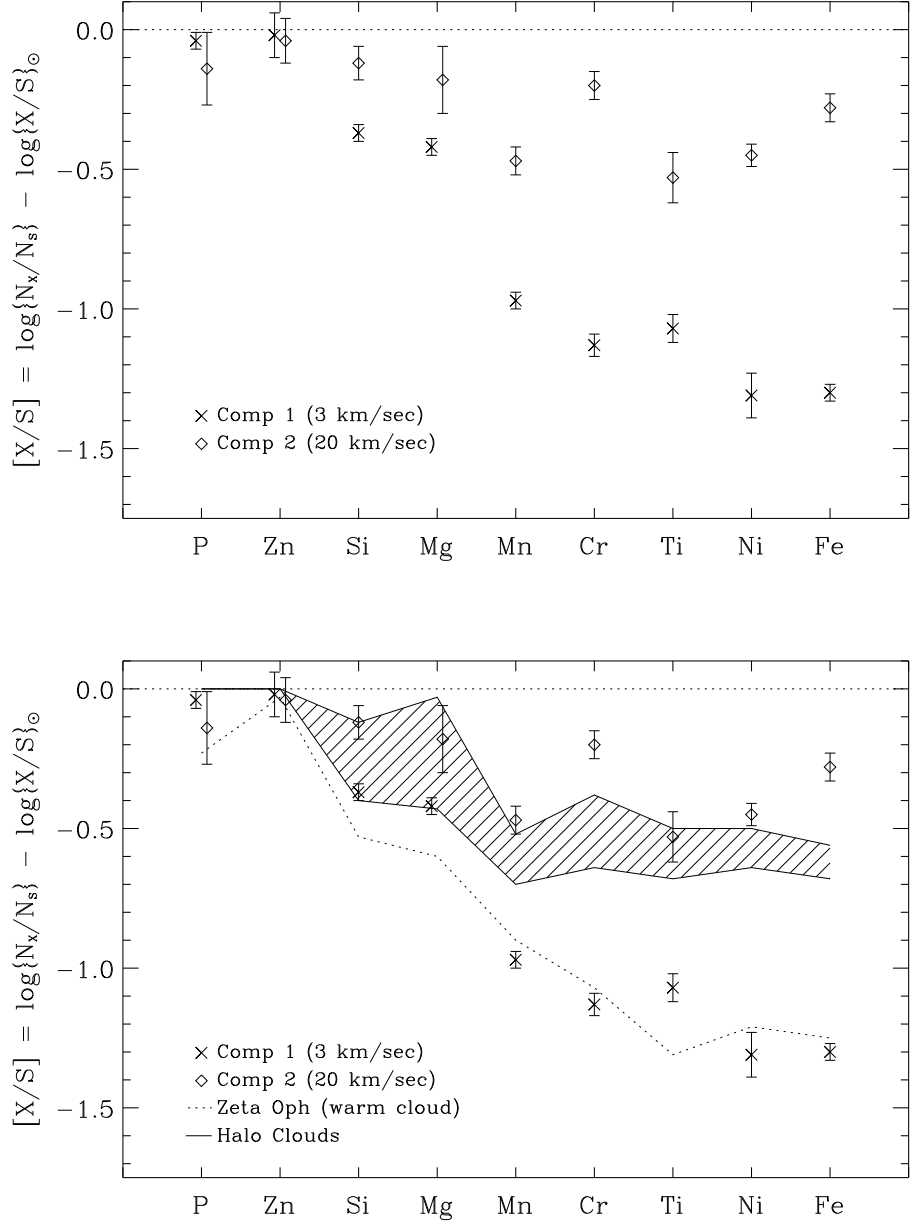}
\figcaption{(a) Plot of the gas-phase abundances of various elements
relative to S (see text).  The $\times$ symbols represent the
integrated abundances of the complex component 1 ($\mvlsr = +3$ km
s$^{-1}$), while $\diamond$ symbols represent the adopted abundances
for component 2 ($\mvlsr = +20$ km s$^{-1}$).  (b) Same as (a) but
with the observed gas-phase abundances of the $\zeta$ Oph warm cloud
included as the dotted line, and the spread of abundances for warm
halo clouds (as taken from Sembach \& Savage 1996) included as the
hatched region.  In both cases we have scaled the \protect\ion{Mg}{2}\
and \protect\ion{Ni}{2}\ abundances to be consistent with our adopted
of the oscillator strengths (see text).
\label{fig:totalabundances}}
\end{figure}

\begin{figure}
\epsscale{0.97} 
\plotone{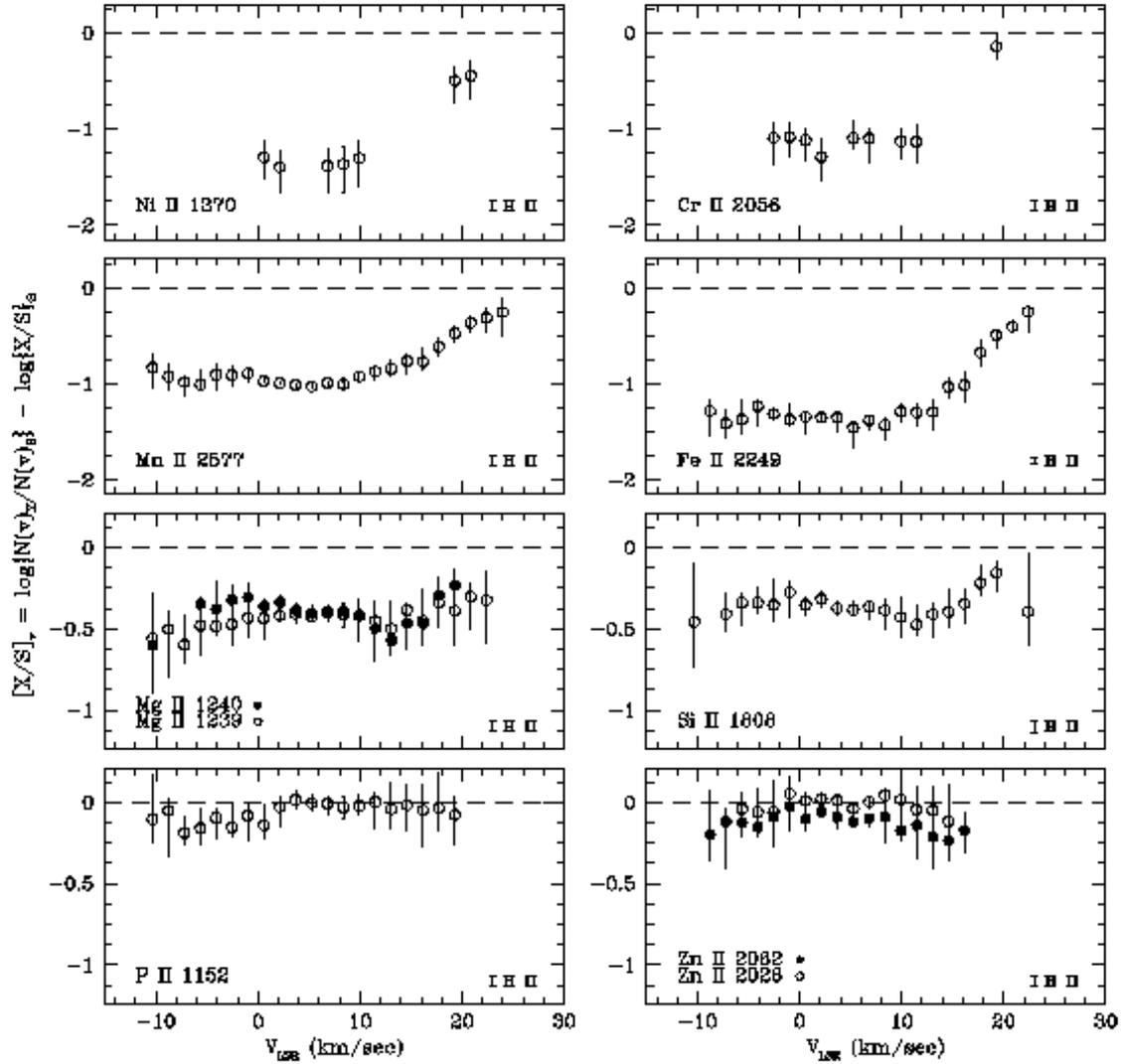}
\figcaption{Normalized gas-phase abundances of several ionic species
relative to S as a function of velocity.  Plotted are the ratios of
\protect\ion{P}{2}, \protect\ZnII, \protect\ion{Mg}{2}, 
\protect\ion{Si}{2}, \protect\ion{Mn}{2}, \protect\ion{Fe}{2}, 
\protect\ion{Ni}{2}, and \protect\ion{Cr}{2} to \protect\ion{S}{2}, 
normalized to the solar system meteoritic abundances of Anders \&
Grevesse (1989), as a function of velocity.  The error bars represent
$1 \sigma$ uncertanties from photon statistics, continuum fits, and
zero point uncertainties as well as a contribution for velocity shifts
between the profiles.  In the lower right of each panel is a bar
representing the estimated size of the uncertainty due to
contamination from material in an H$\,${\small II} region surrounding
$\mu$ Col.  For \protect\ion{Zn}{2} and \protect\ion{Mg}{2} we have
plotted both transitions since there are slight differences between
them.
\label{fig:velabundances}}
\end{figure}

The sub-solar gas-phase abundance patterns seen in Figures
\ref{fig:totalabundances} and \ref{fig:velabundances} likely represent
the imprint of elemental incorporation into interstellar dust grains.
Our arguments from \S \ref{subsec:ionization} suggest that the ionized
gas contribution to the measured total column of each of the species
considered in Figures \ref{fig:totalabundances} and
\ref{fig:velabundances} is minor, certainly less than $\sim 0.1$ dex.
The depletion of elements from the gas- to the solid-phase is a well
known phenomenon and allows us to infer the elemental make-up of dust
grains in the diffuse ISM (Savage \& Sembach 1996a).

The striking similarities between the normalized abundances in the
warm $\zeta$ Oph cloud and the material making up component 1 along
the \mucol\ sightline suggest that this level of depletion, or
incorporation of elements into the dust-phase, is relatively common
among low \HI\ column density clouds that make up the warm neutral
medium in the solar neighborhood (SSC).  The \HI\ column densities of
these two cloud complexes are quite similar [$\log
N(\mbox{\HI})_{\zeta \ {\rm Oph \ A}} \approx 19.74$; Savage \etal\
1992].

Of interest for determining the types of grains present in the ISM is
the dust-phase abundance of each element with respect to hydrogen.
The dust-phase abundance of a species $X$ relative to hydrogen,
$(X/{\rm H})_d$, is given by
\begin{equation}
	(X/{\rm H})_d = (X/{\rm H})_c - (X/{\rm H})_g, 
\label{eqn:xtohdust}
\end{equation}
where the subscripts $d, \ c,$ and $g$ refer to the dust-phase,
cosmic, and gas-phase abundances of $X$, respectively.  Table
\ref{table:dustcontent} gives the dust-phase abundances of a series of
elements relative to H (given in parts per million H) and relative to
Si for components 1 and 2.  We also give the dust-phase abundances of
Mg, Si, and Fe when adopting B-star abundances as determined by Gies
\& Lambert (1992) and Kilian-Montenbruck \etal\ (1994).  Gies \&
Lambert do not derive the abundance of Mg in their work, so we have
adopted the Mg abundance from Kilian-Montenbruck \etal\ in this case.
The values of $(X/{\rm H})_c$ are given for each element in the
different reference systems.  The values $(X/{\rm H})_d$ have been
derived assuming S is present in its cosmic abundance, i.e., is not
depleted into grains.  The cosmic sulfur abundances in the three
systems are: (S/H)$_c = 1.9\times10^{-5}$ (Anders \& Grevesse 1989);
(S/H)$_c =1.6\times10^{-5}$ (Gies \& Lambert 1992); and (S/H)$_c = 9.3
\times 10^{-6}$ (Kilian-Montenbruck {\em et al.}  1994).  We have not
included P or Zn in this table, as neither shows any evidence for
incorporation into dust grains in our data.

Among the elements considered here, the most abundant in dust grains
are Mg, Si, and Fe.  The solid forms of these elements in the ISM are
thought to include silicates, oxides and possibly metallic iron.
Among the silicate forms thought to be most common are various
pyroxenes, (Mg, Fe)SiO$_3$, and olivines, (Mg, Fe)$_2$SiO$_4$
(Ossenkopf \etal\ 1992).  If the Mg- and Fe-bearing dust in the ISM
towards \mucol\ were primarily made up of only these types of
silicates, one would expect a ratio of [(Fe+Mg)/Si]$_d \approx 1 - 2$.
Assuming the solar abundances given in Table \ref{table:dustcontent},
we find:
\begin{eqnarray*}
 \left[{\rm (Fe+Mg)}/{\rm Si}\right] _d & =  & 2.70 \pm 0.11 
	\quad\mbox{in component 1; and} \\	
\left[ {\rm (Fe+Mg)}/{\rm Si} \right] _d & =  & \phn3.3 \pm 1.4 \phn
	  \quad\mbox{in component 2.}
\end{eqnarray*}
If one adopts the Kilian-Montenbruck \etal\ (1994) B-star abundances
as the cosmic reference, these values become: $ [ {\rm (Fe+Mg)}/{\rm
Si} ] _d = 5.06 \pm 0.24$ in component 1; and $ [ {\rm (Fe+Mg)}/{\rm
Si} ] _d = 15 \pm 14$ in component 2.  The abundances of Gies \&
Lambert (1992) yield intermediate values.  The composition of dust
grains containing Mg, Si, and Fe in the material making up component~1
is inconsistent with the sole consituent of this dust being silicate
pyroxenes or olivines, irrespective of the cosmic abundance one
chooses to adopt.  Component 2 may be similar in this regard, but the
large uncertainties make this statement much less certain.

Whittet \etal\ (1997) have recently reported on the detection of O$-$H
stretching modes in OH groups along the diffuse sightline to Cygnus
OB2 No. 12 (VI Cygni No. 12) using the {\em Infrared Space
Observatory}.  Given the lack of ices along this sightline, they
attribute this feature to hydrated silicates.  The values of
(Mg/Si)$_d$ in components 1 and 2, assuming solar system abundances,
is consistent with the incorporation of Mg into a mixture of the
phyllosilicates talc, Mg$_3$Si$_4$O$_{10}$[OH]$_2$, and serpentine,
Mg$_3$Si$_2$O$_5$[OH]$_4$.  However, Whittet \etal\ find a very small
fraction of silicates along that sightline are hydrated.  Thus we
believe it unlikely phyllosilicates can provide for enough of the Mg
to allow Fe-bearing silicates to account for the total dust-phase Fe
abundance.  Some amount of the Mg- and Fe-bearing dust is therefore
likely in the form of oxides or pure iron grains.  Examples of Mg-,
and Fe-bearing oxides include MgO, FeO, Fe$_2$O$_3$, and Fe$_3$O$_4$
(Nuth \& Hecht 1990; Fadeyev 1988).

It is clear from Figures \ref{fig:totalabundances} and
\ref{fig:velabundances} that there is a larger gas-phase abundance of
the refractory elements in component 2 than in component 1.  In
particular, the gas-phase abundances of Fe, Cr, and Ni show much
higher abundances in Figure \ref{fig:totalabundances}.  In principle
higher gas-phase abundances of Fe-peak elements in interstellar clouds
could be interpreted as evidence for enrichment by Type Ia SNe
(Jenkins \& Wallerstein 1996).  We discount this mechanism for
providing the enhanced gas-phase abundances seen in component 2 over
component 1 along the \mucol\ sightline, suggesting instead that the
large enhancements seen in the Fe-peak elements over the other
elements in Figure \ref{fig:totalabundances} are a result of the
return to the gas phase of highly depleted elements.  The increases in
the gas-phase abundances per million H in component 2 over component 1
are: $15.4\pm1.1$ for Fe, $11\pm4$ for Si, $0.018\pm0.006$ for Ti,
$0.27\pm0.04$ for Cr, $0.078\pm0.015$ for Mn, and $0.54\pm0.07$ for
Ni.  If we were to assume gas with abundances similar to component 1
had been enriched by gas from a Type Ia SN, we would expect factors of
2.5 and $3-7$ more Mn and Ni, respectively, relative to Fe using the
nucleosynthetic yields of Nomoto \etal\ (1984) and Thielemann \etal\
(1986).  Also, relative to the models of Type Ia SN nucleosynthesis,
the observed increase in Si relative to Fe is a factor of $\sim2$ too
large, while the increase in Ti to Fe is too high by a factor of 2.5
to 160, depending upon the exact nucleosynthesis result used.
Therefore, we conclude that the increase in gas-phase abundances
observed in component 2 relative to component 1 is likely not due to
enrichment by the nucleosynthetic products of a Type Ia SN.

We interpret the higher gas-phase abundances seen in component 2 as a
return of elements to the gas phase from the solid phase in material
that has been processed by a shock(s).  The dust-phase abundances of
many elements in component 2 have been lowered relative to those in
component 1.  Examining the data in Table \ref{table:dustcontent} we
see that relative to component 1, the dust-phase abundances of Si, Mg,
and Fe have changed by approximately a factor of two.  The values of $
[ {\rm (Fe+Mg)}/{\rm Si} ] _d$ in component 2 are consistent with
those of component 1 given the large errors for this component.  There
is evidence for lower dust-phase abundances of Si, Fe, Ti, Cr, Mn, and
Ni in component 2.  If the dust-phase abundances of the warm $\zeta$
Oph cloud and the clouds making up component 1 are characteristic of a
standard depletion in the low-density WNM of the Galaxy, we can
estimate the fraction of material returned to the gas-phase from dust
in component 2.  The data in Table \ref{table:dustcontent} suggest the
return of $(55\pm20)\%$ of Si, $(50 \pm 4 )$\% of Fe, $(23 \pm8)$\% of
Ti, $(60 \pm9)$\% of Cr, $(26 \pm 5 )$\% of Mn, and $(32 \pm 4)$\% of
Ni to the gas phase from dust grains, when compared to the values
appropriate for component 1.

If one interprets the differences in abundance patterns between
components 1 and 2 as evidence for the stripping of grain mantles from
standard dust-phase abundance of the WNM, the composition of the
mantles may be inferred from a direct comparison of the dust-phase
abundances of components 1 and 2.  The data for the \mucol\ sightline
then suggest that the mantles of dust grains are consistent with
silicate and oxide components.  Our large errors in the values for
component 2 make a detailed determination of the mantle composition
difficult.

	\subsection{Physical Conditions}
	\label{subsec:physicalconditions}

In this subsection we discuss the information about the physical
conditions in the low-velocity material towards \mucol\ contained in
our absorption line data.  The approaches used here are well-discussed
in the works of Spitzer \& Fitzpatrick (1993, 1995) and Fitzpatrick
\& Spitzer (1994, 1997).  

	\subsubsection{Electron Densities in the Ionized Gas}
	
\copernicus\ observations of \ion{N}{2}, \ion{N}{2}$^{*}$, and 
\ion{N}{2}$^{**}$ have yielded estimates for the average electron 
density in the ionized gas, $\langle n_{e,i} \rangle$, along the
sightline to \mucol\ (SY).  These authors' estimates yield $\langle
n_{e,i} \rangle \sim 0.16$ to 0.22 cm$^{-3}$ for the ionized gas.  We
detect weak absorption arising from the excited $^2P_{3/2}$ fine
structure level of \SiII\ at
\wave{1264.738} in our GHRS dataset.  This allows us to verify the
results of SY.  The equation for collisional excitation equilibrium of
the Si$^+$ levels may be written
\begin{equation}
A_{21} n({\rm Si}^{+*}) + \gamma_{21} n_e n({\rm Si}^+) 
	= \gamma_{12} n_e n({\rm Si}^+),
\end{equation}
where $\gamma_{12}$ and $\gamma_{21}$ are the rate coefficient for
collisional excitation and de-excitation, respectively, and $A_{21}$
the spontaneous downward transition probability.  In the case of ion
excitation by electrons we write (Spitzer 1978)
\begin{equation}
\gamma_{12} = \frac{8.63\times10^{-6}}{g_1 T^{1/2}} \Omega_{12} \exp
\left(\frac{-E_{12}}{kT} \right) \ {\rm cm}^3 \ {\rm s}^{-1}.
\end{equation}
For warm gas like that expected in an \HII\ region, exp$(E_{12}/kT)
\approx 1.0$ (to the accuracies being considered here).  We adopt the 
values $A_{21} = 2.17\times10^{-4}$ s$^{-1}$ (Mendoza 1983; his
Appendix), $\Omega_{12} = 5.58$ (Osterbrock 1989; his Table 3.3), and
the statistical weight $g_1 = 2$.  Replacing the particle densities
with column densities we have
\begin{equation}
\langle n_{e,i} \rangle = 9.0 T^{1/2}
	\frac{N({\rm Si}^{+*})}{N({\rm Si}^+)},
\end{equation}
where we have ignored collisional de-excitation.  The effects of
collisional de-excitation are negligible when $n_e \ll n_{cr}$, where
the critical density, $n_{cr}$, is given by $n_{cr} =
A_{21}/\gamma_{21}$.  For $T=8000$ K the critical density is $n_{cr}
\approx 1600$ \percc.  

Using the integrated column density of \SiII\ in component 1 yields
$\langle n_{e,i} \rangle = 0.2$ \percc.  This result assumes $T =
8000$ K, and scales as $T^{1/2}$.  This value is a lower limit to the
true electron density, $n_{e,i}$.  The Si$^+$ absorption includes
contributions from both \HI\ and \HII\ regions, while it is likely
that the Si$^{+*}$ absorption comes primarily from only the densest
gas of \HII -bearing regions.  The column of Si$^+$ associated with
the ionized gas is overestimated, making $n_{e,i}$ too small.  The
results of our photoionization modelling in \S \ref{subsec:ionization}
suggest that the column density of \SiII\ arising in the \HII\ region
about \mucol\ is likely $\log N(\mbox{\ion{Si}{2}})
\approx 14.3$.  If we adopt this value for the column density of \SiII\
associated with the ionized gas and assume that all of the \SiII$^*$
arises in this \HII, we can derive an upper limit for $n_{e,i}$.
Combining these two limits yields
\begin{displaymath}
0.2 \ {\rm cm}^{-3} \, \lesssim n_{e,i} \, \lesssim
	1.2 \ {\rm cm}^{-3}.
\end{displaymath}
The
upper limit is larger than the values we've used for modelling the
\mucol\ \HII\ region, but the conclusions drawn in \S
\ref{subsec:ionization} are not sensitive to the ambient
density.

   \subsubsection{Electron Temperatures and 
	Densities in the Neutral Gas}
	\label{subsubsec:physicalconditions}

We estimate the electron temperatures and densities of the primarily
neutral material in the low-velocity components 1 and 2 by examining
the ionization equilibrium of Ca and Mg in these clouds.  The equation
for ionization equilibrium of the ionic stages $X^i$, $X^{i+1}$ of an
element $X$ may be written
\begin{equation}
\Gamma(X^i) n(X^i) = 
	\alpha(X^i) n_e n(X^{i+1}),
\label{eqn:ionequilibrium}
\end{equation}
where $\alpha(X^i)$ is the recombination coefficient of $X^{i+1}$ to
$X^i$ and $\Gamma(X^i)$ the ionization rate of $X^i$.  The value
$\alpha(X^i)$ is the sum of the radiative recombination coefficient,
$\alpha_{rad}(X^i)$, and the dielectronic recombination coefficient,
$\alpha_{di}(X^i)$, and is a function of temperature.  Using
Equation (\ref{eqn:ionequilibrium}) simultaneously for Ca$^+$/Ca$^{++}$
and Mg$^{\rm o}$/Mg$^+$ ionization equilibrium can yield estimates of
both the electron temperatures and densities.

For the recombination coefficients we adopt the fits suggested in the
compilation of atomic data by D. Verner\footnote{\tt
http://www.pa.uky.edu/$\sim$verner/atom.html}.  Namely for the
Mg$^{\rm o}$ recombination coefficients we use
$\alpha_{di} ({\rm Mg}^{\rm o}) = 4.49\times \e{-4} T^{-3/2} 
  \exp (-5.01\times\e{4}/T) 
  \left[ 1 + 2.1\times \e{-2} \exp (-2.81\times \e{4}/T) \right]$ 
     cm$^3$ s$^{-1}$ and 
$\alpha_{rad} ({\rm Mg}^{\rm o}) = 1.4 \times \e{-13} 
  \left( T/10^4 \right)^{-0.855}$ cm$^3$ s$^{-1}$
from Shull \& Van Steenberg (1982) and Aldrovandi \& P\'{e}quignot
(1973), respectively.  We assume a photoionization rate 
	$\Gamma({\rm Mg}^{\rm o}) = 4.0 \times 10^{-11}$ s$^{-1}$ 
(Frisch \etal\ 1990).  For Ca we adopt the values 
$\alpha({\rm Ca}^{+}) \approx \alpha_{rad}({\rm Ca}^{+}) = 6.78 \times
10^{-13} \left( T/10^4 \right)^{-0.8}$ cm$^3$ s$^{-1}$ 
from Shull \& Van Steenberg (1982)\footnote{ $\alpha_{di}({\rm
Ca}^{+}) \ll \alpha_{rad}({\rm Ca}^{+})$ for $T<10^4$ K.}.  We assume 
$\Gamma({\rm Ca}^+) = 2.0\times10^{-12}$ s$^{-1}$ 
(P\'{e}quignot \& Aldrovandi 1986).  Unfortunately, we do not measure
the dominant ionization stage Ca$^{++}$.  To proceed we assume that
the gas-phase abundances of Ca and Fe are approximately the same in
the WNM towards \mucol\ (see Jenkins 1987).  The profiles of the
\ion{Ca}{2} and \ion{Fe}{2} (e.g., \wave{2249}) transitions are very 
similar for the \mucol\ sightline, suggesting this is a reasonable
assumption.  Using this approximation we can substitute $n({\rm Fe}^+)
\cdot \{ {\rm Ca/Fe} \}_\odot$ for $n({\rm Ca}^{++})$.

\begin{figure}
\epsscale{1.1}
\plotone{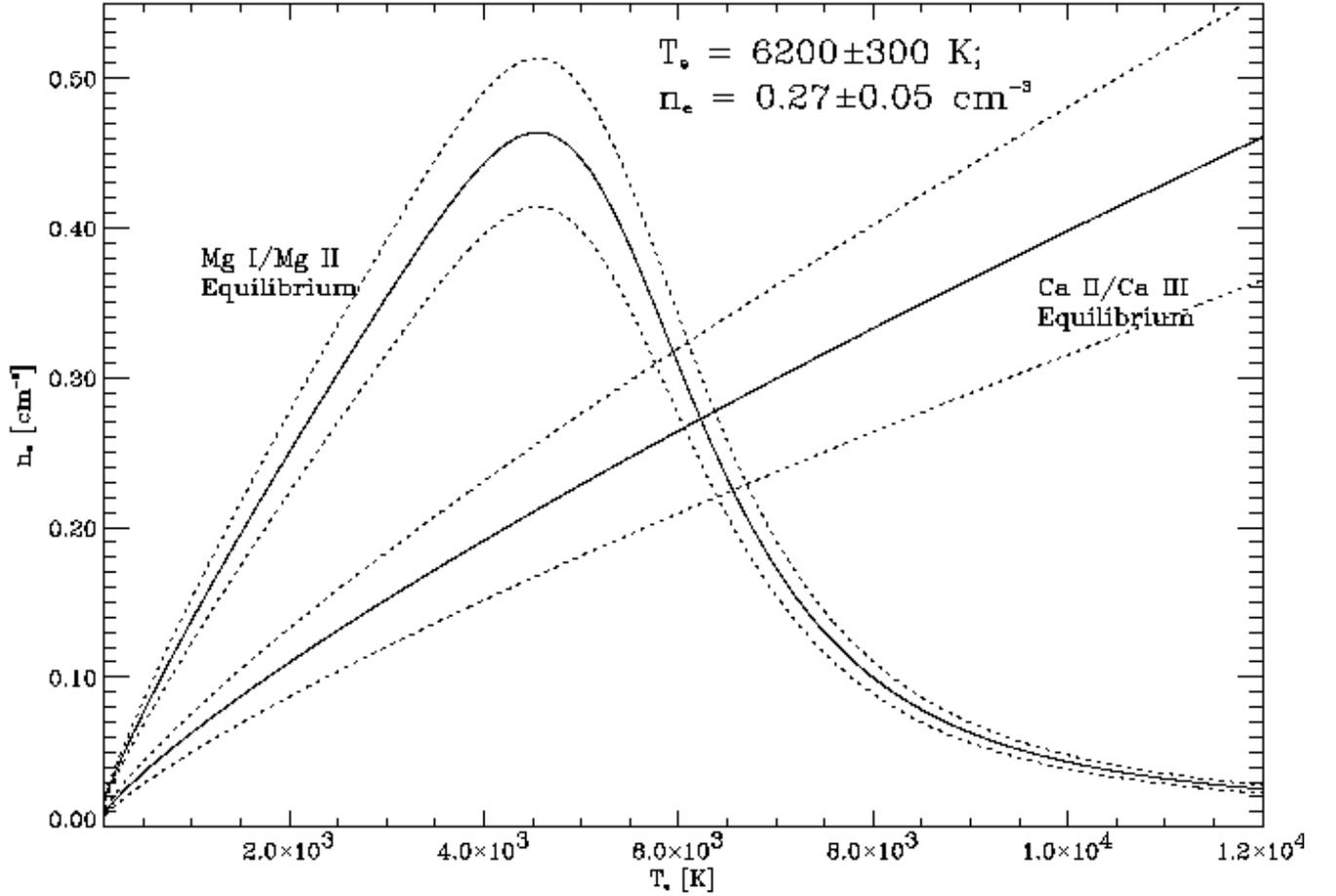}
\figcaption{Plot of the predicted electron density, $n_e$, as a
function of temperature, $T_e$, using
\protect\ion{Mg}{1}/\protect\ion{Mg}{2} and
\protect\ion{Ca}{2}/\protect\ion{Ca}{3} ionization equilibrium.  The
solid lines represent the trends in $n_e$ with $T_e$ for the most
likely ratios of these ionization stages.  The dotted lines represent
the $1\sigma$ uncertainties (not including uncertainties in the atomic
data).  The values predicted by the intersection of the
\protect\ion{Mg}{1}/\protect\ion{Mg}{2} and
\protect\ion{Ca}{2}/\protect\ion{Ca}{3} curves are $\langle T_e
\rangle = 6200\pm300$ K and $\langle n_e \rangle = 0.27\pm0.05$ cm$^{-3}$.
\label{fig:physicalconditions}}
\end{figure}

Figure \ref{fig:physicalconditions} shows the relationship of $n_e$
and $T_e$ for Ca and Mg ionization equilibrium in component 1.  The
solid lines represent the dependence of $n_e$ on $T_e$ for each
diagnostic given the best values of the $N({\rm Ca}^+) / N({\rm
Fe}^{+})$ and $N({\rm Mg^o}) / N({\rm Mg}^+)$ ratios.  The dotted
lines give the $1\sigma$ error limits based upon the sources of error
given in \S \ref{subsec:measurements}.  While there is a formal
solution for very low temperatures, the high-temperature solution is
more applicable to the gas being considered here.  The results of this
approach for components 1 and 2 are given in Table
\ref{table:components}.  Also shown are the electron temperatures and
densities for component 4 (see below).  The treatment of Equation
(\ref{eqn:ionequilibrium}) neglects the charge-exchange reactions
between neutral Mg and ionized H (Allan \etal\ 1988), but we have
found including these reactions makes little difference to the final
results.

In principle one can derive the neutral hydrogen density in WNM clouds
using observations of \ion{C}{1} absorption from the $^3P_0$, $^3P_1$,
and $^3P_2$ levels.  However, our marginal detections of the first
excited $^3P_1$ level and limits on the $^3P_2$ column densities allow
us only to place limits on the neutral hydrogen density.  We find
$\log N_{10} \equiv \log N(\mbox{\ion{C}{1}}\ \,^3P_1) - \log
N(\mbox{\ion{C}{1}}\ \,^3P_0) \la -0.37$ and $\log N_{20} \equiv \log
N(\mbox{\ion{C}{1}}\ \,^3P_2) - \log N(\mbox{\ion{C}{1}}\ \,^3P_1) \la
-0.56$ in component 1.  Interpolating between the values given in the
tabulation of Keenan (1989), and assuming $T\approx6000$ K and $n_e
\approx 0.3$ derived above, we estimate $n_{\rm H} < 10$ \percc\ in
component 1.

\section{IONIZATION AND DUST IN THE INTERMEDIATE-VELOCITY GAS}
\label{sec:highvelocity}

\subsection{Photoionized Gas at Intermediate Positive Velocities}

Components 3 and 4 along the sightline to \mucol\ are examples of low
total column density intermediate-velocity clouds (IVCs).  The
hydrogen column density for component 4 is likely $\log N({\rm H})
\sim 17.3-17.7$ (see below).  With an absorption cross-section of $6.3
\times 10^{-18}$ cm$^2$ for atomic hydrogen (Spitzer 1978), this
implies an optical depth of order unity for this cloud at \wave{912}.
Thus, the assumption that the gas-phase abundances of a cloud are
accurately traced by the dominant stage of ionization in the {\em
neutral} medium may break down for components 3 and 4.  Further, the
effects of ionization will differ from species to species, depending
upon that ionic species' ionization potential and photoionization
cross section, its recombination coefficient, including the effects
dielectronic recombination, and the availability of charge-exchange
reaction pathways.

\begin{figure}
\epsscale{1.0}
\plotone{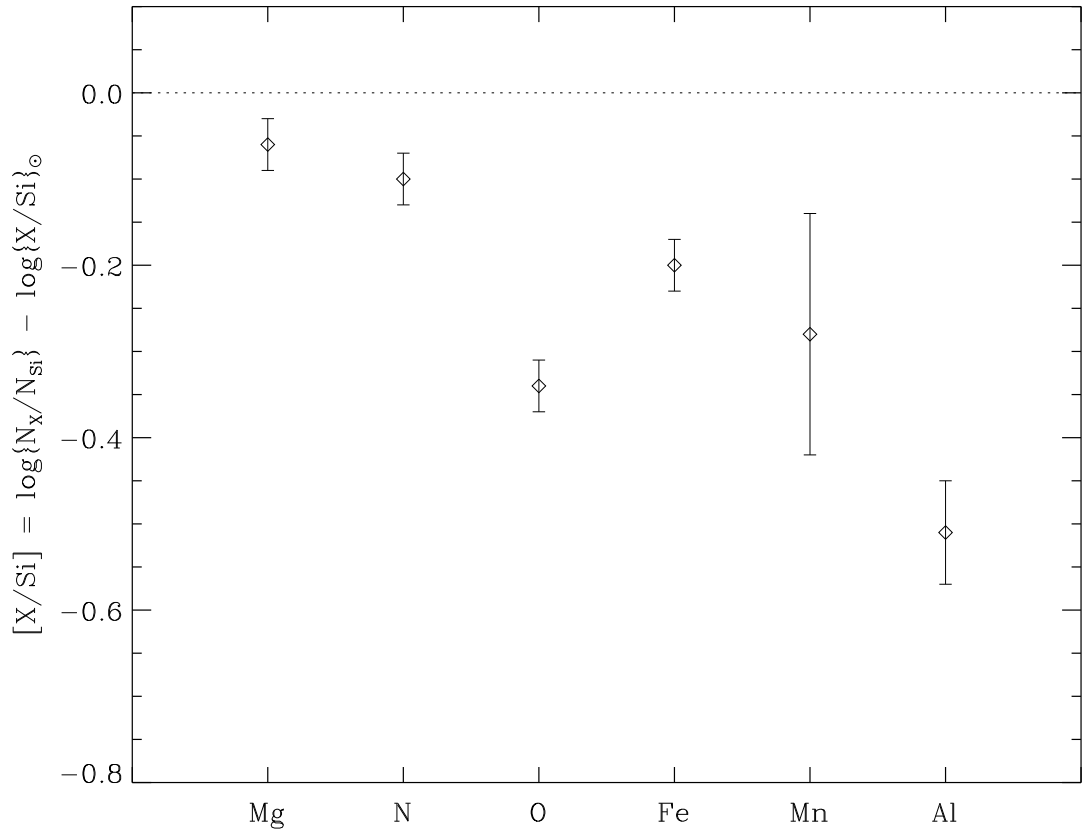}
\figcaption{Normalized gas-phase abundances of several ionic species
in component 4 ($\mvlsr = +41$ km s$^{-1}$).  These species have been
referenced to the column density of Si.
\label{fig:comp4abundances}}
\end{figure}

The species for which we have well-determined column densities for
component 4 include \ion{N}{1}, \ion{O}{1}, \ion{Mg}{1} and {\small
II}, \SiII\ and {\small III}, \ion{Fe}{2}, and \ion{Al}{2}.  In Figure
\ref{fig:comp4abundances} we plot the the values of [$X$/Si] for the
species O, Mg, Si, Al, Mn, and Fe. The values [$X$/Si] for components
3 and 4, derived from the ratios of the usually dominant species, are
also given in Table \ref{table:abundances}.  The unusual nature of
this component can easily be seen in Figure \ref{fig:comp4abundances}.
Typically Si is moderately incorporated into grains in the diffuse ISM
(e.g., in components 1 and 2), while N has only moderately sub-solar
gas phase abundances in the solar neighborhood (Meyer \etal\ (1997)
find $[{\rm N/H}] \approx -0.1$).  Therefore in the WNM we expect to
find $[{\rm N/Si}] > 0$.  In Figure \ref{fig:comp4abundances} we see
that [N/Si] derived from measurements of \ion{N}{1} and \ion{Si}{2} is
sub-solar.  At the same time we find that [Fe/Si] is almost
indistinguishable from that seen in component 2 and similar to halo
cloud values.  We believe that the pattern of gas-phase column
densities observed in component 4 is a result of photoionization of a
low column density cloud with somewhat typical halo-like gas-phase
abundances.  The uncertainties in this ionization, however, hinder our
ability to make firm conclusions regarding the gas-phase abundances of
the intermediate velocity clouds towards \mucol.

For a few species we can place quite stringent limits on the true
gas-phase abundances.  We observe the ionization states \SiII, {\small
III}, and {\small IV} and therefore have a very good measure of the
total column density of Si in component 4, $\log N({\rm Si}) = 12.85
\pm 0.03$, i.e., the contributions from \ion{Si}{3} and {\small IV}
are negligible.  We can place similar constraints on the abundance of
Al in this material.  The $2\sigma$ upper limit $\log
N(\mbox{\ion{Al}{3}}) \lesssim 11.2$ in component 4 implies $-0.51
\lesssim [{\rm Al}/{\rm Si}] \lesssim -0.25$.  If we assume Si is
present in its solar system abundance, the column density of Si
suggests the total hydrogen column density is $\log N({\rm H}) \approx
17.35$, whereas if its abundance is as low as $[{\rm Si/H}] \approx
-0.3$, i.e., having warm disk cloud-like abundances, then the total
hydrogen column density of this cloud is $\log N({\rm H}) \approx
17.65$.

\begin{figure}
\epsscale{0.85}
\plotone{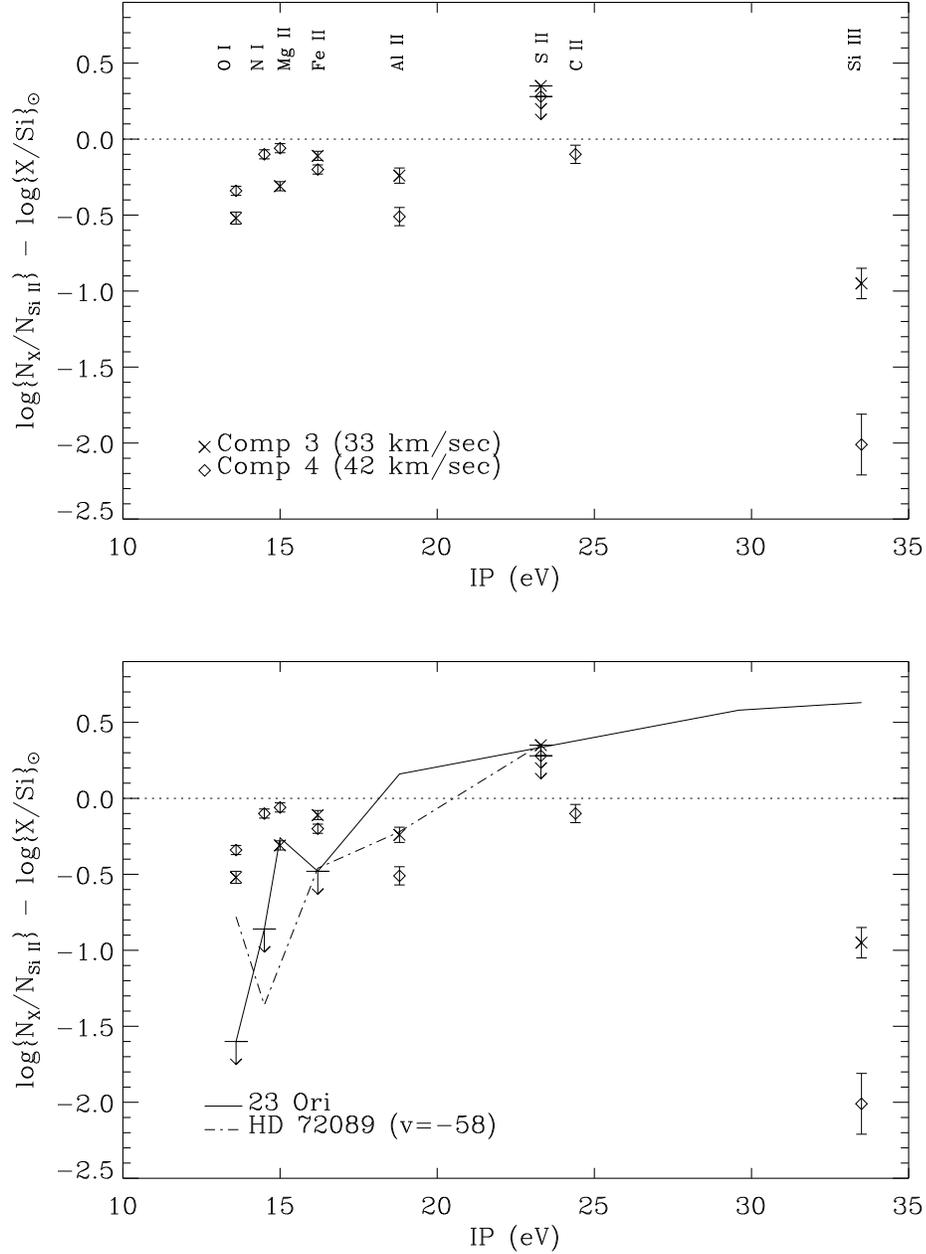}
\figcaption{The ionic abundances of species $X^i$ relative to
Si$\,${\small II} and normalized by the $X$/Si ratio of the solar
system as a function of ionization potential.  The top plot shows the
data for components 3 and 4 towards $\mu$ Col.  The bottom shows the
same data with the data for 23 Ori (Trapero \protect\etal\ 1996) and
the $v_{\rm LSR} = -58$ km s$^{-1}$\ component towards HD 72089
(Jenkins \protect\etal\ 1998) overplotted as the solid and dot-dashed
lines, respectively.  The upper limits plotted for S$\,${\small II}
are $2\sigma$ limits.  The ionization potential of Si$\,${\small II}
is 16.35 eV.
\label{fig:IPabundances}}
\end{figure}

Before discussing the effects of photoionization, we note that the
observed gas-phase abundances are likely not due to collisional
ionization, nor is it likely these clouds have been ionized by the
hard radiation from a strong shock.  Recent works studying
high-velocity material in the Galactic disk (Trapero \etal\ 1996) and
gas associated with the Vela supernova remnant (Jenkins \etal\ 1998)
have suggested these physical mechanisms are responsible for unusual
abundances along the sightlines they studied.  Two lines of evidence
lead us to believe these are unlikely situations for the intermediate
positive velocity gas towards \mucol.  First, the observed $b$-values
for the IVCs towards \mucol\ imply very low temperatures.  We derive a
temperature for component 4 of $T = 4,000\pm700$ K in \S
\ref{subsec:IVphysicalconditions}.  Second, both of these works find
increasing gas-phase abundances as a function of ionization potential
(IP).  In Figure \ref{fig:IPabundances} we show the gas phase
abundances of components 3 and 4 towards \mucol\ referenced to Si.
The bottom plot shows these same data with the results from Jenkins
\etal\ and Trapero \etal\ overlayed.  We see very little trend with
IP, except for a turnover at high IP with \ion{Si}{3}.  Trapero \etal\
find $\log N(\mbox{\ion{Si}{2}}) / N(\mbox{\ion{Si}{3}}) \approx
-0.6$, we find $\log N(\mbox{\ion{Si}{2}}) / N(\mbox{\ion{Si}{3}})
\approx +2.1$ for component 4.

Given the sub-solar abundance [Al/Si] derived above, it is clear that
component 4 contains dust, which is affecting the pattern of column
densties measured in this cloud.  However, the abundances [O/Si] and
[N/Si] derived from measurements of \ion{O}{1} and \ion{N}{1} reveal
that photoionization is also playing a role.  This conclusion is based
on an examination of the relative ionization cross sections and
recombination coefficients for the observed atomic and ionic species.
The ionization equilibrium of a species in its simplest form is
written in a manner similar to Equation (\ref{eqn:ionequilibrium}).
Ignoring for now charge exchange reactions, we write
\begin{equation}
\frac{n(X^i)}{n(X^{i+1})} = \frac{\Gamma(X^i)}{\alpha(X^i)} n_e,
\end{equation}
where $\Gamma(X^i)$ is a function of the ionization cross section and
incident radiation field.  Ions that have higher ratios of the
ionization cross section $\sigma_\nu$ to recombination coefficient
$\alpha(T)$ are more readily ionized by a given photon of frequency
$\nu$.  Sofia \& Jenkins (1998) have discussed in detail the role of
photoionization for \ion{Ar}{1}, which has a large cross section to
ionizing photons.  They find that the observed deficit of Ar relative
to H towards several stars is likely due to the preferential
ionization of \ion{Ar}{1} to {\small II} relative to the ionization of
H.  The results of SJ also show, when using the photoionization cross
sections of Verner \etal\ (1996) and recombination coefficients of
Shull \& Van Steenberg (1982), that \ion{O}{1} and \ion{N}{1} have
significantly larger values $\Gamma(X^i)/\alpha(X^i)$ than most of the
species studied here, while \ion{Al}{2}, \ion{Mg}{2}, and \ion{Si}{2}
were among the lowest.  We suggest the deficits of \NI\ and \OI\
relative to \ion{Si}{2} are the result of photoionization of this low
column density cloud, similar to the deficit of \ion{Ar}{1} relative
to \ion{H}{1} measured by SJ.  The magnitude of the deficits is such
that O would be $\sim50\%$ ionized in component 4.  We expect the
ionization fractions of H to be very similar given the very strong
charge exchange reactions between these elements (Field \& Steigman
1971).

The fraction of C in the form C$^+$ is likely quite large given the
large IP of this ion and the relative similarity of $\Gamma({\rm
C}^+)/\alpha({\rm C}^+)$ to that of Si$^+$ over a large range of
energy.  Sofia \etal\ (1997) have shown the gas-phase abundance of C
is relatively constant at a value $[{\rm C/H}] \approx -0.4$; if we
assume this is also the case in component 4, then we expect $[{\rm
Si/H}] \approx -0.3$.  However, the most important abundance
measurement for this cloud is the value [Al/Si] derived above.  Given
our ability to account for the various stages of ionization of these
elements, our measurements securely show this cloud contains dust
grains.

Component 3 may be even more prone to ionization effects.  The column
density of \ion{Si}{2} in component 3 is similar to that in component
4, though the column density of \ion{O}{1} is lower by a significant
amount, suggesting a higher degree of ionization ($N({\rm
O^{++}})/N({\rm O}) \approx 0.7$).  The total column density of Si in
this component is $\log N({\rm Si}) = 12.83 \pm 0.03$, using the
measured column densities of
\ion{Si}{2} and {\small III} and the upper limit for \ion{Si}{4}:
$\log N(\mbox{\ion{Si}{4}}) \lesssim 11.3$ ($2\sigma$).  The ratio
$\log N(\mbox{\ion{Si}{2}}) / N(\mbox{\ion{Si}{3}}) = +0.95$, while
less than that for component 4, shows that the higher stages of
ionization add relatively little to the abundance of Si.  Again, we
can accurately derive the gas-phase abundance of Al relative to Si
given our upper limits to the column density of \ion{Al}{3}.  We find
$-0.31 \lesssim [{\rm Al}/{\rm Si}] \lesssim -0.11$ for component 3.
Thus the values [Al/Si] may be slightly different in components 3 and
4.  Given our $3\sigma$ upper limits to the column densities of
\ion{S}{2} and {\small III} in this component, [Si/S]$ \gtrsim -0.15$,
or $-0.3$ if we assume the amount of \ion{S}{3} is negligible.  This
suggests component 3 could be similar in its gas-phase composition to
component 2.

\subsection{Collisionally Ionized Gas at Intermediate 
	Negative Velocities} 

Component 5 likely samples a low-column density region of warm,
collisionally ionized gas.  For the species that have well-determined
$b$-values, i.e., \ion{C}{2}, \ion{Mg}{2}, and
\ion{Si}{3},  we derive upper limits to the temperature of $T<36,000$ K,
$<82,000$ K, and $<95,000$ K, respectively.  Unfortunately, our
$b$-values are not well enough constrained to disentangle the thermal
from non-thermal broadening.  For \ion{C}{2} and \ion{Mg}{2}, ions
with a factor of two difference in atomic mass, to have such similar
$b$-values, non-thermal broadening must play a significant, if not
dominant, role in this component.

For component 5 we find $\log N(\mbox{\ion{Si}{2}})/
N(\mbox{\ion{Si}{3}}) = -0.67$, similar to the value observed towards
23~Ori by Trapero \etal\ (1996).  This value is also similar to a gas
in collisional ionization equilibrium having a temperature slightly
less than $T \approx 25,000$ K (Sutherland \& Dopita 1993), which
consistent with our upper limit of $36,000$ K. If Si is not depleted
in this gas and all of the Si is found in the Si$^{+}$ and Si$^{+2}$,
then the column density of hydrogen (${\rm H^o} + {\rm H}^+$) in this
component should be $\log N({\rm H}) \ga 16.8$.  We quote a lower
limit since some depletion is likely.

Photoionization models using stellar input spectra or models for the
interstellar radiation field are unable to match the observed ratio of
\ion{Si}{2} to \ion{Si}{3} (e.g., Howk \& Savage 1999) unless the gas
is directly exposed to a hard radiation field.  It is thus unlikely
that this component represents \HII\ region gas or gas associated with
the WIM, even if it is made up of several overlapping clouds.  Howk \&
Savage (1999), for example, find $\log N(\mbox{\ion{Si}{2}})/
N(\mbox{\ion{Si}{3}}) \ga 0.0$ for \HII\ regions surrounding stars
having $T_{eff} \la 39,000$ K.

If component 5 is indeed in collisional ionization equilibrium with
$T\approx 25,000$ K, the fraction of C in stages other than C$^+$
should be minimal given the high IP of C$^+$ (Sutherland \& Dopita
1993).  Using $\log [ N(\mbox{\ion{Si}{2}}) + N(\mbox{\ion{Si}{3}})] =
12.37\pm0.03$ as the column density of Si in this component, we find
$[{\rm C / Si}] = 0.0\pm0.1$.  Various authors have found gas-phase
abundances of $[{\rm C / H}] \approx -0.4$ in the local ISM
(referenced to solar system values) with very little intrinsic spread
(Sofia \etal\ 1997; Cardelli \etal\ 1993).  Thus, if we assume this
value also holds in component 5, our derived [C/Si] suggests [Si/H]
similar to that found in component 1.

We also note the presence of an absorbing component at $\mvlsr =
-49.5\pm0.3 $ \kms\ visible only in \ion{Si}{3}.  This component may
be similar to component 5.  Component fitting analysis yields: $\log
N(\mbox{\ion{Si}{3}}) = 11.62\pm 0.03$, and $b = 5.9\pm1.0$ \kms,
constraining the temperature $T \la 59,000$.

The relative ionization levels of component 5 are similar to the
clouds observed towards 23 Ori by Trapero \etal\ (1996), as evidenced
by the relative column densities of \ion{Si}{2} and {\small III}.
There are two main observational differences between the
collisionally-ionized material along the \mucol\ and 23 Ori
sightlines, which may or may not imply true physical differences
between these clouds.  First, the collisionally-ionized gas towards 23
Ori is at high velocities relative to the LSR of $\mvlsr \approx -100$
to $-125$ \kms.  However, while component 5 does not have as extreme a
velocity as the ionized material seen towards 23 Ori, its velocity is
inconsistent with models of Galactic rotation in this direction.  This
alone, however, is not extremely unusual given that components 2, 3,
and 4 towards \mucol\ are also at odds with simple models of Galactic
rotation.  These clouds make up only $\sim10\%$ of the total column
density along the \mucol\ sightline.

A second difference between the material making up our component 5 and
the HVCs observed by Trapero \etal\ is the observed component
structure of the gas.  Trapero \etal\ report the existence of four
well-separated components in the HVC seen towards 23 Ori, each of
which is characterized by a $b$-value between 2 and 3 \kms.  Thus they
derive temperature limits $T \la 12,000$ K for each of their
components.  These authors suggest the gas is not in collisional
ionization equilibrium, but rather that the ionization state of the
gas is ``frozen in'' with relative ionization levels more
representative of $T \sim 25,000$ K.  The gas temperature suggested by
comparing the amounts of \ion{Si}{2} and {\small III} with ionization
equilibrium calculations for component 5 ($\sim 25,000$ K) is
consistent with the $b$-values derived from our component-fitting
analysis (suggesting $T \la 36,000$ K).  We must caution, however,
that although our observations of component 5 are made with the
echelle-mode gratings of the \ghrs, with a resolution of $\approx 3.5$
\kms, there may be unresolved component structure present in the
profile.  Though this component is reasonably well fit by a model
cloud represented by a single gaussian, this does not exclude the
presence of multiple components within this velocity range.  Indeed,
the similarity of the \ion{C}{2} and \ion{Mg}{2} $b$-values suggests
that the temperature of the gas in this component is much less than
the upper limit derived from our the \ion{C}{2} $b$-value.

Thus the intermediate negative velocity gas along the \mucol\
sightline appears to be an example of low column density,
collisionally ionized material.  The two clouds at $\mvlsr \approx
-29$ and $-49$ \kms\ may be similar to the HVCs observed towards the
disk stars 23 Ori and $\tau$ CMa.  Though the negative velocity IVCs
towards \mucol\ have slightly lower column densities and appear at
lower velocities, the ionization state of these IVCs appear to be
similar to the HVCs presented by Trapero \etal\ (1996).

\subsection{Physical Conditions in the Intermediate-Velocity Gas}
	\label{subsec:IVphysicalconditions}

\begin{figure}
\epsscale{1.0}
\plotone{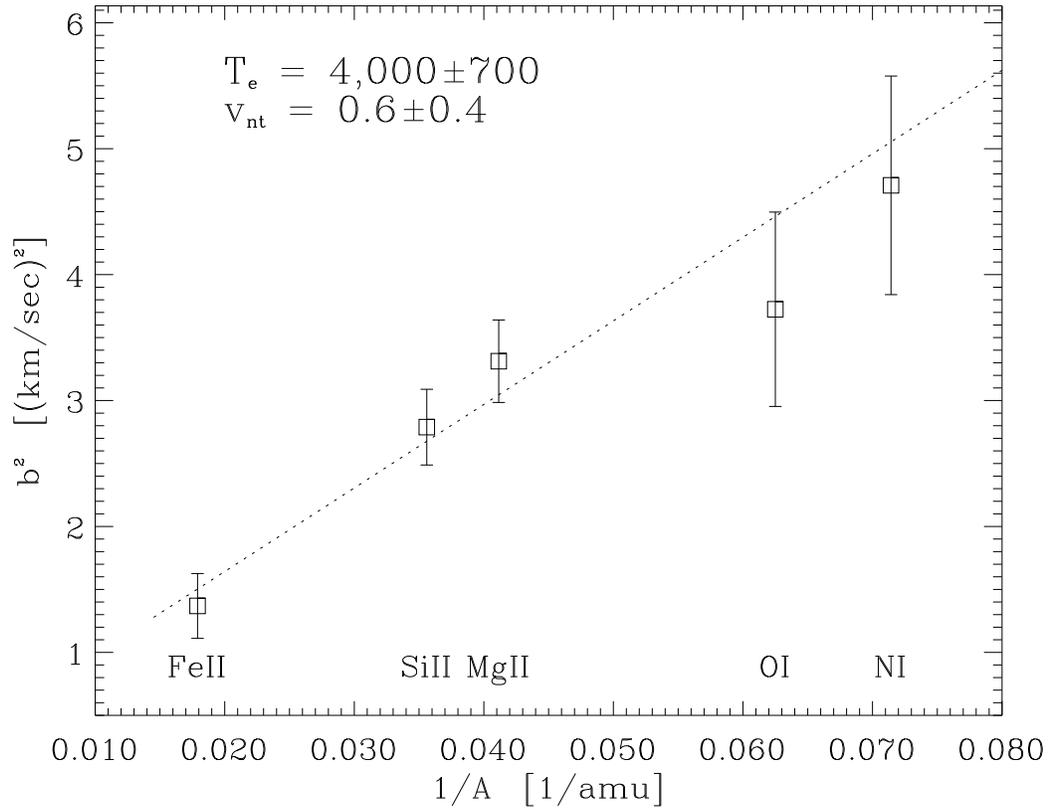}
\figcaption{The square of the Dopper parameter, $b^2$, derived for
component 4 versus the inverse of the atomic mass, $1/A$, for the
species \protect\ion{N}{1}, \protect\ion{Mg}{2}, \protect\ion{Si}{2},
and \protect\ion{Fe}{2}.  The line represents the best fit, yielding a
temperature $T = 4,000 \pm 700$ K with a contribution from non-thermal
velocities $v_{nt} = 0.6\pm0.4$ km s$^{-1}$. \label{fig:tempcomp4}}
\end{figure}

Our component fitting results for component 4 include good $b$-values
for species over a wide range of atomic mass, from \ion{N}{1} to
\ion{Fe}{2}.  Figure \ref{fig:tempcomp4} shows a plot of $b^2$
vs. $1/A$, where $A$ is the ion mass in atomic mass units, for
component 4.  The best linear fit to these data is also overplotted.
Theoretically this fit is a sum of the expected broadening from
thermal and non-thermal sources such that:
\begin{equation}
   b^2 = \frac{2kT}{Am_{\rm H}} + v_{nt}^2,
\label{eqn:bval}
\end{equation}
where $m_{\rm H}$ is the mass of hydrogen and $v_{nt}$ is the
non-thermal velocity dispersion.  We determine a temperature $T =
4,000\pm700$ K with a non-thermal velocity component $v_{nt} =
0.6\pm0.4$ \kms\ for component 4.  While the fit to the data is quite
good, this temperature is slightly lower than typically found in
diffuse interstellar clouds (e.g., $T\approx6,000$ K by Spitzer \&
Fitzpatrick 1993).  If we use only the measurements for \ion{Fe}{2},
\ion{Si}{2}, and \ion{Mg}{2} in this fit, we derive $T = 4,300
\pm700$, in agreement with the values derived when including the less
certain \ion{O}{1} and \ion{N}{1} $b$-values.

Spitzer \& Fitzpatrick (1993) have discussed the population of the
upper fine-structure level of the C$^+$ ground state via electron
collisions.  Their Equation (7) relates the value of $n_e$ in diffuse
clouds to the column of C$^{+*}$ in the same way we have approached
the excitation of the Si$^+$ lines in \S
\ref{subsec:physicalconditions}.  The extreme strength of the C$^+$
line at \wave{1334} makes the column density of the lower-lying level
uncertain.  Adopting the atomic constants used by Spitzer \&
Fitzpatrick (1993), namely: $A_{21} = 2.29 \times 10^{-6}$ s$^{-1}$
(Mendoza 1983) and $\Omega_{12} = 2.90$ (Osterbrock 1989), we find
\begin{equation}
\frac{N({\rm C}^{+*})}{N({\rm C^+})} = 5.5 \frac{n_e}{T^{1/2}}.
\label{eqn:carbon}
\end{equation}
We have assumed that the column densities may be used in place of the
particle densities. Using Equation (\ref{eqn:carbon}) with the
temperature $T = 4,000\pm700$ K derived from our component fitting,
and assuming $\log N({\rm C^+}) = 13.75\pm0.1$, gives $n_e =
0.47\pm0.14$ \percc.  Analysis of the Mg$^{\rm o}$/Mg$^+$ ionization
equilibrium using Equation (\ref{eqn:ionequilibrium}) and this
temperature yields $n_e = 0.64\pm0.14$ \percc, in surprising agreement
with the density derived from C$^{+*}$.

The cooling per nucleon via the \ion{C}{2} $^2P_{3/2} \rightarrow
^2P_{1/2}$ radiative transition at $\lambda158 \ \mu$m can be written
\begin{equation}
l_c = h \nu_{12} A_{12} N(\mbox{\ion{C}{2}}^*)/N({\rm H}),
\end{equation}
where $h \nu_{12} = 1.26\times10^{-4}$ ergs is the energy of the
emitted photon, and $A_{21} = 2.29 \times 10^{-6}$ s$^{-1}$ is the
Einstein $A$-value for the transition (Mendoza 1983).  If $\log N({\rm
H}) \approx 17.3 - 17.7$ in this component, our \ion{C}{2}$^*$ column
density yields $l_c = (3-4)\times 10^{-25}$ ergs s$^{-1}$
nucleon$^{-1}$.  This value is similar to that derived for sightlines
in the Galactic disk (Pottasch \etal\ 1979; Gry \etal\ 1992), and
higher than observed along most halo sightlines (Savage \& Sembach
1996b).

The charge exchange reaction between O$^+$ and H$^{\rm o}$ is large
enough to effectively keep the ionization fractions of O and H
coupled.  If we assume that O and Si in components 3 and 4 have the
same intrinsic gas-phase abundances (Si in low-density regions may be
slightly more abundant), we can use the observed deficit of \OI\ to
estimate the ionization fraction of H in these clouds.  Using the
values [O/Si] in Table \ref{table:abundances}, and again using $x_e
\equiv n_e / n_{\rm H} \approx N({\rm H}^+)/N({\rm H})$, we find $x_3
\approx 0.7$ and $x_4 \approx 0.5$.  If we assume $n_e \approx 0.5$
\percc\ in component 4, we estimate $n_{\rm H} \approx 1.0$ \percc,
assuming He is mostly neutral.  The corresponding thermal pressure is
$P/k \approx 6,000$ K \percc, similar to other pressure estimates for
diffuse gas.  The assumption of an intrinsic value [O/Si]$\, = 0.0$
likely makes the estimated ionization fractions upper limits and the
densities and pressures lower limits.

If we assume $x \approx 0.5$ and $n_e \approx 0.5$ \percc\ for
component 4 and adopt the column density for H derived above ($\log
N({\rm H}) \approx 17.3 - 17.7$), the thickness of the absorbing
region is $\sim 0.04 - 0.1$ pc.  For the HVC towards $\tau$ CMa,
Trapero \etal\ (1996) suggest the absorbing region is likely $\sim
0.065$ pc.

\section{DISCUSSION}
\label{sec:discussion}

\subsection{Interstellar Reference Abundances}
\label{subsec:refabundance}

An important result of this paper is our conclusion that the relative
abundances of S, P, and Zn are consistent with solar system relative
abundances, i.e., $[{\rm Zn/S}] \approx [{\rm P/S}] \approx 0.0$.
Furthermore, $[{\rm S/H}]=+0.08\pm0.02$, $[{\rm P/H}]=+0.05\pm0.02$,
and $[{\rm Zn/H}]=+0.06\pm0.08$ using the integrated sightline column
densities with no correction for the presence of ionized gas along the
sightline.  There is no evidence for saturation in our \ion{S}{2}
observations of the weaker two lines (\twowave{1250.534 and 1253.811};
see Figure \ref{fig:navprofiles}), and the strongest line shows only
moderate saturation effects.  In \S \ref{subsec:ionization} we showed
that the contribution of ionized gas in an \HII\ region to these
species was in the range 0.04 to 0.05 dex.  An additional correction
of $\sim-0.04$ dex may need to be applied, in particular, to the \SII\
measurements to correct for the presence of partially-ionized material
in the predominantly neutral \HI-bearing clouds.  Thus relative to H,
the abundances of S, P, and Zn are consistent with solar (within the
errors).

The conclusion that S, P, and Zn relative to H are consistent with
their solar system abundances (or greater) is significant because we
believe we understand the effects of ionized gas on these and other
species along this sightline.  Abundance measurements of [S/H] derived
from \SII\ and \HI\ are particularly susceptible to contamination from
H$^+$-containing regions.  In ionized regions, which are not included
in the derived \HI\ reference column density, the dominant ion of S
can be S$^+$ given its high ionization potential (23.3 eV; see Howk \&
Savage 1999).  Thus the derived gas-phase abundance [S/H] may be
unreliable if significant amounts of H$^+$ are present along a
sightline.  Comparing our quality measurements of \HI\ with
\ion{Zn}{2} and \ion{P}{2}, which have lower ionization potentials
(18.0 and 19.7 eV, respectively) than \ion{S}{2}, allows us to firmly
conclude that we are seeing gas with solar system abundances in S, Zn,
and P.

The derived solar system relative abundances of the gas towards
\mucol\ are not tracing initially sub-solar metallicity material
contaminated by a SN or other enrichment event.  The measurement of S,
an $\alpha$-process element, Zn, a tracer of Fe-peak elements, and P,
an odd-Z element produced primarily in the hydrostatic O- and
Ne-burning shells of massive stars (Timmes \etal\ 1995), shows no
evidence for a different nucleosynthetic signature in the gas towards
\mucol\ than that responsible for producing the observed metals in the
solar system.  

Recent studies by Sembach \etal\ (1995) and Roth \& Blades (1995) have
shown the abundance of Zn in the ISM is a function of the sightline
value of $\log N(\mbox{\ion{H}{1}})$ as well as the fractional
abundance of molecular hydrogen, $f({\rm H}_2) \equiv 2 N({\rm
H}_2)/[2 N({\rm H}_2) + N(\mbox{\ion{H}{1}})]$.  This dependence on
sightline parameters such as column density, average sightline neutral
density, and molecular content may be interpreted as evidence for the
incorporation of Zn into grains in the ISM (e.g., Jenkins 1987).  The
\mucol\ sightline samples low density WNM gas ($\langle n_{\rm H I}
\rangle \equiv N(\mbox{\ion{H}{1}})/d \approx 0.06$ \percc).
Therefore it would not be surprising if this sightline were to show
higher gas-phase abundances of Zn than a sample including high density
sightlines.  However, both Sembach \etal\ (1995) and Roth \& Blades
(1995) find values $\langle [{\rm Zn/H}] \rangle \approx -0.2$ in
their sample of Milky Way disk and halo stars with little molecular
gas [$\log f({\rm H}_2) \la -3$].  This is significantly lower than
the observed [Zn/H] towards \mucol, which has $\log f({\rm H_2}) \sim
-4.35$ according to the results of SY.  The \HI\ column density
towards \mucol\ is significantly lower than any in the Sembach \etal\
sample and lower than all but one of the stars studied by Roth \&
Blades.  The few stars from the latter sample with average sightline
neutral densities similar to the sightline to \mucol\ still show
$[{\rm Zn/H}] \sim -0.1$, though the errors are likely of order 0.08
dex.

Quality, high-resolution measurements of [S/H] are somewhat more
sparse in the literature, in part because the transitions of \SII\ are
stronger than those of \ZnII\ and therefore saturate along even
relatively low column density sightlines.  We have collected from the
literature the following quality integrated sightline measurements of
[S/H] using the \SII\ transitions: $[{\rm S/H}] = 0.00 \pm 0.09$
towards $\gamma^2$ Vel (Fitzpatrick \& Spitzer 1994); $[{\rm S/H}] =
-0.05 \pm 0.04$ towards HD~93521 (Spitzer \& Fitzpatrick 1993); and
$[{\rm S/H}] = -0.09 \pm 0.09$ towards HD~215733 (Fitzpatrick \&
Spitzer 1997).  These values were derived assuming \HI\ column
densities from Diplas
\& Savage (1994), except for the $\gamma^2$ Vel [S/H] value, which
uses an \HI\ column density that is the weighted average of the Diplas
\& Savage (1994) measurement with those of Bohlin \etal\ (1978) and
York \& Rogerson (1976).  All of these measurements are roughly
consistent with a solar system abundance of sulfur.  The number of
measurements is too few and the errors on many of the measurements are
too large to provide good estimates of the intrinsic scatter in the S
abundance.  We do note that the high-quality \mucol\ and HD~93521
measurements are inconsistent with one another at $\ga 2\sigma$
significance.  However, we have not accounted for the effects of
ionized gas.  We have seen that towards \mucol\ the contribution from
ionized gas to the \ion{S}{2} measurements can be of order 0.05--0.08
dex.  The values given above should be viewed as upper limits to the
true gas-phase abundance of sulfur.  Disentangling the influence of
ionized gas on the sulfur abundance for many sightlines, where the
fraction of ionized gas may vary considerably, may be non-trivial.

Our measurements of solar-like abundance ratios along the \mucol\
sightline argue for a solar reference abundance along at least one
sightline.  Comparing the [Zn/H] abundance towards \mucol\ with [Zn/H]
along the low-density sightlines studied by Roth \& Blades (1995) and
Sembach \etal\ (1995) suggests possible abundance variations in the
low-density ISM.  Our measurements of solar system relative abundances
along this sightline are in striking contrast to the results of Meyer
\etal\ (1997, 1998) and Cardelli \& Meyer (1997) regarding the
constancy of (sub-solar) O/H, N/H, and Kr/H abundances in the ISM
within $\la1.5$ kpc of the sun.  Our data are inconsistent with $[{\rm
Zn/H}] = -0.2$ or $[{\rm P/H}] = -0.2$ (for comparison with the
results of Sembach \etal\ and Roth \& Blades, as well as the Meyer
\etal\ [O/H] results) at the $2.5\sigma$ and $10\sigma$ levels,
respectively, after subtracting off the expected contribution from
ionized gas (0.04 to 0.05 dex).

There has been growing evidence in the literature to support the
choice of a sub-solar reference abundance in the local ISM (Meyer
\etal\ 1998; Mathis 1996; Snow \& Witt 1995, 1996).  The arguments for
sub-solar abundances appear to be supported by observations of local B
star abundances (e.g., Gies \& Lambert 1992; Kilian-Montenbruck \etal\
1994) and \HII\ region abundances (e.g., Afflerbach, Churchwell, \&
Werner 1996; Simpson \etal\ 1995; Osterbrock, Tran, \& Veilleux 1992;
Shaver \etal\ 1983).  However our results suggest the issue is not
resolved.  There may be intrinsic abundance variations in the ISM near
the sun.  For many sightlines the solar system abundances may actually
be the appropriate reference abundance system.  Some caution is
warranted given that we have identified only one sightline with solar
system abundances.  However, the \mucol\ sightline is exceptionally
well observed, and our understanding of the ionization state of the
gas along this path through the ISM is far better than for most
others.  The interpretation of the measured O/H and N/H ratios are
simplified by the strong charge exchange reactions between O$^{\rm o}$
and H$^{\rm o}$, and to a lesser extent N$^{\rm o}$ and H$^{\rm o}$.
This keeps the relative ionization fractions of these species locked
to that of H.  Though our measurements are of species less strongly
tied to hydrogen, the fact that P and Zn appear in their solar system
ratios relative to H, and that our photoionization modelling can
account for the observed super-solar abundance of S/H, gives us
confidence that we are indeed seeing solar system abundances along
this sightline.

The identification of the proper reference abundance for the local
ISM, or for an individual sightline, is likely not an easy task.
Furthermore, the observed abundances for the B-star and \HII-region
reference systems are less well constrained at this time than for that
of the sun.  Indeed, the spread in the B-star abundances (0.2-0.7 dex)
seen by Kilian-Montenbruck \etal\ (1994) may be evidence that there
are intrinsic variations in the local reference abundance.  However,
these authors see 0.2 to 0.7 dex abundance spreads for stars {\em
within the same clusters}, possibly suggesting very small scale
abundance inhomogeneities in the material from which the stars were
formed or inhomogeneities introduced by the star formation process
itself.  At this point we caution that it is not clear that either
solar system abundances or some constant fraction thereof should be
adopted out of hand as the cosmic reference abundance.

\subsection{Implications for High-Redshift Absorption Systems}

Our work has important implications for the study of high-redshift
quasar absorption line systems, particularly the damped \lya\ systems
(DLAs).  Among the most important aspects of this paper for
interpreting absorption line spectroscopy of higher-redshift gas are
the solar system abundances of undepleted elements as discussed in \S
\ref{subsec:refabundance}, and in particular the measurement $[{\rm
P/Zn}] \approx 0.0$, and the application of new \ion{Ni}{2} oscillator
strengths to the study of the gaseous medium of the Galaxy.

The new high-quality laboratory measurements of \ion{Ni}{2} absolute
oscillator strengths imply that previously derived Ni column densities
should be increased by $\sim 0.3 - 0.4$ dex, depending on the source
of the $f$-values.  In particular, column densities derived using the
$f$-values of Zsarg\'{o} \& Federman (1998) should be increased by
$+0.272$ dex, although there may still be uncertainties in the
oscillator strengths of the lower-wavelength transitions (e.g.,
\twowave{1317 and 1370}; see \S \ref{subsec:measurements}).  Nickel
can be an extremely important element for studies of DLAs because the
Ni/Fe ratio is set by nuclear statistical equilibrium in the sites of
Ni and Fe production.  Thus the ratio of Ni/Fe (dust+gas) is expected
to be the same no matter the nucleosynthetic history of the material.
Observations of disk and halo stars over a large range of
metallicities confirm this expectation (Edvardsson \etal\ 1993;
Gratton \& Sneden 1988, 1991; see also discussion in Lu \etal\ 1996),
although there has been disputed evidence for a slight increase in
[Ni/Fe] at very low metallicities (see discussion in Wheeler, Sneden,
\& Truran 1989 and McWilliam 1997 and references therein).  Edvardsson
\etal\ (1993) derive $[\langle {\rm Ni/Fe} \rangle ] = +0.02 \pm 0.05$
for their sample of 189 galactic disk G and F dwarfs, which cover a
range of abundances $-1.1 \la [{\rm Fe/H}] \la +0.25$.  Gratton \&
Sneden (1991) similarly find $[\langle {\rm Ni/Fe} \rangle ] = -0.05
\pm 0.08$ for their sample of 22 metal-poor stars ($[{\rm Fe/H}] \la
-0.8$).

Measurements of the gas-phase [Ni/H] in disk and halo clouds by
numerous authors have suggested $[{\rm Ni/Fe}] < 0.0$ (see Savage \&
Sembach 1996a and references therein) when using the \ion{Ni}{2}
$f$-values from Morton (1991).  Several groups have found sub-solar
Ni/Fe ratios in DLAs as well (e.g., Lu \etal\ 1996; Prochaska \& Wolfe
1999; Pettini \etal\ 1999).  This depletion of Ni relative to Fe in
DLAs has been interpreted to imply the existence of dust grains in
these systems, given that one expects $[{\rm Ni/Fe}] \approx 0$ in the
absence of dust.  Lu \etal\ (1996), however, pointed out the possible
uncertainties in the $f$-values of the \ion{Ni}{2} transitions and
adopted to reject this line of evidence for presence of dust given the
uncertainties in the atomic data.  The recent Fedchak \& Lawler
measurements of the \ion{Ni}{2} resonance oscillator strengths revise
all of the previous DLAs Ni abundance measurements upwards by $\sim
0.3$ dex.  This is typically enough to bring the Ni/Fe ratio near the
solar system abundance in DLAs.  Combining the 20 DLAs with {\em
detected} Fe and Ni absorption from the observations of Lu \etal\
(1996) and Prochaska \& Wolfe (1999) we find $[ \langle {\rm Ni/Fe}
\rangle ] = 0.15\pm0.25$ in these DLAs (typical measurement errors are
$\sim0.1$ dex).  The absorbing systems in this combined dataset cover
a range in redshift $1.8 \la z \la 3.9$; the [Ni/Fe] measurements seem
not to be a function of redshift.

The revisions in the \ion{Ni}{2} oscillator strengths imply that
[Ni/Fe] measurements no longer offer a straightforward indicator of
the presence of dust.  We calculate a revised value $[{\rm Ni/Fe}] =
+0.04\pm0.05$ for cloud A along the \zoph\ sightline (using both
revised \ion{Ni}{2} and \FeII\ oscillator strengths).  Coupled with
our own measurements of [Ni/Fe]$\, = +0.02\pm0.06$ towards \mucol\
(see Table \ref{table:abundances}), this implies the ratio Ni/Fe can
be consistent with solar system abundances, {\em even in the presence
of dust}.  Many of the existing claims for dust in DLAs that use
sub-solar Ni/Fe ratios as evidence should be reconsidered given the
new absolute $f$-value scale for the \twowave{1710 and 1741}
transitions, and the suggested relative scale for many other
transitions.  This result once again emphasizes the need for
high-quality oscillator strength measurements.

Another possibly important aspect of our work for studies of high-{\em
z} gas is the observation [P/Zn]$\, \approx 0$ along the \mucol\
sightline.  The elements S and Zn have often been used to search for
enhancements of $\alpha$-process elements over Fe-peak elements in the
distant universe, particularly because they are not expected to be
incorporated into dust grains in large amounts.  These comparisons
offer clues to the nucleosynthetic history of the gas, which can be
compared with that of low-metallicity Galactic stars.  Unfortunately,
the strength of the \ion{S}{2} transitions relative to the \ion{Zn}{2}
lines often means that the S lines are saturated where the Zn lines
are strong enough to be detectable.  The \ion{P}{2} \wave{1152}
transition is expected to have a strength closer to the \ion{Zn}{2}
(e.g., see Table \ref{table:eqwidths}) and might therefore be better
choice for comparing abundances of an elemental produced by type II
SNe (P) with one tracing Fe-peak elements (Zn).

The interpretation of the P/Zn ratio is not as straightforward as that
of S/Zn, however.  First, $\alpha$-elements, such as S, are known to
show enhancements relative to Fe-peak elements, traced by Zn, in
low-metallicity Galactic halo stars (e.g., Wheeler \etal\ 1989 and
references therein).  This is understood as a result of the time delay
between the explosions of massive stars as type II SNe (which produce
the $\alpha$-process elements) and the explosions of white dwarfs in
binary systems as type Ia SNe (which in turn produce much of the
Fe-peak elements).  Unfortunately the emperical information about the
chemical evolution of P is not as well constrained given the lack of
abundance measurements of P in Galactic stars (see Timmes, Woosley, \&
Weaver 1995).  The production of $^{31}$P primarily occurs in the
hydrostatic O- and Ne-burning shells of pre-supernova massive stars
(Woosley \& Weaver 1995; Trimble 1991).  Unlike many elements, the
amount of $^{31}$P produced in a type II SN is relatively insensitive
to the exact mechanism of the explosion (Timmes \etal\ 1995).  Timmes
\etal, in their detailed chemical evolution calculations, find the
abundance [P/Fe] should be significantly sub-solar in low metallicity
environments ($[{\rm P/Fe}] \ga -0.8$), and rise to super-solar values
($[{\rm P/Fe}] \la +0.15$) as the metallicity [Fe/H] increases.
Therefore, the yield of P in type II SN explosions is expected to be a
function of the initial metallicity.  This is similar to the expected
behavior of [N/Fe] if the production rate of nitrogen depends on the
initial CNO abundances, i.e., if the synthesis of N has a secondary
origin.  The measurement of [P/Zn] in high-redshift DLAs thus probes
the chemical evolution of an element with a secondary origin in
massive stars and an Fe-peak element produced primarily through
low-mass stars.

The validity of conclusions drawn from P and Zn comparisons in DLAs
depends on the assumption that P and Zn are undepleted in the gas, or
have similar levels of depletion.  Our measurements suggest that the
former is true along a low-density WNM sightline in the Galactic disk.
The abundance of P in the diffuse ISM of the galaxy has not been well
characterized.  Dufton, Keenan, \& Hibbert (1986) used {\em
Copernicus} data to study the abundance of P; they concluded that the
abundance of P in low-density clouds was consistent with solar.
However, there were only a handful of \ion{P}{2} measurements in their
sample of warm, low-density sightlines.  Savage \etal\ (1992) find
$[{\rm P/Zn}] \approx -0.2$ in the warm cloud A towards \zoph\ (using
our adopted oscillator strengths), which otherwise shows abundances
similar to component 1 along the \mucol\ sightline.  While the \mucol\
data suggest that P and Zn trace each other well in the Galactic ISM,
the contradictory results for a similar cloud towards \zoph\ suggest a
better sample of P measurements is needed to constrain the relative
behavior of P and Zn in the Milky Way.  Measurements of the P to Zn
abundance in the ISM of the Magellanic Clouds and other local
low-metallicity environments may also be helpful in this regard.
Without this fundamental knowledge of the properties of low-redshift
gas, the usefulness of P and Zn as tracers of high-redshift chemical
evolution may be limited.

\subsection{Composition of Dust in the Diffuse ISM}

In Table \ref{table:silicates} we have compiled data for the
dust-phase abundances of Mg, Si, and Fe in well-studied diffuse
interstellar clouds.  The data are taken from the works of Fitzpatrick
\& Spitzer (1997), Savage \etal\ (1992), Sofia \etal\ (1994), and
Spitzer \& Fitzpatrick (1993, 1995) as well as this work.  References
to the abundance measurements are given in the table.  We only include
clouds for which quality measurements of Mg, Si, and Fe exist.  The
data have been adjusted to reflect our choice of \MgII\ oscillator
strengths (Fitzpatrick 1997), and reference abundance (see Savage \&
Sembach 1996a).  The table has two sections: one showing the
dust-phase abundances of several ``warm disk'' clouds for comparison
with the dust-phase abundances in component 1, and the other
containing information on ``halo'' clouds for comparison with
component 2 (see Sembach \& Savage 1996).  Also given are the weighted
mean of each of these samples and the standard error of the mean.  The
sample presented here does not represent an unbiased selection, but
the averages are useful for judging the significance of variations in
the dust-phase abundances within these clouds.

Our selection of clouds for each group (halo or disk) is based upon
the Sembach \& Savage (1996) categorizations and the similarity of
sightline environments.  As such we have chosen clouds with similar
properties for each category.  The small spread in (Mg/H)$_d$,
(Si/H)$_d$, and (Fe/H)$_d$ in the warm disk clouds listed in Table
\ref{table:silicates} is quite interesting.  The dust phase abundances
of Mg, Si, and Fe show a spread of $\la \pm10\%$ from the mean values,
with the exception of the (Fe/H)$_d$ measurement for HD~18100.  This
sightline, for which we have used the sightline-integrated values of
[X/Zn] in deriving the dust-phase abundances (Savage \& Sembach
1996b), likely contains a larger fraction of low-density material than
the others categorized as warm disk clouds.  Even given the large
difference in (Fe/H)$_d$ for HD~18100 [and correspondingly large
excursion of (Fe/Si)$_d$ from the mean], the value of [(Mg+Fe)/Si]$_d$
is still only $\sim10\%$ from the mean value.  Thus it would appear
that a population of warm clouds in the Galactic disk can be
identified with very similar dust composition.  Component 1 towards
\mucol\ seems to lie on the low end of the (Mg/H)$_d$ and (Si/H)$_d$
values, which is consistent with the slightly higher [Mg/S] and [Si/S]
values in this blend compared with cloud A towards \zoph.  The
dust-phase abundance of Fe is close to that observed in other warm
disk clouds.  The dust-phase abundances associated with the disk
clouds in Table \ref{table:silicates} may be representative of grains
in the WNM of the disk, particularly given the relatively small
variation in the abundances.

The data given in Table \ref{table:silicates} suggest there is a quite
small spread in the dust-phase abundances for clouds having
``halo''-like abundances, although the errors in the available data
are larger than for typical disk clouds.  The dust-phase abundances of
Fe, Mg, and Si relative to H are consistently lower in the halo clouds
than the disk clouds of Table \ref{table:silicates}.  Again, \mucol\
seems to lie near the lower end of the dust-phase abundances.  This is
consistent with the findings of SSC.  In Figure 6 of Sembach \& Savage
(1996), which shows gas-phase abundances for many sightlines, the SSC
results for component 2 along the \mucol\ sightline typically show the
highest abundance [$X$/H] for a given element. Though the errors are
large, it does not appear that the ratios of Si, Mg, and Fe inclusions
in the grains associated with component 2 are different than for the
other halo clouds listed in Table \ref{table:silicates}.  This cloud
is therefore characterized by a greater return of elements to the
gas-phase than the other halo clouds with no fundamental change in the
dust-phase composition.

Table \ref{table:silicates} shows that the dust-phase abundances of
both typical WNM disk clouds and halo clouds are inconsistent with the
sole constituent of Galactic dust being pure olivine or pyroxene
grains.  Fitzpatrick (1997) has also pointed out that the gas-phase
abundances from works of Fitzpatrick \& Spitzer (1997) and Spitzer \&
Fitzpatrick (1993, 1995) are consistent with non-silicate inclusions
of Fe and/or Mg into grains, possibly in the form of oxides, if
adopting the \MgII\ $f$-values he derives in that work.  It has been
suggested that the dust-phase abundances in the halo clouds are
representative of the composition of the resilient cores of grains,
and in the warm disk clouds of the cores plus mantles (see Savage \&
Sembach 1996a).  If this is the case then the data in Table
\ref{table:silicates} suggest the mantles contain Si, Fe, and Mg in
the ratio Si:Fe:Mg$\, \approx \,$7:5:4.  Therefore,
[(Mg+Fe)/Si]$_{mantle} \approx 1.4$, which {\em is} consistent with a
mixture of the common silicates mentioned above.  This is slightly
larger than the value [(Mg+Fe)/Si]$_{mantle} \approx 0.94$ derived by
Savage \& Sembach (1996a).  This difference is primarily due to the
assumed \ion{Mg}{2} oscillator strengths.  The data in Table
\ref{table:silicates} imply the postulated resilient grain cores must
include a greater fraction of non-silicate Fe and Mg inclusions than
the more fragile mantles.

The fact that the different oscillator strengths used for the \MgII\
doublet near \wave{1240} variously imply no Mg depletion (e.g.,
Spitzer \& Fitzpatrick 1993, 1995), moderate Mg depletion similar to
that of Si (this work), and relatively large Mg depletion similar to
that of Mn (e.g., Sembach \& Savage 1996) shows the need for an an
accurate laboratory measurements of the $f$-values for these important
transitions.

\section{SUMMARY}
\label{sec:summary}

We present absorption line profiles from the GHRS for many atomic and
ionic species along the line of sight to the nearby star \mucol.  The
majority of these data were taken with the echelle-mode resolution of
$\approx 3.5$ \kms, which we analyze with a combination of the
apparent column density method and component fitting techniques.  This
dataset is the most extensive interstellar absorption line database
for studying the WNM of the Galaxy.  The principal results of this
study are:

\begin{enumerate}

\item The presence of ionized gas near $\mvlsr \approx 0$ \kms\ is
suggested by the presence of \ion{S}{3}, \ion{Al}{3}, \ion{Fe}{3},
\ion{Si}{3} and {\small IV}, and \ion{Si}{2}$^*$.  \copernicus\
observations of \ion{N}{2} and \ion{N}{2}$^*$ and \wham\ spectra of
H$\alpha$ emisson also imply ionized gas at these velocities.  Based
upon CLOUDY photoionization models and our measurements of the doubly
ionized species, we estimate the contribution from gas associated with
ionized hydrogen to the column densities of singly ionized species
used as abundance tracers in the neutral gas.  We find that the
relative contribution from ionized gas to these tracers is minimal.

\item We find no evidence for sub-solar gas-phase abundances of Zn or
P relative to S over the velocity range for which we are able to
measure these species, i.e., [Zn/S]$ \approx $[P/S]$ \approx 0.0$.
The gas-phase abundances [Zn/H] and [P/H] are also solar within the
errors.  The [S/H] abundance derived from \SII\ and \HI\ is
super-solar by $+0.08$ dex, though our photoionization model, which
corrects for S$^+$ in ionized gas reduces this by $-0.05$ to $-0.08$
dex.  This suggests the proper reference abundances for investigating
the gas along the sightline to \mucol\ are the solar system abundances
(e.g., Anders \& Grevesse 1989).

\item Adopting the $f$-values derived for \ion{Mg}{2}
\twowave{1239.925 and 1240.395} by Fitzpatrick (1997), we find the
abundance patterns of Mg and Si in the WNM towards \mucol\ as a
function of velocity are very similar.  This result is in accord with
the trends noted by Fitzpatrick (1997).

\item The gas-phase abundances for the warm neutral absorbing complex
centered at $\mvlsr = +3$ \kms\ are similar to those found in the warm
cloud towards $\zeta$ Oph and other warm disk clouds (Savage \&
Sembach 1996a).  Table \ref{table:abundances} gives our derived
gas-phase abundances which are shown in Figures
\ref{fig:totalabundances} and \ref{fig:velabundances}.  This level of
gas-phase abundances seems to be typical of WNM clouds in the diffuse
ISM of the Galactic plane.

\item Component 2 at $\mvlsr = +20.1$ \kms\ shows higher gas-phase
abundances of the refractory elements compared to component 1,
suggesting grain processing has played a part in the evolution of this
cloud.  Comparison of the gas-phase abundances of components 2 to 1
suggests the liberation from the the solid phase $(55\pm20)\%$ of Si,
$(50 \pm 4 )$\% of Fe, $(23 \pm8)$\% of Ti, $(60 \pm9)$\% of Cr, $(26
\pm 5 )\%$ of Mn, and $(32 \pm4)$\% of Ni.  We rule out enrichment
from Type Ia SNe as the cause of these enhanced gas-phase abundances.

\item The values of the dust-phase abundances of Mg and Fe relative to
Si derived for the low-velocity material towards \mucol,
[(Mg+Fe)/Si]$_d = 2.7 - 3.3$, assuming solar relative abundances, are
inconsistent with incorporation of Fe and/or Mg solely into olivine
and pyroxene-type silicates, which predict [(Mg+Fe)/Si]$_d = 1 - 2$.
Oxides are a likely component of the Mg- and Fe-bearing grains in the
neutral ISM towards this star.  This result is even more pronounced if
one adopts B-star reference abundances.

\item The low-velocity gas (components 1 and 2) along this sightline
is characterized by $T_e \approx 6,000 - 7,000$ K and $n_e \approx
0.3$ \percc, derived from ionization equilibrium of Mg and Ca.  The
temperature is typical for warm diffuse interstellar clouds of the
Galactic disk and low halo (e.g., Spitzer \& Fitzpatrick 1993).  In
the densest ionized gas regions at low velocities we find $0.2 \la n_e
\la 1.2$ \percc\ from excitation equilibrium analysis of the
$^2P_{3/2}$ and $^2P_{1/2}$ fine structure levels of Si$^+$.  This
ionized gas may trace an \HII\ region about \mucol.

\item The relative column densities of atomic and ionic species in the
IVCs towards \mucol, components 3 and 4 at $v_{\rm LSR} = +31.0$ and
$+41.2$ \kms, respectively, are dominated by the effects of
(photo)ionization and elemental incorporation into grains. Our
conclusions regarding these clouds are limited by the uncertain
ionization state of the gas.  However, we can bracket $-0.31 \lesssim
[{\rm Al}/{\rm Si}] \lesssim -0.11$ in component 3 and $-0.51 \lesssim
[{\rm Al}/{\rm Si}] \lesssim -0.26$ in component 4.  These
measurements are based upon several stages of ionization and should
therefore be quite robust.  Component 5 at $\mvlsr = -30$ km s$^{-1}$
traces collisionally ionized gas and is only measured in five species.

\item From our component fitting results for component 4 we derive $T
= 4,000\pm700$ K.  Using collisional excitation equilibrium and the
column densities of \ion{C}{2} in the $^2P_{3/2}$ and $^2P_{1/2}$ fine
structure levels, we derive $n_e = 0.47\pm0.14$ \percc.  From
ionization equilibrium of \ion{Mg}{1} and {\small II} we derive $n_e =
0.64\pm0.14$ \percc.  The cooling in the \ion{C}{2} $\lambda158 \
\mu$m line via radiative decay from the $^2P_{3/2}$ level is likely in
the range $(3-4)\times10^{-25}$ ergs s$^{-1}$ nucleon$^{-1}$.
Component 5 has $T\la 36,000$ K.  The ratio of \ion{Si}{2} to {\small
III} is similar to that expected for gas in collisional ionization
equilibrium at $T\approx 25,000$ K.

\end{enumerate}

\acknowledgements 

We thank J. Mathis for conversations on many aspects of this project.
We also extend thanks to M. Haffner and R. Reynolds for providing
their \wham\ spectrum, and to B. Welsh and N. Craig for their
\ion{Ti}{2} profile.  Our thanks also to E. Fitzpatrick for sharing
his component-fitting software with us.  We thank R. Robinson for
sharing data on the post-COSTAR GHRS LSF for the LSA.  We also thank
J. Fedchak \& J. Lawler for sharing their important work on
\ion{Ni}{2} oscillator strengths prior to publication.  The work of
G. Ferland at the University of Kentucky and his collaborators on the
CLOUDY code has provided us with a great resource, and we thank these
workers for their years of effort.  Our analysis has made use of the
SIMBAD database, operated at CDS, Strasbourg, France.

Support for this work was provided by NASA through grant number
AR07538.01-96A from the Space Telescope Science Institute, which is
operated by AURA, Inc., under NASA contract NAS 5-26555.  JCH also
recognizes support from a NASA Graduate Student Researcher Fellowship
under grant number NGT-5-50121.

\pagebreak 

\appendix
\section{DERIVATION OF $N(\mbox{H$\,${\small I}})$ TOWARDS $\mu$ COL}
\label{appendix:lya}
	
In this appendix we derive the column density of neutral hydrogen
along the sightline towards \mucol.  The \HI\ column density for this
sightline has been determined previously by Diplas \& Savage (1994)
and Bohlin \etal\ (1978) from {\em IUE} and \copernicus\ satellite
spectra.  However, the high signal to noise observations taken of the
region near Lyman-$\alpha$ (\lya) with the intermediate resolution
gratings of the \ghrs, coupled with their exceptional scattered light
properties, make a new determination of \NHI\ a useful endeavor given
our final goal of deriving very accurate gas-phase abundances along
this sightline.

To derive the \HI\ column density towards \mucol\ we use the continuum
reconstruction method (Bohlin 1975; Diplas \& Savage 1994; Sofia
\etal\ 1994).  The optical depth of an interstellar absorption line as
a function of wavelength, $\tau_\lambda$, is (Spitzer 1978):
\begin{equation}
\tau_\lambda (N) = \frac{\pi e^2}{m_e c} f N \phi(\Delta \lambda) \equiv
	\sigma_\lambda N,
\end{equation}
where $N$ is the column density of the absorbing species, $f$ is the
oscillator strength, $\phi$ is the line profile function, and $\Delta
\lambda \equiv \lambda - \lambda_o$, where $\lambda_o$ is the
wavelength at line center.  The continuum reconstruction method finds
the correct column density of an ionic species by multiplying the
observed flux, $F_\lambda$, by $\exp[+\tau_\lambda(N)]$.  The column
density $N$ is chosen such that the reconstructed continuum best
matches the non-absorbed stellar continuum, $F_{\lambda, o}$.

This method is best applied to very strong lines with cores near zero
flux (very large optical depths) and strong Lorentzian damping wings.
We rely on these wings to provide us with an accurate estimate of
$\tau_\lambda (N)$.  The Voigt line profile function is a convolution
of natural and Doppler broadening components:
\begin{equation}
\phi_V (\Delta \lambda) = \phi_L(\Delta \lambda) \otimes 
	\phi_D(\Delta \lambda).
\end{equation}
The Lorentzian profile, $\phi_L(\Delta \lambda)$, is 
\begin{equation}
\phi_L(\Delta \lambda) = \frac{\Sigma A/4 \pi^2}{(\Sigma A/4 \pi)^2 
	+ [(c / \lambda_o^2)\Delta \lambda]^2},
\end{equation}
where $\Sigma A$ is the summed Einstein probabilities for spontaneous
radiative decay.  Using the $f$-values and transition probabilities
from Morton (1991), we derive the following numerical values in
$\sigma_\lambda$ for the \lya\ absorption:
\begin{equation}
\sigma_\lambda({\rm cm^2})
	= \left\{  \frac{4.26\times \e{-20}}
	{6.04 \times \e{-10} + \Delta \lambda^2} \right\}
	\otimes \phi_D (\Delta \lambda),
\end{equation}
where $\Delta \lambda$ is in \AA.  The Dopper profile, $\phi_D(\Delta
\lambda)$, is written
\begin{equation}
\phi_D(\Delta \lambda) = \frac{\lambda_o}{\pi^{1/2} b} 
	\exp[-(c\Delta \lambda/\lambda_o b)^2].
\end{equation}
Here the Dopper spread parameter, $b$, includes thermal and
non-thermal components added in quadrature [e.g., equation
(\ref{eqn:bval})].  To derive the thermal component we assume $T =
6,000$ K, equivalent to the temperature for component 1 derived in \S
\ref{subsec:physicalconditions}.  For the non-thermal component we
estimate the $b$-value of a single gaussian component fit to the
\ion{S}{2} \Nav\ profile.  Assuming the spread is primarily due to
non-thermal broadenings, we derive $v_{nt} \approx 8$ \kms
.\footnote{For gas with a temperature of 6000 K, we expect a $b$-value
$b \la 2$ \kms\ for S II, suggesting that the non-thermal motions do
dominate the observed \Nav\ profile of this ion.}  The total Dopper
spread parameter is $b = 12.8$ \kms\ [in good agreement with the value
$b = 12^{+3}_{-2}$ derived by York \& Rogerson (1976)].

To derive the reconstructed continuum flux, $F_\lambda^R$, we
therefore perform the following
\begin{equation}
F_\lambda^R = \frac{F_\lambda }
	{\exp[- \sigma_\lambda N]\otimes \phi_I(\Delta \lambda)}.
\end{equation}
The additional term $\phi_I(\Delta \lambda)$ in the denominator is
the instrumental spread function.

We adopt an instrumental spread function that consist of two Gaussian
components: a narrow ``core'' and more extended ``halo.''  The width
of the core component was taken from the plots shown in the \ghrs\
Instrument Handbook (Soderblom \etal\ 1995).  The adopted resolutions
of the G140M and G160M gratings at \wave{1215.67} are $\Delta v =
14.4$ \kms\ and $19.3$ \kms\ (FWHM), respectively.  The halo component
of the adopted spread function is due to effects at the detector (see
Cardelli, Ebbets, \& Savage 1990).  We characterise it as a Gaussian
with a FHWM of 5 diodes (Spitzer \& Fitzpatrick 1993).  The relative
contribution of these two components was derived following Spitzer \&
Fitzpatrick (1993) by fitting a spline to the relative contributions
as a function of wavelength observed in pre-flight testing of the
instrument (Cardelli \etal\ 1990).  At \wave{1215.67} the contribution
of the halo component is 7.5\% of the total spread function.

\begin{figure}
\plotone{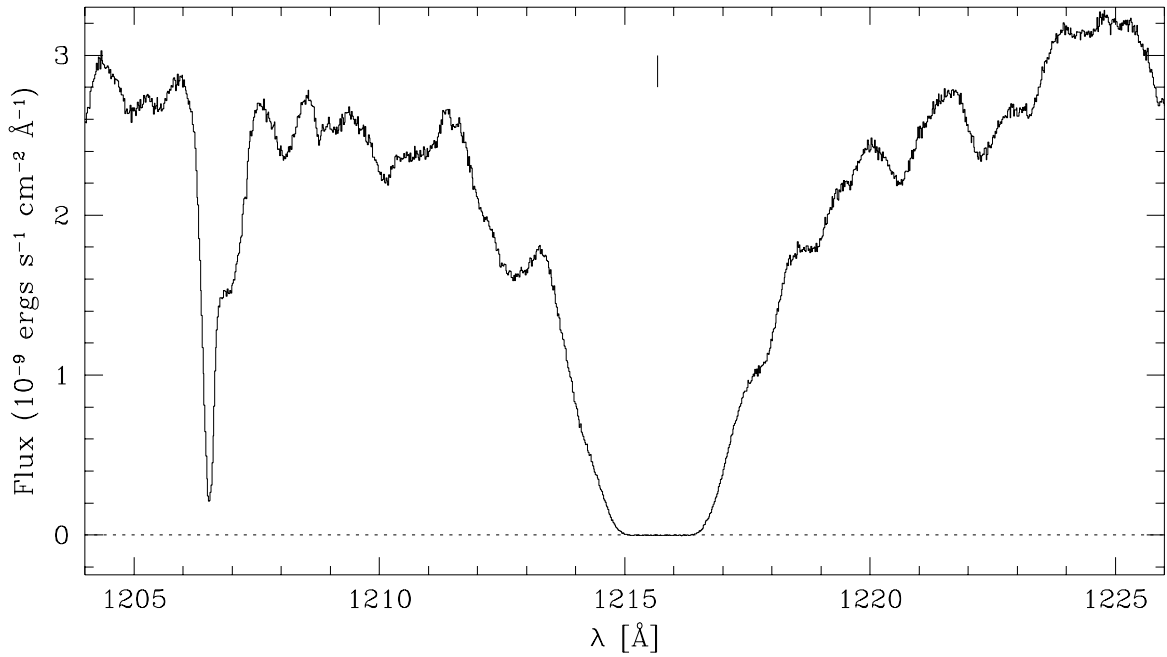}
\figcaption{Observed flux as a function of wavelength (LSR) in the
G160M observations of the region containing interstellar \lya\ at
$\lambda1215.67$ \AA.  These data have a signal to noise ratio of
$\sim 100:1$.  The tick marks the expected position of the line
center.  Absorption due to interstellar the \protect\ion{Si}{3}
$\lambda1206.5$ \AA\ transition is also visible in this spectrum.  All
of the broad features in this spectrum are due to stellar lines.
\label{fig:lya}}
\end{figure}

The \ghrs\ dataset for \mucol\ includes observations of the region
containing interstellar \lya\ absorption ($\lambda_o = 1215.670$) made
through the SSA and LSA with both the G140M and G160M intermediate
resolution gratings.  To derive the \HI\ column density we have used
the G160M SSA data ($S/N \ga 100$) with the LSA and SSA G140M data
($S/N \ga 60$ and 30, respectively).  We use observations taken
through both apertures with the G140M data because the signal to noise
of the SSA data do not allow us to probe as far into the core as we
would like, while the LSA data are complicated by the poorly known
instrumental spread function.  The scattered light properties of the
intermediate resolution \ghrs\ first-order gratings are superb.
Figure \ref{fig:lya} shows the wavelength region near \lya\ observed
with the G160M grating.  The core of the deep \lya\ profile is flat
bottomed at zero flux level, consistent with no contribution from
scattered light.  The use of observations taken with both gratings
reassures us that the complicating influences of spread functions,
variations in the sensitivity at different points on the blaze
function, and other instrumental effects do not impact our derived
column density.  In performing our analysis we follow the
reconstructed continuum to a point where the observed signal to noise
ratio falls below 10:1.  Thus for the G160M observations we exclude
information between $1215.0 \la
\lambda \la 1216.3$ \AA.  This corresponds to an optical depth (before
instrumental smearing) of $\tau \ga 5.0$.  The two observations with
the G140M grating do not allow us to trace the continuum to quite as
high optical depths.

\begin{figure}
\plotone{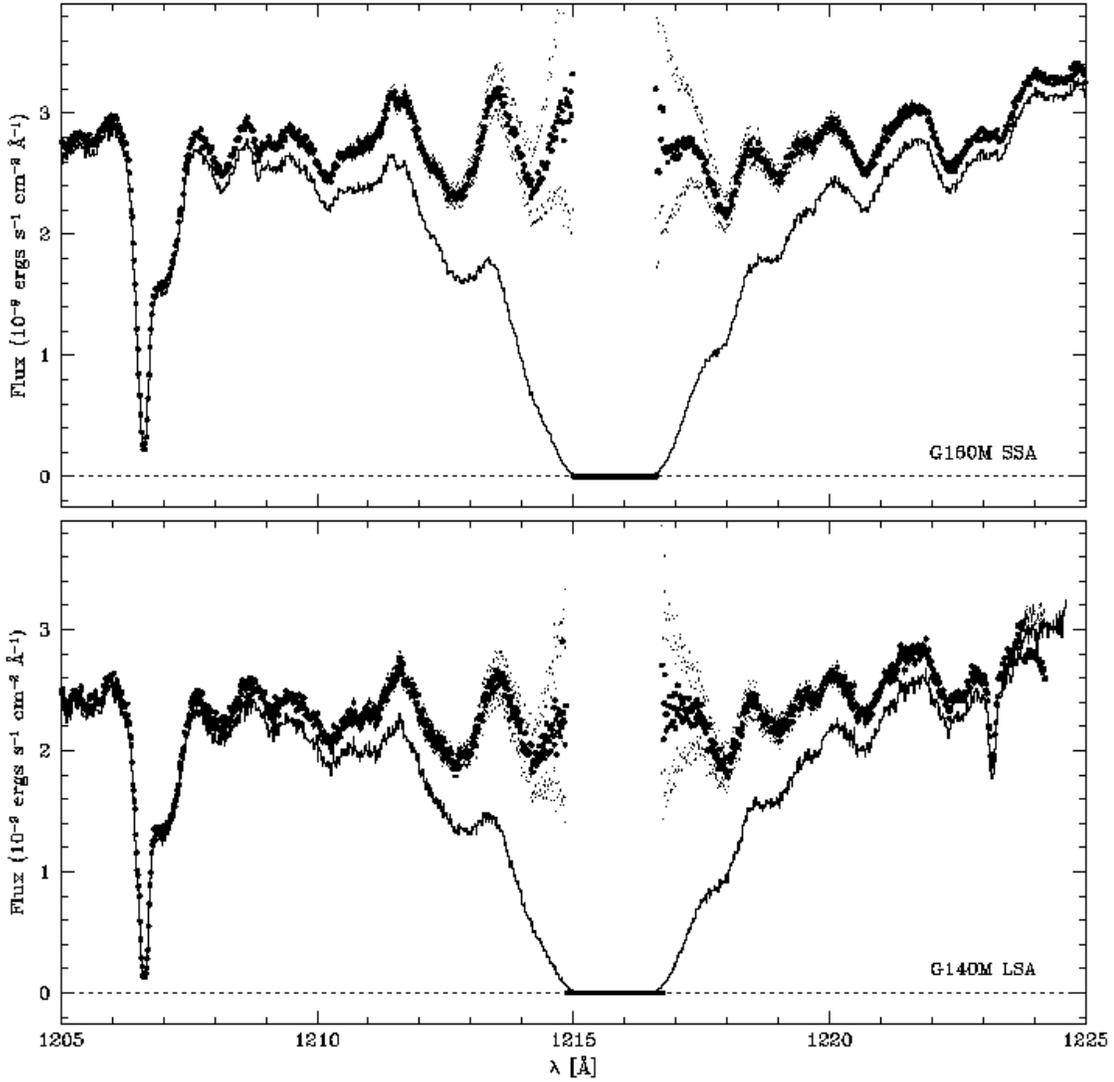}
\figcaption{The best fit (filled circles) and $\pm2 \sigma$ (points)
values for the reconstructed continuum in the G160M and G140M
observations (top and bottom, respectively).  The best-fit column
density of \HI\ along this sightline is $\log N({\rm H\; \mbox{\small
I}}) = 19.87\pm0.015$.  The thick line at the center of the line
profile shows the range over which we have not reconstructed the
profile, i.e., the range over which the signal to noise is less than
10.
\label{fig:reconstruct}}
\end{figure}

From our analysis of the \ghrs\ observations of \lya\ towards \mucol,
we derive a column density of neutral hydrogen $\log
N(\mbox{\ion{H}{1}}) = 19.87 \pm 0.015$ ($\pm1 \sigma$ systematic).
Figure \ref{fig:reconstruct} shows the observed and reconstructed
intensities for the G160M observations.  We also show the $\pm
2\sigma$ (systematic) reconstructed continua.  The derivation of \NHI\
is relatively subjective.  Given the complex undulating stellar
continuum against which we are viewing the \lya\ absorption, we feel a
formal fitting procedure is unwarranted.

Estimating the errors in our column density determination is not a
straightforward or easily quantifiable task.  The undulations in the
reconstructed continuum are primarily due to stellar absorption lines.
The signal to noise ratios of our \lya\ spectra are high enough that
the systematic errors of the continuum reconstruction method (i.e.,
continuum placement, central velocity, etc.)  almost certainly
dominate the statistical errors (e.g., photon statistics, background
subtraction, etc.) in our derivation.  We have estimated $\pm2\sigma$
systematic errors for the values of \NHI\ we judged to give the best
continuum reconstruction by increasing and decreasing \NHI\ until we
judge the resulting continuum reconstructions as highly implausable.
In Figure \ref{fig:reconstruct} we display the best fit continuum
reconstruction along with the $\pm2\sigma$ reconstructions.  One can
see that these $\pm2 \sigma$ results are truly unacceptable.  That the
range of appropriate values is so small is a result of the high signal
to noise and exceptional scattered light properties of the
observations.

One problem in deriving the \HI\ column density is determining the
velocity alignment of the spectrum.  We have used the lines of the
\ion{N}{1} triplet near \wave{1200} (e.g., see Figure \ref{fig:norm2})
to give us a range of acceptable values for the velocity zero-point of
our spectrum, but we have used the reconstructed wings of the \HI\
\lya\ profile themselves to fine-tune this determination.  The
assumption in this approach is that the absorption is symmetric.  This
may not be the case given complex component structure along the line
of sight.  Though the \ion{S}{2} profile is reasonably approximated by
a Gaussian fit (i.e., our turbulent broadening discussed above), we
have in deriving our final number derived the thermal Gaussian profile
with a detailed model of the component structure along the sightline.
We use the results of our component fitting to the \SII\ profile, with
additional approximations for components 3, 4, and 5, as the kernel
for the non-thermal structure along the sightline.  The results of
using this detailed model are almost indistinguishable from the simple
Gaussian turbulent broadening model described above.  

Because the derivation of \NHI\ relies heavily upon the very broad
wings of the absorption profile, the precise details of the
convolutions described here make little difference to the final
result.  As mentioned above, using a detailed component model as
opposed to a simple Gaussian model for the non-thermal broadening
gives indistinguishable results.  Adopting temperatures in the range
$100 < T < 10000$ K makes virtually no difference to the final
continuum reconstruction.  Even neglecting the smearing by the
instrumental spread function makes only very negligible changes to the
reconstructed continuum.  Not only are the results we present quite
robust to errors in these facets of our approach, but future works at
comparable resolutions need not worry so much about the details of
such complications when attempting to derive the \HI\ column density
along a given line of sight.  Again, the keys to our ability to
precisely derive the \HI\ column density are the high signal-to-noise
ratios of our data, the exceptional scattered light properties of the
\ghrs\ first-order gratings, and an appropriate (though not overly
high) resolution.

Our derived value of \NHI\ is quite similar to past determinations.
From {\em IUE} measurements of \lya, Diplas \& Savage (1994) derive
$\log N(\mbox{\ion{H}{1}}) = 19.88 \pm 0.08$. Bohlin \etal\ (1978)
derive $\log N(\mbox{\ion{H}{1}}) = 19.85 \pm 0.08$ from {\em
Copernicus} observations of \lya, and from {\em Copernicus}
measurements of Lyman-$\beta$ and higher Lyman series absorption, York
\& Rogerson (1976) have estimated $\log N(\mbox{\ion{H}{1}}) = 19.9
\pm 0.1$.  The primary advantage of our derivation is that the errors
are almost completely due to the systematics of the continuum
reconstruction method applied to a star with undulating stellar
absorption features $\sim 5 - 30 \%$ deep.

The derived column density $\log N(\mbox{\ion{H}{1}}) = 19.87 \pm
0.015$ contains contributions from the stellar photosphere as well as
the interstellar medium.  We must account for the stellar component if
we are to very accurately derive the interstellar \HI\ column density.
Savage \& Panek (1974) have shown that the observed equivalent widths
of stellar \lya\ absorption for nearby early-type stars are in
agreement with the non-LTE predictions of Mihalas (1972a,b).  Their
work was used by Diplas \& Savage (1994) to derive stellar
contributions to the \lya\ equivalent width based upon an empirical
calibration using the Str\"{o}mgren $[c_1]$ index.  They suggest a
contribution from the photosphere of \mucol\ of order $W_\lambda
(\mbox{\lya}) = 2.95$ \AA, equivalent to $\log N_{\rm H I} \equiv
18.27 + \log W_\lambda (\mbox{\lya}) = 18.74$.

In this work we will estimate the contribution of stellar \lya\
absorption by using the Mihalas (1972a,b) models presented in Savage
\& Panek (1974) to constrain the expected ratio of
$W_\lambda(\mbox{\lya}) / W_\lambda ({\rm H} \beta)$.  Using
measurements of the H$\beta$ equivalent width by Buscombe (1969) we
will then estimate the stellar contribution to
$W_\lambda(\mbox{\lya})$.  Using the data in Table 3 of Savage \&
Panek (1974) we find the ratio of these two hydrogen lines is well fit
by a third order polynomial such that:
\begin{equation}
W_\lambda(\mbox{\lya}) / W_\lambda ({\rm H} \beta) = 
14.89 - 38.33\left( \frac{T_{eff}}{30,000 \ {\rm K}} \right)
+ 39.53 \left( \frac{T_{eff}}{30,000 \ {\rm K}} \right)^2
-15.12 \left( \frac{T_{eff}}{30,000 \ {\rm K}} \right)^3.
\end{equation}
For \mucol\ we predict $W_\lambda(\mbox{\lya}) / W_\lambda ({\rm H}
\beta) \approx 0.418$.  The measured H$\beta$ equivalent width from
the photosphere of \mucol\ is $W_\lambda ({\rm H} \beta) = 1.77$ \AA\
(Buscombe 1969), suggesting a stellar contribution to the \lya\
equivalent width of $W_\lambda(\mbox{\lya}) \approx 0.74$ \AA.  This
is equivalent to $\log N_{\rm H I} \approx 18.14$.  Thus the stellar
contribution should represent $\approx 0.01$ dex of the total.  We
adopt a final interstellar column density to the star \mucol\ of $\log
N_{\rm H I} = 19.86\pm 0.015$ ($1 \sigma$ systematic).

\section{DERIVATION OF THE LARGE SCIENCE APERTURE LINE SPREAD
FUNCTION}
\label{appendix:lsf}

\begin{figure}
\epsscale{0.82} 
\plotone{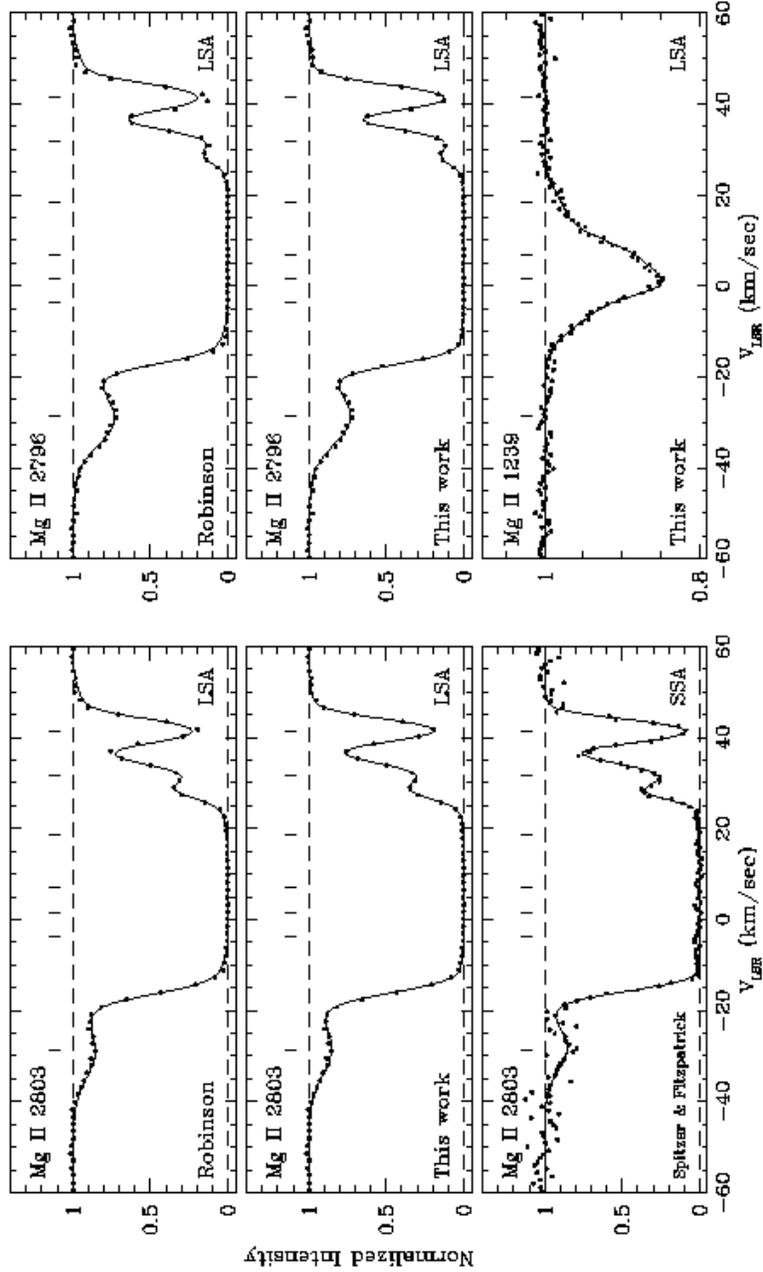}
\figcaption{A comparison of the best-fit component models for the
\protect\ion{Mg}{2} $\lambda \lambda2796$ and 2803 \AA\ transitions
using the Robinson \etal\ LSF (1998; top row), that derived in this
work (middle row), and the SSA data plus Spitzer \& Fitzpatrick (1993)
LSF.  Also shown in the bottom row is the best fit model for
\protect\ion{Mg}{2} $\lambda1239$ \AA\ transition derived
simultaneously with the stronger transitions using our derived LSF.
The central velocities of the individual components are marked as
ticks above the profiles.
\label{fig:mgcompare}}
\end{figure}

In the process of attempting to fit component models to our
interstellar line profiles, we have come to the conclusion that the
LSF presented by Robinson \etal\ (1998) for the LSA is not appropriate
at most wavelengths covered by the GHRS, evidenced by our inability to
adequately fit model profiles to the narrow, often deep, component 4.
Figure \ref{fig:mgcompare} shows the best fit model for
\ion{Mg}{2}  \twowave{2796 and 2803} in the top row when using the LSF
suggested by Robinson \etal\ (1998).  Our best fit component models
for \ion{Mg}{2} at \twowave{2796 and 2803} using the Robinson \etal\
LSF were unable to match the depth of the core of component 4, while
at the same time they provided too much absorption in the wings of the
profile.  Furthermore, when we fit the pre-COSTAR observations of
\ion{Mg}{2} \wave{2803} taken through the SSA using the spread
function derived by Spitzer \& Fitzpatrick (1993) and Fitzpatrick
(priv. comm.), we found significant discrepancies in the derived
column densities for components 3 and 4, which are both very narrow
and deep.  Using the Robinson \etal\ LSF we derived a column density
$\log N_4 (\mbox{\ion{Mg}{2}}) = 13.22\pm0.09$; using the SSA
observations and the LSF of Fitzpatrick \& Spitzer we derived $\log
N_4 (\mbox{\ion{Mg}{2}}) = 12.86\pm0.05$, more than a factor of two
different.  Also, the derived $b$-values were slightly different: $b_4
= 2.0\pm0.3$ \kms\ using the SSA observations and $b_4 = 1.54\pm0.15$
\kms\ using the LSA observations.  We found similar discrepancies
comparing the SSA and LSA observations of \ion{O}{1}.  However, the
SSA and LSA results agreed relatively well when comparing fits of the
\ion{Si}{2} \wave{1304} transition.  We believe this is because 
component 4 is not as deep in this line as the others.  Thus the
differences in the LSF seem to give significantly different results
only for strong, very narrow absorption features.

\begin{figure}
\epsscale{0.85} 
\plotone{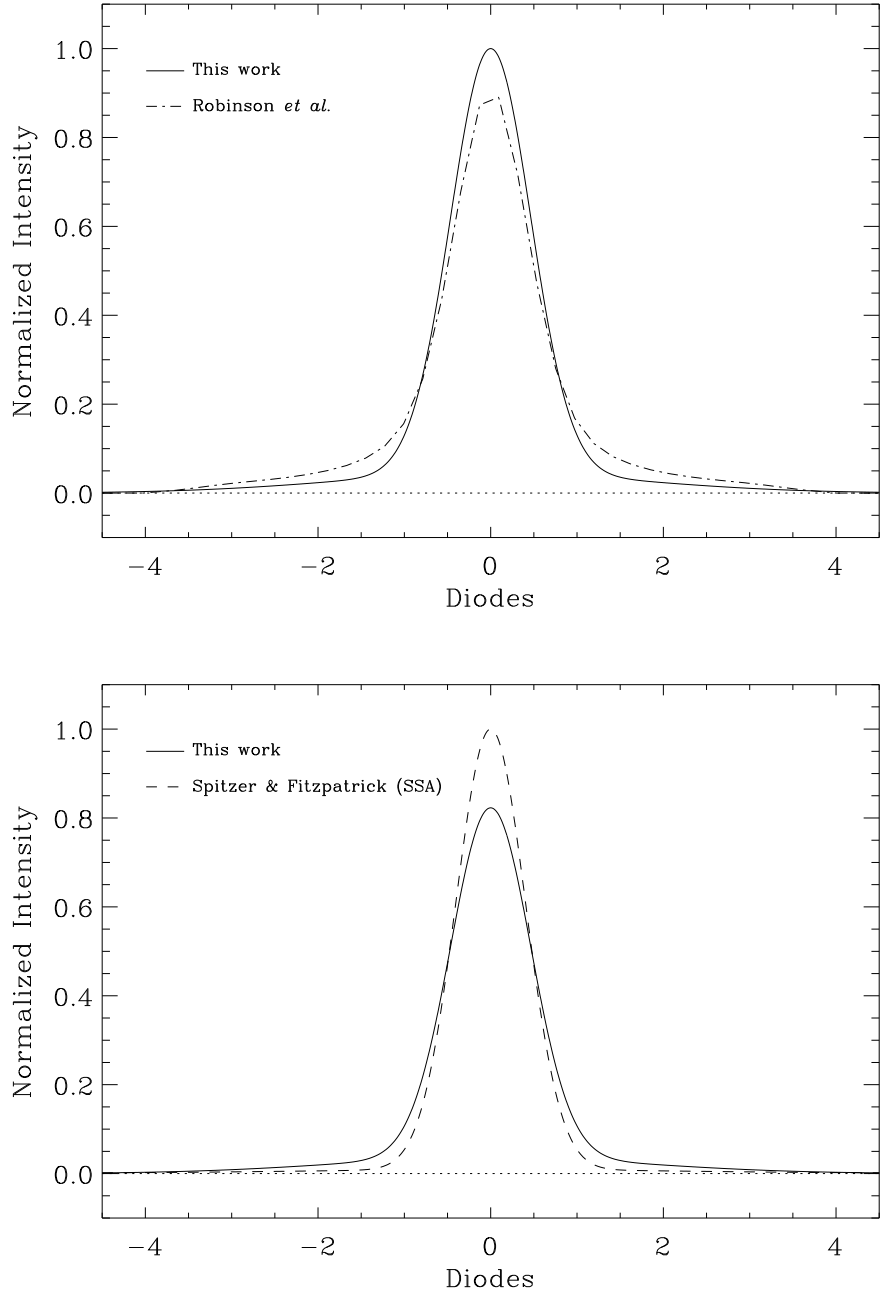}
\figcaption{A comparison of the instrumental LSFs for the GHRS at
$\lambda1900$ \AA.  The top panel shows our derived LSF with the
Robinson \etal\ (1998) LSF shown as the dash-dotted line, both for the
LSA.  The bottom panel shows our LSF compared with that derived for
the SSA by Spitzer \& Fitzpatrick (1993).  The LSFs in a given
plot are scaled so that each has equal area.
\label{fig:lsf}}
\end{figure}

We have attempted to rederive the line spread function for the LSA.
Though the problem is poorly constrained by the current dataset, we
have pieced together a spread function that is able to match the SSA
results and provide good fits to the data.  To begin we have fit the
Robinson \etal\ (1998) spread function with two Gaussians: a narrow
``core'' and a broad ``halo'' component.  Table \ref{table:lsf} gives
the FWHMs (in diodes) and relative areas of these Gaussians.  Our
approach has been to assume a two-component Gaussian model is a valid
approximation to the LSA spread function, and that the values of the
FWHM from the Robinson \etal\ LSF are reasonable.  We assume the
widths of the core and halo Gaussians in diode-space are constant.
The effective resolution changes with placement of the diode array in
the echelle due to differences in the mapping of wavelengths to diode
space.  Our free parameter in the fit is the relative area of the core
and halo components.  We have proceeded by fitting the SSA data for
\ion{O}{1}, \ion{Si}{2}, \ion{Mg}{2}, and \ion{Fe}{2} using the
Fitzpatrick \& Spitzer LSF.  Then we have held the results of these
fits constant and applied them to our LSA observations using varying
values of the relative power of the core and halo components of our
LSF.  For each transition of these species that showed component 4
strongly, we have derived a best fit value for the relative power of
the halo to the LSF.  Figure \ref{fig:lsf} shows our derived spread
function with the Robinson \etal\ (1998) and Spitzer \& Fitzpatrick
(1993) LSFs at $\lambda1900$ \AA.

The extended wings of the LSF in the LSA are a result of the broad
wings of the point spread function from the aberrated primary (plus
COSTAR) being allowed to enter the spectrograph through the $1\farcs
74 \times 1\farcs 74$ LSA.  For the SSA, which effectively blocks the
far wings of the telescope point spread function, the broad component
adopted by Fitzpatrick \& Spitzer is a result of aberrations in the
internal optics of the spectrograph and effects in the detector.  The
relative power of the broad component is much greater in the spread
function for the LSA than for the SSA.  We have fit the fractional
contribution of the halo component, $f_{halo}$, to the total as a
function of wavelength with the following function:
\begin{equation}
f_{halo} = 0.486 - (2.89\times 10^{-4}) \lambda + 
	(5.52\times 10^{-8}) \lambda^2 .
\end{equation}
We find that this prescription gives results that agree in column
density and $b$-value with fits to the SSA data using the Fitzpatrick
\& Spitzer SSA spread function.  The best fit profile using this
profile is also better able to reproduce the data than the Robinson
\etal\ LSF.  Figure \ref{fig:mgcompare} shows the results of the
component fits for the \ion{Mg}{2} lines.  The top row shows the best
fit component model using the Robinson spread function for the
\ion{Mg}{2} lines near \wave{2800}, the middle row
shows the best fit using our derived spread function, and the bottom
row shows the fit to the SSA data using the Spitzer \& Fitzpatrick
(1993) spread function.  Also shown in the bottom row is the fit
obtained simultaneously for the \ion{Mg}{2} \wave{1239} transition
using our derived LSF.  The tick marks show the locations of our model
interstellar components.  When examining the fit to component 4 near
$v_{\rm LSR} = +41$ km s$^{-1}$, it is clear that our LSF provides a
better fit to this strong, narrow component.  Furthermore, it provides
results consistent with those derived using the SSA data.

While our empirically-derived LSF is able to match the SSA data, we do
not wish to suggest the LSA spread function is well understood.  Our
fit is consistent with the available data, though the problem is
ill-constrained, particularly in the wavelength range $1600 < \lambda
< 2300$ \AA, where we have no data.  The true LSF of the LSA is likely
more complex than the two Gaussian model we have adopted.  Further,
the true spread function may be a function of the positioning of the
star within the LSA.  It is important to realize that the adoption of
a specific shape for the instrumental profile can have important
consequences when trying to derive accurate column densities for
narrow, deep lines.  When possible, investigators interested in
deriving the column densities of such components should use the SSA,
for which the spread function has been better characterized (e.g.,
Spitzer \& Fitzpatrick 1993) and is better behaved.  When using data
taken through the LSA for such work, one should investigate the
consequences of adopting slightly different spread functions.  In our
case the values derived for the column densities of \ion{O}{1} and
\ion{Mg}{2} were different by factors of 1.5 to 2.5 between the
different spread functions.

\clearpage
\pagebreak


\begin{deluxetable}{cclccc}
\tablewidth{0pt} 
\tablecolumns{6} 
\tablecaption{Velocities and Conditions in the Absorbing Regions
Towards $\mu$ Columbae \label{table:components}}  
\tablehead{
\colhead{Region} &
\colhead{$v_{\rm LSR}$\tablenotemark{a}} & 
\colhead{$\Delta v$\tablenotemark{b}} &
\colhead{$T_e$\tablenotemark{c}} & 
\colhead{$n_e$\tablenotemark{d}} &
\colhead{Previous}\nl
\colhead{Number} & 
\colhead{[km s$^{-1}$]} & 
\colhead{[km s$^{-1}$]} &
\colhead{[K]}  & 
\colhead{[cm$^{-3}$]}  &
\colhead{ID} 
}
\startdata 
1 &\phs 3    & $-$17.0 to 15.5   & $6,200\pm300$ & $0.27\pm0.05$ & A \nl 
2 &\phs 20.1 & \phs 15.5 to 29.0 & $6,900\pm600$ & $0.31\pm0.13$ & B \nl
3 &\phs 31.0 & \phs 29.0 to 37.5 & $\lesssim10,000$ &  \nodata   & D \nl 
4 &\phs 41.2 & \phs 37.5 to 47.0 & $4,000\pm700$ & $0.64\pm0.12$ & C \nl 
5 & $-28.8$  & $-45.0$ to $-17.0$ & $\lesssim36,000$ & \nodata 
	& \nodata\tablenotemark{e} \nl
\enddata
\tablenotetext{a}{Approximate central velocity of the absorbing region in the
LSR frame.  For components 2, 3, and 4 these values come from fitting
a model interstellar line profile to the \ion{Si}{2} lines, which are
absolutely calibrated using the nearby \ion{O}{1}$^*$ atmospheric
line.}
\tablenotetext{b}{The velocity range over which we have integrated to
obtain the gross values of equivalent width and apparent column
density for these absorbing regions.} 
\tablenotetext{c}{Electron temperatures as derived from ionization
equilibrium and component fitting analyses.  For components 1 and 2,
these values were derived from ionization equilibrium of Ca and Mg (\S
4.4.2).  For components 3, 4 and 5 the temperatures are derived from
our component fitting results (\S 5.1).}
\tablenotetext{d}{Electron densities as derived from various
combinations of ionization equilibrium.  For components 1 and 2, these
values were derived from ionization equilibrium of Ca and Mg (\S
4.4.2).  For component 4 these values were derived from ionization
equilibrium of Mg assuming the temperature suggested from component
fitting methods (\S 5.1).}
\tablenotetext{e}{Shull \& York (1977) refer to this component as
``trailing absorption'' in their discussion of the \ion{Si}{3}
profile.}
\end{deluxetable}

\clearpage

\begin{deluxetable}{clcccll}
\tablewidth{0pc}
\tablecolumns{7}
\tablecaption{Log of GHRS Observations \label{table:log}}
\tablehead{\colhead{Spectral Range} &
	\colhead{Rootname\tablenotemark{a}} & 
	\colhead{Exp.\tablenotemark{b}}  & 
	\colhead{Mode} & \colhead{Aper.\tablenotemark{d}} &
	\colhead{d$_c$\tablenotemark{e}} & \colhead{Species}\\
\colhead{[\AA]} &  \colhead{} & \colhead{[sec]} & 
	\colhead{\& Order\tablenotemark{c}} & \colhead{} &
	\colhead{} & \colhead{Detected}}
\startdata
1116-1124
	& Z2X30108T & 81.6 & G140M/1 & LSA & 0.0 & \FeII, \ion{Fe}{3} \nl
	& Z2X30109T & 81.6 & G140M/1 & LSA &  & \nl
1131-1137 
	& Z2C0011AP & 108.8  & A/49 & LSA & 0.011 & \NI, \FeII \nl
1142-1148
	& Z2AF011AT & 108.8 & A/49 & LSA & 0.061: & \FeII \nl
	& Z2C0030RT & 108.8 & A/49 & LSA &  &  \nl	
	& Z308010OT & 108.8 & A/49 & LSA &  &  \nl	
	& Z3JN010OT & 108.8 & A/49 & LSA &  &  \nl
1149-1153 
	& Z2AF020RT & 108.8 & A/49 & LSA & 0.061: & \ion{P}{2}\nl
	& Z2C0010QP & 108.8 & A/49 & LSA &  & \nl
	& Z2C0021AP & 108.8  & A/49 & LSA &  & \nl
1154-1161 
	& Z2C00119P & 108.8 & A/48 & LSA & 0.058: & \CI \nl
1157-1163 
	& Z2C0041AT & 108.8 & A/49 & LSA & 0.011: & \CI \nl
1184-1191 
	& Z2AF010PT & 108.8 & A/47 & LSA & 0.000 & \CI, \SiII,
						\ion{S}{3} \nl
	& Z2C0020PP & 108.8 & A/47 & LSA &       & \nl
1191-1197 
	& Z2AF0118T & 108.8 & A/47 & LSA & 0.038 & \SiII \nl
	& Z2AF021CT & 108.8 & A/47 & SSA & & \nl
	& Z2C0030PT & 108.8 & A/47 & LSA & & \nl
1195-1230
	& Z2AN020AT & 54.4  & G160M/1 & SSA & 0.0 & \ion{H}{1} \nl
1197-1225
	& Z2XE010DT & 27.2  & G140M/1 & LSA & 0.0 & \ion{H}{1} \nl
1197-1225
	& Z2XE0117T & 27.2  & G140M/1 & SSA & 0.0 & \ion{H}{1} \nl
1197-1200 
	& Z2AF020PT & 108.8 & A/47 & LSA & 0.038: & \NI, \ion{Mn}{2}\nl
1200-1203
 	& Z2C00218P & 108.8 & A/47 & LSA & 0.035 & \NI \nl
1206-1212 
	& Z2AZ010MT\tablenotemark{f} 
		& 1740.8 & A/46 & SSA & 0.027 & \ion{Si}{3} \nl
%
1237-1244
	& Z2AZ010CT\tablenotemark{f} 
		& 1088.0 & A/45 & LSA & 0.049: &\ion{N}{5}, 
						\ion{Mg}{2} \nl
	& Z2AZ010OT\tablenotemark{f} 
		& 1088.0 & A/45 & SSA & & \nl
1249-1256 
	& Z2AF020MT & 27.2 & A/45 & LSA & 0.049: & \ion{S}{2}\nl
	& Z2C00215P & 27.2 & A/45 & LSA & & \nl
1259-1262
	& Z2C0040MT & 27.2 & A/45 & LSA & 0.033 & \ion{S}{2} \nl
1259-1267
	& Z2C00114P & 27.2 & A/44 & LSA & 0.026 & \SiII$^*$ \nl
1271-1278 
	& Z2AF0114T & 27.2 & A/44 & LSA & 0.046: & \CI\nl
1301-1308 
	& Z2AF0113T & 27.2 & A/44 & LSA & 0.050 & \ion{O}{1}, \SiII,
					\ion{P}{2}\nl
	& Z2AF0217T & 27.2 & A/44 & LSA &  & \nl
	& Z2C0030KT & 27.2 & A/44 & LSA &  & \nl	
	& Z308010HT & 27.2 & A/44 & LSA &  & \nl
1313-1320 
	& Z2C0010JP & 27.2 & A/43 & LSA & 0.050: & \ion{Ni}{2} \nl
	& Z2C0040KT & 27.2 & A/43 & LSA & & \nl
1326-1333 
	& Z2AF010JT & 27.2 & A/42 & LSA & 0.041: & \CI \nl
  	& Z2C0020JP & 27.2 & A/42 & LSA & & \nl
1332-1339
	& Z2AF0112T & 27.2 & A/42 & LSA & 0.041 & \ion{C}{2}, 
					\ion{C}{2}$^*$ \nl
	& Z2AF0216T & 27.2 & A/42 & SSA & & \nl
 	& Z2C0030JT & 27.2 & A/42 & LSA & & \nl
	& Z308010GT & 27.2 & A/42 & LSA & & \nl
1364-1372 
	& Z2AF0111T & 27.2 & A/41 & LSA & 0.039: & \ion{Ni}{2} \nl
	& Z2AF0215T & 27.2 & A/41 & SSA & & \nl
	& Z2C0030IT & 27.2 & A/41 & LSA & & \nl
	& Z308010FT & 27.2 & A/41 & LSA & & \nl
1526-1534
	& Z2C0040DT & 54.4 & A/37 & LSA & 0.055 & \SiII \nl
	& Z2C0040ET & 27.2 & A/37 & LSA & & \nl
1554-1562
	& Z2AF010VT & 108.8 & A/36 & LSA & 0.031: & \CI \nl
	& Z2AF020ZT & 108.8 & A/36 & SSA \nl
	& Z2C0030CT & 54.4  & A/36 & LSA \nl
1606-1614
	& Z2AF020BT & 108.8 & A/35 & LSA & 0.030: & \FeII \nl
	& Z2C0020UP & 54.4  & A/35 & LSA & \nl
1653-1662
	& Z2AF020AT & 108.8 & A/34 & LSA & 0.030: & \CI \nl
	& Z2C0020TP & 54.4  & A/34 & LSA \nl
1668-1676
	& Z2C0040TT & 54.4 & A/34 & LSA & 0.030 & \ion{Al}{2} \nl
1706-1712
	& Z2AF0209T & 108.8 & A/33 & LSA & 0.029: & \ion{Ni}{2} \nl
	& Z2CX011KT  & 27.2 & B/33 & LSA & 0.045: \nl
	& Z2C0020SP & 54.4  & A/33 & LSA & 0.029: \nl
	&  Z2D4021JT & 54.4 & B/33 & LSA & 0.045: \nl
1740-1749
	& Z2D4020MT & 54.4 & B/32 & LSA & 0.043: & \ion{Ni}{2} \nl 
1805-1809
	& Z2CX010MT & 27.2 & B/31 & LSA & 0.041: & \SiII \nl
	& Z2D40211T & 54.4 & B/31 & LSA & \nl
1846-1857
	& Z2D40118T & 54.4 & B/30 & LSA & 0.039: & \ion{Al}{3} \nl
1860-1866
	& Z2CX010LT & 27.2 & B/30 & LSA & 0.039: & \ion{Al}{3} \nl
	& Z2D4020KT & 54.4 & B/30 & LSA & \nl
2017-2027
	& Z2D4030IT & 54.4 & B/28 & LSA & 0.035: & \ion{Mg}{1},
					\ion{Zn}{2}\nl
2052-2061
	& Z2D40115T & 54.4 & B/27 & LSA & 0.033: & \ion{Cr}{2} \nl
2062-2063
	& Z2D40115T & 54.4 & B/27 & LSA & 0.033: & \ion{Cr}{2} \nl
	& Z2D4020HT & 54.4 & B/27 & LSA \nl
2063-2074
	& Z2D4020HT & 54.4 & B/27 & LSA & 0.033: & \ion{Cr}{2} \nl
2242-2254
	& Z23L040FT & 57.6 & B/25 & LSA & 0.030: & \FeII \nl
	& Z29I060FT & 57.6 & B/25 & LSA \nl
	& Z2D4021BT & 54.4 & B/25 & LSA \nl
	& Z308020DT & 115.2 & B/25 & LSA \nl
	& Z3JN020DT & 115.2 & B/25 & LSA \nl
2259-2264
	& Z2CX011CT & 27.2 & B/25 & LSA & 0.030: & \FeII \nl
	& Z2D4030FT & 54.4 & B/25 & LSA \nl
2335-2348
	&  Z23L040ET & 57.6 & B/24 & LSA & 0.031  & \FeII \nl
	&  Z308020CT & 115.2& B/24 & LSA \nl
	&  Z3JN020CT & 115.2& B/24 & LSA \nl
2364-2376 
	& Z2D4030UT & 54.4 & B/24 &  LSA & 0.028: & \FeII \nl
2567-2581
	& Z2D4030CT & 54.4 & B/22 & LSA & 0.017: & \ion{Mn}{2} \nl
2575-2590 
	& Z2D4030ST & 54.4 & B/22 &  LSA & 0.017: & \FeII \nl
2590-2603
	& Z2D40318T & 54.4 & B/22 & LSA & 0.017 & \ion{Mn}{2}, 
					\FeII \nl  
2792-2802
	&  Z23L0413T & 28.8 & B/20 & LSA & 0.015 & \ion{Mg}{2}\nl
	& Z29I0613T & 28.8 & B/20 & LSA \nl
2802-2807
	& Z23L0413T & 28.8 & B/20 & LSA & 0.015 & \ion{Mg}{2} \nl
	& Z23L0414T & 28.8 & B/20 & LSA \nl
	& Z23L040AT & 57.6 & B/20 & LSA \nl
	& Z29I060AT & 57.6 & B/20 & LSA \nl
	& Z29I0613T & 28.8 & B/20 & LSA \nl
	& Z29I0614T & 28.8 & B/20 & LSA \nl
2852-2857
	& Z23L0418T & 28.8 & B/20 & LSA & 0.015: & \ion{Mg}{1} \nl
	& Z23L0419T & 28.8 & B/20 & LSA \nl
	& Z29I0618T & 28.8 & B/20 & LSA \nl
\enddata
\tablenotetext{a}{STScI archival rootname.}
\tablenotetext{b}{Total exposure time given in seconds.  Typically
	94\% of this time was spent on-spectrum, while 6\% was used to
	measure the interorder echelle background.}
\tablenotetext{c}{Echelle mode and spectral order used for the
	observation.} 
\tablenotetext{d}{Aperture used for the observation.  The LSA subtends
	$1\farcs 74 \times 1\farcs 76$ on the sky, the SSA $0\farcs 22
	\times 0\farcs 22$, for post-COSTAR observations.}
\tablenotetext{e}{Second-order scattered light correction (see
	Cardelli et al. 1993).  For spectral regions containing 
	strongly saturated lines we have derived $d_c$ directly.
	Those values marked with a colon are regions for which we have
	not independently confirmed the value of $d_c$.  In several
	cases we have applied our derived values of $d_c$ to spectral
	regions in the same echelle order that lack strongly saturated
	lines; in regions where no independent information was
	available we have used the values suggested by Cardelli et
	al. (1993) for pre-COSTAR SSA observations.} 
\tablenotetext{f}{This observation was taken using the FP-SPLIT=4
	observing routine.  The fixed-pattern noise spectrum of this
	observation has been removed from the observations.}
\end{deluxetable}

\clearpage

\newcommand{\bad}{\tablenotemark{g}}

\begin{landscape}
\begin{deluxetable}{lcccccccccc}
\tablecolumns{11}
\tablewidth{0pc}
\tablecaption{Integrated Equivalent Widths \label{table:eqwidths}}
\tablehead{
\colhead{}         &
\colhead{IP}       & \colhead{IP}            &  \colhead{}          &
\colhead{}         & \colhead{}              &  
\multicolumn{4}{c}{W$_{\lambda}$ [m\AA] \tablenotemark{c}}  & 
\colhead{} \\
\cline{7-10}
\colhead{Species}  & \colhead{$\chi^{i-1}$}  & \colhead{$\chi^{i}$} & 
\colhead{$\lambda_{vac}$\tablenotemark{a}}   & 
\colhead{log $\lambda f$}                    &  
\colhead{Ref.\tablenotemark{b}}              & \colhead{Comp 1}     &
\colhead{Comp 2}   & \colhead{Comp 3}        & \colhead{Comp 4}     &
\colhead{S/N\tablenotemark{d}} \\
\colhead{}         & \colhead{[eV]}          & \colhead{[eV]}       & 
\colhead{[\AA]}    & \colhead{}              & \colhead{}           & 
\colhead{(3 km s$^{-1}$)}         & \colhead{(20 km s$^{-1}$)}      & 
\colhead{(31 km s$^{-1}$)}        & \colhead{(41 km s$^{-1}$)}      & \\
}
\startdata
 \ion{C}{1}        & \nodata & 11.26 & 1328.833 & 1.887 & 1
   & $7.2\pm0.5$  & $<2.2$\tablenotemark{e}  
   & $<1.9$  &  $<1.9$ & 48  \nl
   & & & 1560.309 & 2.099 & 1 
   & $12.0\pm0.7$ & $2.7\pm0.5$ & $<1.7$  &  $<1.7$ & 64 \nl
   & & & 1656.928 & 2.367 & 1
   & $18.8\pm2.4$ & $3.6\pm1.5$ & $<3$  & $<3$ & 26 \nl
 \ion{C}{1}$^*$    & \nodata & 11.26 & 1656.267 & 1.987 & 1
   & $2.4\pm0.9$ & $<3$ & $<3$ & $<3$ & 42 \nl
   & & & 1657.907 & 1.890 & 1
   & $4.8\pm2.0$ & $<4$ & $<4$ & $<4$ & 32 \nl
 \ion{C}{1}$^{**}$ & \nodata & 11.26 & 1277.550 & 2.016 & 1
   & $<3$ & $<2.0$ & $<1.7$ & $<1.7$ & 51 \nl
   & & & 1561.438 & 2.023 & 1
   & $<4$ & $<3$ & $<3$ & $<3$ & 42 \nl
   & & & 1658.121 & 1.765 & 1
   & $<5$ & $<4$ & $<4$ & $<4$ & 31 \nl
%
 \ion{C}{2} & 11.26 & 24.38 & 1334.532 & 2.232 & 1 
   & $155\pm9$ & $56.0\pm1.1$ & $65.0\pm1.3$ & $28.3\pm0.6$ & 56\nl  
 \ion{C}{2}$^*$ & 11.26 & 24.38 & 1335.708 & 2.186  & 1
   &$132\pm3$  & $23.6\pm0.6$ & $2.3\pm0.6$
	& $3.5\pm0.5$ & 31\nl
 \ion{N}{1} & \nodata & 14.53 & 1134.165 & 1.182 & 1
 & $79.2\pm2.0$ & $28.0\pm1.0$ & \phn \phd$<3$
	& \phn \phd$<3$ & 12 \nl
 & & & 1134.415 & 1.483 & 1
 & $96.2\pm2.1$ & $24.9\pm1.0$ & $<3.3$ & $2.9\pm1.0$ & 12 \nl
 & & & 1134.980 & 1.660 & 1
 & $106.0\pm2.2$ & $23.8\pm0.9$ & $<2.4$ & $5.0\pm0.8$ & 14 \nl
 & & & 1199.550 & 2.202  & 1
 & $118\pm3$ & $34.9\pm0.8$ & \nodata\tablenotemark{f} &
 	$14.8\pm0.8$ & 27 \nl 
 & & & 1200.223 & 2.026 & 1
 & $116.0\pm2.3$ & $34.3\pm0.7$ & \nodata\tablenotemark{f} &
	 $11.3\pm0.4$ & 46\nl 
 & & & 1200.710 & 1.725 & 1
 & $109.2\pm2.2$ & $32.2\pm0.7$ & \nodata\tablenotemark{f} &
         $7.7\pm0.4$\tablenotemark{f} & 49\nl
\ion{O}{1} & \nodata & 13.62 & 1302.169 & 1.804 & 1
  & $137\pm3$ & $52.6\pm1.1$ & $16.4\pm0.4$ & $22.8\pm0.5$ & 54 \nl
  & & & 1355.598 & -2.772 & 1
  & $<3$ & $<2.3$ & $<2.0$ & $<2.0$ & 46 \nl
\ion{Mg}{1} & \nodata & 7.65 & 2026.477 & 2.369\bad & 1
  & $16.5\pm1.7$ & $2.0\pm0.5$ & $<1.5$ & $<1.5$ & 44 \nl
  & & & 2852.964 & 3.718\bad & 1 
  & $207\pm4$ & $44.8\pm1.0$ & $2.2\pm0.4$ & $3.9\pm0.5$ & 80\nl
\ion{Mg}{2} & 7.65 & 15.04 & 1239.925 & -0.100\bad & 2
 & $9.82\pm0.24$ & $0.79\pm0.09$ & \phn$<0.3$ & \phn$<0.3$ & 195 \nl
 & & & 1240.395 & -0.401\bad & 2
 & $5.47\pm0.18$ & $0.36\pm0.09$ & $<0.18$ & $<0.18$ & 217 \nl
 & & & 2796.352 & 3.234 & 1
 & $307\pm6$ & $117\pm3$ & $59.4\pm1.2$ & $49.9\pm1.1$ & 111 \nl
 & & & 2803.531 & 2.933 & 1
 & $297\pm6$ & $112.1\pm2.3$ & $46.3\pm1.0$ & $44.2\pm0.9$ & 136 \nl
\ion{Al}{2} & 5.99 & 18.83 & 1670.787 & 3.486 & 1
 & $148\pm3$ & $55.7\pm1.3$ & $9.3\pm1.0$ & $6.5\pm1.0$ & 16 \nl
\ion{Al}{3} & 18.83 & 28.45 & 1854.716 & 3.017 & 1
 & $15.0\pm2.0$ & $<3.0$ & $<3.0$ & $<3.0$ & 30 \nl
 & & &1862.790 & 2.716 & 1 
 & $9.9\pm1.5$ & $<2.4$ & $<2.4$ & $<2.4$ & 35 \nl
\tablebreak
\ion{Si}{2} & 8.15 & 16.35 & 1190.416 & 2.474 & 1
  & $116.0\pm2.3$ & $48.3\pm1.0$ & $12.1\pm0.3$ & $12.3\pm0.3$ & 54 \nl
 & & &1193.290 & 2.775 & 1
  & $124\pm3$ & $50.0\pm1.0$ & $19.0\pm0.5$ & $17.0\pm0.4$ & 48 \nl
 & & &1304.370 & 2.050 & 3,4
  & $121\pm3$ & $42.5\pm0.9$ & $6.91\pm0.24$ & $8.5\pm0.3$ & 57 \nl
 & & &1526.707 & 2.225 & 3,4 
  & $148\pm3$ & $56.3\pm1.2$ & $11.9\pm0.5$ & $12.4\pm0.6$ & 27 \nl
 & & &1808.013 & 0.596 & 5,4 
  &$61.2\pm1.6$ & $6.5\pm0.2$ & $<1.5$ & $<1.5$ & 36 \nl
\ion{Si}{2}$^*$ & 8.15 & 16.35 & 1264.738 & 3.058 & 1 
  & $3.60\pm1.16$ & $<1.8$ & $<1.8$ & $<1.8$ & 33 \nl
\ion{Si}{3} & 16.35 & 33.49 & 1206.500 & 3.304 & 1
  & $126\pm3$ & $40.9\pm0.9$ & $10.7\pm0.5$\tablenotemark{f} &
  $1.4\pm0.5$\tablenotemark{f} & 76 \nl
\ion{P}{2} & 10.49 & 19.72 & 1152.818 & 2.435\bad & 1
  & $53.2\pm1.4$ & $4.9\pm0.5$ & $<1.5$ & $<1.5$ & 32 \nl
 & & & 1301.874 & 1.351\bad & 1	
 & $6.6\pm0.6$ & \nodata\tablenotemark{h} &\nodata\tablenotemark{h} 
	&\nodata\tablenotemark{h} & 55 \nl
\ion{S}{2} & 10.36 & 23.33 & 1250.534 & 0.834 & 1 
  & $68.6\pm1.4$ & $ 7.4\pm0.4$ & $<0.9$ & $<0.9$ & 40 \nl
  & & &1253.811 & 1.135 & 1
  & $90.9\pm1.8$ & $11.3\pm0.4$ & $<0.9$ & $<0.9$ & 43\nl
  & & &1259.519 & 1.311 & 1
  & $104.6\pm2.1$ & $20.0\pm0.6$ & $<1.3$ & $<1.2$ & 36 \nl
\ion{S}{3} & 23.33 & 34.83 & 1190.208 & 1.421 & 1
  & $16.3\pm0.6$ & $<0.9$ & $<0.9$ & \nodata\tablenotemark{h} & 58 \nl
\ion{Cr}{2} & 6.77 & 16.50 & 2056.254 & 2.334 & 6
  & $12.2\pm1.3$ & $4.2\pm0.7$ & $<1.5$ & $<1.5$ & 43 \nl
  & & & 2062.234 & 2.206 & 6
  & $6.7\pm1.3$ & $3.5\pm0.6$ & $<1.2$ & $<1.2$ & 73 \nl
  & & & 2066.161 & 2.027 & 6 
  & $6.0\pm0.9$ & $2.6\pm0.5$ & $<1.2$ & $<1.2$ & 48 \nl
\ion{Mn}{2} & 7.44 & 15.64 & 1197.184 & 2.273 & 1
  & $5.6\pm0.4$ & $1.0\pm0.3$ & $<0.6$ & $<0.6$ & 42 \nl
  & & & 2576.877 & 2.956 & 1
  & $55.8\pm1.3$ & $10.9\pm0.4$ & $1.6\pm0.4$ & $1.1\pm0.4$ & 68 \nl
  & & & 2594.499 & 2.847 & 1
  & $43.6\pm1.1$ & $8.7\pm0.5$ & $<1.2$ & $<1.2$ & 60 \nl
\ion{Fe}{2} & 7.87 & 16.16 & 1133.665 & -1.222\bad & 1
  & $9.5\pm2.2$ & $4.5\pm1.1$ & $<3$ & $<3$ & 13 \nl
  & & & 1143.226 & 1.182\bad & 1
  & $24.7\pm1.0$ & $10.1\pm0.4$ & $<2.2$ & $<2.2$ & 35 \nl
  & & & 1144.938 & 2.080 & 1 
  & $75.1\pm1.7$ & $32.1\pm0.8$ & $3.6\pm0.6$ & $2.1\pm0.6$ & 31 \nl
  & & & 1608.451 & 1.998 & 7,8 
  & $100.3\pm2.1$ & $36.6\pm0.8$ & $3.6\pm0.5$ & $3.9\pm0.5$ & 35 \nl
  & & & 2249.877 & 0.612 & 9,10
  & $10.2\pm0.3$ & $3.96\pm0.20$ & $<0.6$ & $<0.6$ & 218 \nl
  & & & 2260.780 & 0.742 & 9,10
  & $15.0\pm0.5$ & $6.9\pm 0.3$ & $<0.9$ & $<0.9$ & 97 \nl
  & & & 2344.214 & 2.410\bad & 1 
  & $204\pm4$ & $70.0\pm1.4$ & $16.0\pm0.3$ & $11.7\pm0.3$ & 236 \nl
  & & & 2374.461 & 1.889\bad & 7,10 
  & $125\pm3$ & $47.7\pm1.0$ & $5.3\pm0.3$ & $4.1\pm0.3$ & 65 \nl
  & & & 2586.650 & 2.248 & 7,10
  & $196\pm4$ & $74.0\pm1.5$ & $11.4\pm0.4$ & $9.8\pm0.4$ & 60 \nl
  & & & 2600.173 & 2.765 & 1
  & $238\pm5$ & $98.9\pm2.0$ & $28.7\pm0.7$ & $22.1\pm0.6$ & 51 \nl
\tablebreak
\ion{Fe}{3} & 16.16 & 30.65 & 1122.526 &1.947 & 1
 & $19\pm4$\tablenotemark{i} & \nodata & \nodata & \nodata & 24 \nl
\ion{Ni}{2} & 7.64 & 18.17 & 1317.217 & 2.012\bad & 1,11
  & $6.8\pm0.7$ & $3.6\pm0.4$ & $<1.2$ & $<1.2$ & 44 \nl
  & & & 1370.132 & 2.023\bad & 11,12 
  & $8.1\pm0.5$ & $3.4\pm0.3$ & $<0.6$ & $<0.6$ & 68 \nl
  & & & 1709.600 & 1.784 & 12
  & $4.9\pm1.0$ & $3.4\pm0.5$ & $<1.2$ & $<1.2$ & 44 \nl
  & & & 1741.569 & 1.858 & 12
  & $14.5\pm1.6$ & $5.5\pm0.9$ & $<2.4$ & $<2.4$ & 21 \nl
\ion{Cu}{2} & 7.73 & 20.29 & 1358.773 & 2.713 & 1 
  & $<3$ & $<2.2$ & $<1.9$ & $<1.9$ & 50 \nl
\ion{Zn}{2} & 9.39 & 17.96 & 2026.136 & 2.996\bad & 6
  & $52.5\pm1.6$ & $4.5\pm0.8$ & \nodata\tablenotemark{g} & 
	\nodata\tablenotemark{g} & 46 \nl
  & & & 2062.664 & 2.723\bad & 6
  & $26.5\pm1.1$ & $2.2\pm0.5$ & $<1.2$ & $<1.2$ & 65 \nl
\enddata
 \tablenotetext{a}{Vacuum wavelengths from Morton (1991).}  
 \tablenotetext{b}{Reference codes for $f-$values.}
 \tablenotetext{c}{Measured equivalent widths and $\pm 1 \sigma$
  errors, given in m\AA.}
 \tablenotetext{d}{Empirical signal-to-noise ratio derived from the
 fit to the continuum.}
 \tablenotetext{e}{All upper limits are 3 $\sigma$ limits.}
 \tablenotetext{f}{The values derived for these components are rather
  uncertain  due to the nature of the continuum placement and the
  possibility of fixed-pattern noise contamination (see text).}
\tablenotetext{g}{Denotes oscillator strengths that we feel {\em may} have
  uncertainties approaching or exceeding our measurement errors.  This 
  classification is based on the current dataset and our experience
  with the current oscillator strength measurements and other
  absorption line datasets.} 
 \tablenotetext{h}{Confused with adjacent line.}
 \tablenotetext{i}{Based upon G140M data.  This line has been
  integrated over the velocity range -20 km s$^{-1}$ to +23 km
  s$^{-1}$.  Though the line \ion{C}{1} $\lambda$1122.518 \AA\
  potentially overlaps the \ion{Fe}{3} absorption, we believe its
  contribution to the value of $W_\lambda$ quoted here is negligible
  based upon the absence of the stronger line of \ion{C}{1} at
  1122.438 \AA.}
\tablerefs{ (1) Morton (1991); (2) Fitzpatrick (1997); (3) Spitzer \&
  Fitzpatrick (1993); (4) Dufton, {\em et al.} (1983, 1992); (5)
  Bergeson \& Lawler (1993a); (6) Bergeson \& Lawler (1993b); (7)
  Cardelli \& Savage (1995); (8) Bergeson, {\em et al.}  (1996a); (9)
  Bergeson, {\em et al.} (1994); (10) Bergeson, {\em et al.} (1996b);
  (11) Fedchak \& Lawler (1999); (12) Zsarg\'{o} \& Federman (1997)}
\end{deluxetable}
\end{landscape}

\clearpage 


\newcommand{\sat}{$\times \times \times$}

\begin{deluxetable}{lccccc}
\tablecolumns{6}
\tablewidth{0pc}
\tablecaption{Integrated Apparent Column Densities\label{table:coldens}}
\tablehead{
\colhead{} & \colhead{} & 
	\multicolumn{4}{c}{ Log(N$_a$) [cm $^{-2}$] \tablenotemark{b}}\\
\cline{3-6}
\colhead{Species} & \colhead{$\lambda_{vac}$\tablenotemark{a}} & 
	\colhead{Comp 1} & \colhead{Comp 2} & 
	\colhead{Comp 3} & \colhead{Comp 4} \\
\colhead{} & \colhead{[\AA]} &  \colhead{(3 km s$^{-1}$)} & 
	\colhead{(20 km s$^{-1}$)} & \colhead{(31 km s$^{-1}$)} & 
	\colhead{(41 km s$^{-1}$)} }
\startdata
\ion{C}{1} & 1328.833 & $12.92\pm0.04$ & \nodata &
	 	\nodata & \nodata \nl
\ion{C}{1} & 1560.309 & $12.87\pm0.03$ & $12.21\pm0.09$ &
	 	\nodata & \nodata \nl
\ion{C}{1} & 1656.928 & $12.79\pm0.07$ & \nodata &
	 	\nodata & \nodata \nl
\ion{C}{1}$^*$ & 1656.267 & $12.46^{+0.11}_{-0.15}$ & \nodata &
	 	\nodata & \nodata \nl
\ion{C}{1}$^*$ & 1657.907 & $12.65^{+0.15}_{-0.24}$ & \nodata &
	 	\nodata & \nodata \nl
\nl
\ion{C}{2}     & 1334.532 & \sat \tablenotemark{c} & \sat & \sat & 
	$>13.5$ \nl
\ion{C}{2}$^*$ & 1335.708 & \sat & $>13.3$ &
	$12.13\pm0.10$ & $12.32\pm0.14$ \nl
\nl
\ion{N}{1} & 1134.165 & \sat & \sat & \nodata & \nodata \nl
\ion{N}{1} & 1134.415 & \sat & \sat & \nodata & 
		$13.02^{+0.13}_{-0.18}$\phn \phd \nl
\ion{N}{1} & 1134.980 & \sat & \sat & \nodata & $13.08\pm0.08$\nl
\ion{N}{1} & 1199.550 & \sat & \sat & \nodata & 
		$12.95\pm0.10$\tablenotemark{d} \nl
\ion{N}{1} & 1200.223 & \sat & \sat & \nodata & $13.09\pm0.02$\nl
\ion{N}{1} & 1200.710 & \sat & \sat & \nodata & 
		$13.19\pm0.10$\tablenotemark{d} \nl
\nl
\ion{O}{1} & 1302.169 & \sat & \sat & $13.49\pm0.02$ &
		$13.72\pm0.02$ \nl
\nl
\ion{Mg}{1} & 2026.477 & $12.63\pm0.05$ & $11.72\pm0.11$ \phn
	\phd & \nodata & \nodata \nl
\ion{Mg}{1} & 2852.964 & $12.54\pm0.02$ & $11.68\pm0.02$ & 
	$10.23\pm0.09$ & $10.48\pm0.06$ \nl
\nl
\ion{Mg}{2} & 1239.925 & $15.08\pm0.02$ & $13.96\pm0.05$ &  
	\nodata & \nodata \nl
\ion{Mg}{2} & 1240.395 & $15.11\pm0.03$ 
	& $13.92^{+0.10}_{-0.12}$\phn \phd &  
	\nodata & \nodata \nl
\ion{Mg}{2} & 2796.352 & \sat & \sat & $12.44\pm0.02$ &
	$12.35\pm0.02$ \nl
\ion{Mg}{2} & 2803.531 & \sat & \sat & $12.52\pm0.02$ &
	$12.53\pm0.02$ \nl
\nl
\ion{Al}{2} & 1670.787 & \sat & \sat & $11.30\pm0.05$ & 
	$11.21\pm0.06$\nl
\nl
\ion{Al}{3} & 1854.716 & $11.97\pm 0.07$ &\nodata & \nodata & \nodata \nl
\ion{Al}{3} & 1862.790 & $12.08\pm 0.07$ & \nodata & \nodata & \nodata \nl
\nl
\ion{Si}{2} & 1190.416 & \sat & \sat & $12.70\pm0.02$ & 
	$12.70\pm0.03$ \nl
\ion{Si}{2} & 1193.290 & \sat & \sat & $12.70\pm0.02$ &
	$12.60\pm0.02$ \nl
\ion{Si}{2} & 1304.370 & \sat & \sat & $12.79\pm0.03$ &
	$12.87\pm0.03$ \nl
\ion{Si}{2} & 1526.707 & \sat & \sat & $12.80\pm0.03$ &  
	$12.82\pm0.03$ \nl
\ion{Si}{2} & 1808.013 & $15.10\pm0.02$ & $14.05\pm0.06$ & 
	\nodata & \nodata \nl
\ion{Si}{2}$^*$ & 1264.738 & $11.47^{+0.12}_{-0.17}$\phn \phd 
	& \nodata & \nodata & \nodata \nl
\nl
\ion{Si}{3} & 1206.500 & \sat & $>12.67$ & 
   	$11.81\pm0.03$\tablenotemark{e} &
	$10.84^{+0.16}_{-0.21}$\tablenotemark{e}\phn \phd \nl 
\nl
\ion{P}{2} & 1152.818 & $13.46\pm0.02$ & $12.29\pm0.05$ &
	\nodata & \nodata \nl
\ion{P}{2} & 1301.874 & $13.42\pm0.04$ & \nodata &
	\nodata & \nodata \nl
\ion{S}{2} & 1250.534 & $15.19\pm0.02$ & $14.04\pm0.03$& 
	\nodata & \nodata \nl
\ion{S}{2} & 1253.811 & $15.17\pm0.02$ & $13.97\pm0.03$& 
	\nodata & \nodata \nl
\ion{S}{2} & 1259.519 & $15.13\pm0.02$ & $14.08\pm0.03$& 
	\nodata & \nodata \nl
\nl
\ion{S}{3} & 1190.208 & $13.82\pm0.02$ & \nodata & \nodata & \nodata\nl
\nl
\ion{Cr}{2} & 2056.254 & $12.51\pm0.05$ & $12.05\pm 0.07$ & 
	\nodata & \nodata \nl
\ion{Cr}{2} & 2062.234 & $12.37\pm0.09$\phn \phd 
	&  \nodata & \nodata & \nodata \nl
\ion{Cr}{2} & 2066.161 & $12.50\pm0.07$ 
	& $12.14^{+0.07}_{-0.09}$\phn \phd &
	\nodata & \nodata \nl
\nl
\ion{Mn}{2} & 1197.184 & $12.44\pm0.04$ & \nodata & 
	\nodata & \nodata \nl
\ion{Mn}{2} & 2576.877 & $12.50\pm0.02$ & $11.75\pm0.03$ & 
	$10.89^{+0.09}_{-0.12}$ \phn \phd & 
	 $10.67^{+0.13}_{-0.20}$ \phn \phd \nl
\ion{Mn}{2} & 2594.499 & $12.48\pm0.02$ & $11.75\pm0.03$ & 
	\nodata & \nodata \nl
\nl

\ion{Fe}{2} & 1133.665 & $14.18\pm 0.09$ & $13.85^{+0.10}_{-0.12}$ & 
	\nodata & \nodata \nl
\ion{Fe}{2} & 1143.226 & $14.27\pm0.07$ & $13.87\pm0.05$ & 
 	\nodata & \nodata \nl
\ion{Fe}{2} & 1144.938 & $14.15\pm0.03$ & $13.70\pm0.02$ &
 	\nodata & \nodata \nl
\ion{Fe}{2} & 1608.451 & $14.10\pm 0.02$ & $13.63\pm0.02$ & 
	$12.43\pm0.06$ & $12.46\pm0.07$ \nl
\ion{Fe}{2} & 2249.877 & $14.11\pm0.02$ & $13.70\pm0.02$ & 
	\nodata & \nodata \nl
\ion{Fe}{2} & 2260.780 & $14.15\pm0.02$ & $13.81\pm0.02$ & 
	\nodata & \nodata \nl
\ion{Fe}{2} & 2344.214 & \sat           & $13.52\pm0.02$ & 
	$12.55\pm0.02$ & $12.41\pm0.02$ \nl
\ion{Fe}{2} & 2374.461 & $14.09\pm0.03$ & $13.67\pm0.02$ &
	$12.54\pm0.04$ & $12.43\pm0.04$ \nl
\ion{Fe}{2} & 2586.650 & $14.10\pm0.02$ & $13.61\pm0.02$ & 
	$12.51\pm0.03$ & $12.43\pm0.03$ \nl
\ion{Fe}{2} & 2600.173 & \sat & \sat & $12.50\pm0.02$ & 
            $12.35\pm0.02$ \nl
\nl
\ion{Fe}{3} & 1122.526 & 13.37$^{+0.09}_{-0.11}$\tablenotemark{f}\phn \phd 
	& \nodata & \nodata & \nodata \nl 
\nl
\ion{Ni}{2} & 1317.217 & $12.77\pm0.05$ & $12.50\pm0.05$ & 
	\nodata & \nodata \nl
\ion{Ni}{2} & 1370.132 & $12.82\pm0.03$ & $12.46\pm0.04$ & 
	\nodata & \nodata \nl
\ion{Ni}{2} & 1709.600 & $12.73\pm0.09$\phn \phd
	 & $12.58\pm0.07$ & \nodata & \nodata \nl
\ion{Ni}{2} & 1741.569 & $13.16\pm0.06$ & $12.71\pm0.07$ & 
	\nodata & \nodata \nl
\nl
\ion{Zn}{2} & 2026.136 & $12.56\pm0.03$ 
	& $11.35\pm0.09$\phn \phd & 
	\nodata & \nodata \nl
\ion{Zn}{2} & 2062.664 & $12.48\pm0.03$ & $11.38\pm0.09$ & 
	\nodata & \nodata \nl
\enddata
  \tablenotetext{a}{Vacuum wavelengths from Morton (1991).}  
  \tablenotetext{b}{Integrated column densities for the components
   considered here with $\pm 1 \sigma$ errors, given in atoms cm$^{-2}$.} 
  \tablenotetext{c}{The \sat\ indicates the profile is strongly saturated
   in this wavelength region, yielding no useful information.}
  \tablenotetext{d}{The $\lambda 1199.550$ \AA\ and $\lambda 1200.710$
   \AA\ lines of \ion{N}{1} each show evidence for possible contamination
   from fixed-pattern noise as well as lying in regions where the
   continuum placement is uncertain.  The errors quoted here are
   estimates based upon experiments with various reasonable continuum
   fits.} 
  \tablenotetext{e}{The values quoted here for the higher-velocity 
   components of \ion{Si}{3} should be considered more uncertain than the
   formal errors suggest due to uncertainties in the level of the
   continuum.}
  \tablenotetext{f}{Based upon G140M data.  This line has been
  integrated over the velocity range -20 km s$^{-1}$ to +23 km
  s$^{-1}$.}
\end{deluxetable}

\clearpage


\begin{landscape}
\begin{deluxetable}{lccccccccccccc}
\tablecolumns{14}
\tablewidth{0pc}
\tablecaption{Component Fitting Column Densities and 
	$b$-values\label{table:compfitting}}
\tablehead{
\colhead{} & 
	\multicolumn{1}{c}{Comp 1} & &
	\multicolumn{2}{c}{Comp 2} & & 
	\multicolumn{2}{c}{Comp 3} & &
	\multicolumn{2}{c}{Comp 4} & &
	\multicolumn{2}{c}{Comp 5} \nl
\cline {2-2} \cline{4-5} \cline{7-8} \cline{10-11} \cline{13-14}
\multicolumn{1}{c}{Species} & 
\colhead{$\log N(X)$\tablenotemark{a}} & &
\colhead{$b$\tablenotemark{c}} &
	\colhead{$\log N(X)$\tablenotemark{d}} & \colhead{}&
\colhead{$b$\tablenotemark{c}} &
	\colhead{$\log N(X)$\tablenotemark{d}} & \colhead{}&
\colhead{$b$\tablenotemark{c}} &
	\colhead{$\log N(X)$\tablenotemark{d}} & \colhead{}&
\colhead{$b$\tablenotemark{c}} &
	\colhead{$\log N(X)$\tablenotemark{d}} \nl
\multicolumn{1}{c}{$X$} & 
\colhead{[cm$^{-2}$]} & &
\colhead{[km s$^{-1}$]} &
\colhead{[cm$^{-2}$]} & &
\colhead{[km s$^{-1}$]} &
\colhead{[cm$^{-2}$]} & &
\colhead{[km s$^{-1}$]} &
\colhead{[cm$^{-2}$]} & &
\colhead{[km s$^{-1}$]} &
\colhead{[cm$^{-2}$]} 
}
\startdata
\ion{C}{2}
	& \sat &
	& \sat & \sat  &
	& \sat & \sat  &
	& $3.0\pm0.5$ & $13.75\pm0.05$  & 
	& $7.0\pm0.9$ & $13.37\pm0.10$ \nl
\ion{C}{2}$^*$ 
	& \sat\tablenotemark{e} &
	&  $ 3.5\pm0.5$    &  $13.23\pm0.03$ &  
	&  \nodata         &  \nodata        &
	&  $ 2.4\pm1.5$    &  $12.36\pm0.05$ & 
	& \nodata & \nodata \nl
\ion{N}{1} 
	& \sat &
	&  $ 4.1\pm0.6$    &  $14.25\pm0.08$ &  
	& \nodata          & \nodata         &
	&  $ 2.17\pm0.20$  &  $13.17\pm0.02$ & 
	& \nodata & \nodata \nl

\ion{O}{1} 
	& \sat &
	& $3.9\pm0.6$      &  $14.55\pm0.05$ &
	&  $ 4.2\pm0.7$    &  $13.58\pm0.03$ &  
	& $2.58\pm0.12$    &  $13.83\pm0.02$ & 
	& \nodata & \nodata \nl
\ion{Mg}{1} 
	& $12.59\pm0.05$  &
	&  $ 2.60\pm0.18$  &  $11.72\pm0.02$ &  
	&  \nodata         &  $10.28\pm0.11$ &
	&  \nodata         &  $10.52\pm0.04$ & 
	& \nodata & \nodata \nl  
\ion{Mg}{2} 
	& $15.09\pm0.20$  &
	&  $ 4.9 \pm1.5 $  &  $13.95\pm0.12$ &  
	&  $ 2.6 \pm0.3 $  &  $12.50\pm0.02$ &  
	&  $ 1.82\pm0.09$  &  $12.82\pm0.02$ & 
	& $7.5\pm1.1$ & $11.98\pm0.02$ \nl
\ion{Al}{2}
	& \sat &
	&  $3.5\pm0.8$            &  $12.64\pm0.05$ &  
	&  2.65\tablenotemark{f}  &  $11.47\pm0.05$ &  
	&  1.74\tablenotemark{f}  &  $11.27\pm0.06$ & 
	& \nodata & \nodata \nl
\ion{Si}{2} 
	&  $15.13\pm0.11$  &
	&  $ 3.6 \pm0.6 $  &  $13.98\pm0.05$ &  
	&  $ 2.7 \pm0.4 $  &  $12.78\pm0.02$ &  
	&  $ 1.67\pm0.09$  &  $12.85\pm0.02$ & 
	&  $4.4^{+4.9}_{-2.3}$ &  $11.64\pm0.10$ \nl
\ion{Si}{3}
	& \sat & 
	& $4.6\pm2.0$      & $12.65\pm0.08$ &
	& $3.34\pm1.7$     & $11.83\pm0.10$ &
	& \nodata          & $10.67\pm0.22$ & 
	& $7.5\pm1.0$      & $12.29\pm0.03$ \nl
\ion{P}{2} 
	& $13.50\pm0.15$ &
        & 3.6\tablenotemark{g}  & $11.98\pm0.13$ &
	& \nodata & \nodata & 
	& \nodata & \nodata & 
	& \nodata & \nodata \nl
\ion{S}{2} 
	& $15.24\pm0.05$&
	&  $5.0 \pm0.7$ & $13.82\pm0.02$ &
	& \nodata & \nodata & 
	& \nodata & \nodata & 
	& \nodata & \nodata \nl
\ion{Cr}{2} 
	& $12.48\pm0.07$  &
	& 3.0\tablenotemark{h} & $12.03\pm0.05$ & 
	& \nodata & \nodata & 
	& \nodata & \nodata & 
	& \nodata & \nodata \nl
\ion{Mn}{2} 
	& $12.52\pm0.10$  &
	&  $ 3.0\pm1.0$          &  $11.61\pm0.05$ &  
	&  1.76\tablenotemark{i} &  $10.82\pm0.09$ &  
	&  1.18\tablenotemark{i} &  $10.55\pm0.14$   & 
	& \nodata & \nodata \nl
\ion{Fe}{2} 
	& $14.13\pm0.02$  &
	&  $ 2.75\pm0.07$  &  $13.78\pm0.02$ &  
	&  $ 1.76\pm0.14$  &  $12.63\pm0.02$ &  
	&  $ 1.18\pm0.11$  &  $12.61\pm0.02$ & 
	&  $ 3.9^{+5.0}_{-2.2}$ &  $11.34\pm0.10$ \nl
\ion{Ni}{2} 
	& $12.86\pm0.08$ &
	& 2.79\tablenotemark{j} & $12.35\pm0.04$ &
	& \nodata & \nodata & 
	& \nodata & \nodata & 
	& \nodata & \nodata \nl
\ion{Zn}{2} 
	& $12.55\pm0.08$  &
	&  2.79\tablenotemark{j}  &  $11.16\pm0.08$ &  
	& \nodata & \nodata & 
	& \nodata & \nodata & 
	& \nodata & \nodata \nl
\enddata
\tablenotetext{a}{Total column density contained within the component
	1 blend.  We have fit this region with three individual
	components, but we choose only to display the sums given the
	lack of uniqueness often involved with fitting this central
	region.  The values given here should be compared with the
	values of the integrated column densities given in Table 4.}
\tablenotetext{c}{Doppler spread parameters and $\pm1\sigma$ errors
	(in km s$^{-1}$) derived through component fitting for the
	absorbing components 2, 3, and 4.  These values were derived
	through component fitting techniques as described in \S
	3.3.2.}
\tablenotetext{d}{Column densities and $\pm1\sigma$ errors (in atoms
	cm$^{-2}$) derived through component fitting for the absorbing
	components 2, 3, and 4.}
\tablenotetext{e}{The \sat\ indicates the profile is strongly
	saturated in this wavelength region, yielding little or no
	reliable information.}
\tablenotetext{f}{The $b$-values adopted for components 3 and 4 in the
	\ion{Al}{2} fit are the weighted means of the \ion{Mg}{2} and
	\ion{Si}{2} fits for the same components.}
\tablenotetext{g}{The $b$-values adopted for component 2 in \ion{P}{2}
	were set equal to that derived for \ion{Si}{2}: $b =
	3.6\pm0.6$ km s$^{-1}$.}
\tablenotetext{h}{The $b$-values for component 2 in this ion were set
	equal to that derived for \ion{Mn}{2}: $b = 3.0\pm1.0$ km
	s$^{-1}$.}
\tablenotetext{i}{The $b$-values for components 3 and 4 for the
	\ion{Mn}{2} fit were set equal to the derived $b$-values of
	\ion{Fe}{2} in these components: $b_3 = 1.76\pm0.14$ km
	s$^{-2}$ and $b_4 = 1.18\pm0.11$ km s$^{-2}$ .}
\tablenotetext{j}{The $b$-values for component 2 in these ions were
	set equal to that derived for \ion{Fe}{2}: $b = 2.75\pm0.07$ km
	s$^{-1}$.  }
\end{deluxetable}

\end{landscape}

\clearpage


\begin{deluxetable}{lccccccc}
\tablecolumns{8}
\tablewidth{0pc}
\tablecaption{Component Fitting Velocities\label{table:compvelocities}}
\tablehead{
\colhead{} & 
\colhead{Comp 2} & & 
\colhead{Comp 3} & & 
\colhead{Comp 4} & & 
\colhead{Comp 5} \nl
\cline{2-2} \cline{4-4} \cline{6-6} \cline{8-8}
\colhead{Species} &
\colhead{$v_{\rm LSR}$\tablenotemark{a}} & &
\colhead{$v_{\rm LSR}$\tablenotemark{a}} & &
\colhead{$v_{\rm LSR}$\tablenotemark{a}} & &
\colhead{$v_{\rm LSR}$\tablenotemark{a}} \nl
\colhead{$X$} &
\colhead{[km s$^{-1}$]} & &
\colhead{[km s$^{-1}$]} & &
\colhead{[km s$^{-1}$]} & &
\colhead{[km s$^{-1}$]}   }
\startdata
\ion{C}{2} 
	& \nodata &
	& \nodata &
	& $41.25\pm0.12$ & 
	& $-28.0\pm0.05$ \nl
\ion{C}{2}$^*$ 
	& $20.18\pm0.15$ & 
	& \nodata &         
	& $41.7\pm0.3$   & 
	& \nodata \nl
\ion{N}{1} 
	& $18.2\pm0.5$     & 
	& \nodata   &        
	& $41.10\pm0.05$  & 
	& \nodata \nl

\ion{O}{1} 
	& $20.38\pm0.12$ & 
	& $30.5 \pm0.3 $ & 
	& $41.20\pm0.03$ & 
	& \nodata \nl
\ion{Mg}{1} 
	& $20.46\pm0.05$ & 
	& $30.8\pm0.5$   &  
	& $41.20\pm0.21$ & 
	& \nodata \nl
\ion{Mg}{2} 
	& $18.2 \pm1.0 $ &   
	& $31.34\pm0.10$  &  
	& $41.20\pm0.03$  & 
	& $-28.37\pm0.19$ \nl
\ion{Al}{2}
	& $19.88\pm0.23$  &  
	& $30.4\pm0.4$    &  
	& $40.9\pm0.3$    & 
	& \nodata \nl
\ion{Si}{2} 
	& $20.1 \pm0.3 $  &  
	& $31.03\pm0.08$  &  
	& $41.20\pm0.05$  & 
	& $-30.2\pm0.8$ \nl
\ion{Si}{3} 
	& $21.6 \pm0.5 $  &  
	& $30.9 \pm0.6 $  &  
	& $41.1 \pm2.0 $  & 
	& $-28.3\pm0.4$ \nl
\ion{P}{2} 
        & $19.0\pm1.2$  & 
	& \nodata  & 
	& \nodata & 
	& \nodata \nl
\ion{S}{2} 
	& $20.1$\tablenotemark{b} & 
	& \nodata  & 
	& \nodata & 
	& \nodata \nl
\ion{Cr}{2} 
	& $20.4\pm0.3$ & 
	& \nodata    &  
	& \nodata    & 
	& \nodata \nl
\ion{Mn}{2} 
	& $20.4\pm0.3$ & 
	& $31.2\pm0.6$   & 
	& $40.2\pm0.8$  & 
	& \nodata \nl
\ion{Fe}{2} 
	& $21.16\pm0.03$ &   
	& $31.24\pm0.05$ &   
	& $41.21\pm0.05$  & 
	& $-30.1\pm0.9$ \nl
\ion{Ni}{2} 
	& $20.7\pm0.3$ &    
	& \nodata  & 
	& \nodata & 
	& \nodata \nl
\ion{Zn}{2} 
	& $20.6\pm0.6$ & 
	& \nodata  & 
	& \nodata & 
	& \nodata \nl
\enddata
\tablenotetext{a}{Central velocities in the LSR restframe and
	$\pm1\sigma$ errors (in km s$^{-1}$) for the absorbing
	components 2--5.  These values were derived through
	component fitting techniques as described in \S 3.3.2.}
\tablenotetext{b}{The velocity for component 2 in the \ion{S}{2} fit
	was held equal to that of \ion{Si}{2} (see \S 3.3.2).}
\end{deluxetable}


\begin{deluxetable}{lccccc}
\tablecolumns{6}
\tablewidth{0pc}
\tablecaption{Adopted Column Densities\label{table:finalcoldens}}
\tablehead{ 
\colhead{} & \colhead{} & 
 \multicolumn{4}{c}{ $\log N_X$ [cm $^{-2}$] \tablenotemark{b}} \nl
\cline{3-6} \colhead{Species}  & \colhead{$\log N_X$\tablenotemark{a}} & 
	\colhead{Comp 1} & \colhead{Comp 2} & 
	\colhead{Comp 3} & \colhead{Comp 4} \nl
\colhead{} &  \colhead{[cm $^{-2}$]} &
	\colhead{(3 km s$^{-1}$)}  & \colhead{(20 km s$^{-1}$)} & 
	\colhead{(31 km s$^{-1}$)} & \colhead{(41 km s$^{-1}$)} }
\startdata
\ion{H}{1}
  & $19.86\pm0.015$\tablenotemark{c} &
	\nodata & \nodata & \nodata & \nodata \nl
\ion{C}{1}  & $12.96\pm0.03$ &
	$12.87\pm0.03$ & 
	$12.21\pm0.09$ & 
	\nodata & 
	\nodata \nl
\ion{C}{1}$^*$    & $12.50^{+0.10}_{-0.12}$\phn \phd  & 
	$12.50^{+0.10}_{-0.12}$\phn \phd  & 
	\nodata & \nodata & \nodata \nl
\ion{C}{1}$^{**}$ & $<12.16$       & $<12.16$       & 
	\nodata & \nodata & \nodata \nl
\ion{C}{2} &  \nodata &
	\nodata & \nodata & \nodata & $13.75\pm0.05$ \nl
\ion{C}{2}$^*$ & \nodata &
	\nodata & 
	$13.23\pm0.03$ &
	\nodata        & 
	$12.36\pm0.05$ \nl
\ion{N}{1} &  \nodata &
	\nodata & 
	$14.25\pm0.08$ & 
	\nodata & 
	$13.17\pm0.02$ \nl
\ion{O}{1}  &  $<17.0$\tablenotemark{d} &
	\nodata & 
	\nodata & 
	$13.58\pm0.03$ & 
	$13.83\pm0.02$ \nl
\ion{Mg}{1} & $12.61\pm0.02$ &
	$12.55\pm0.02$ & 
	$11.72\pm0.02$ & 
	$10.28\pm0.11$ & 
	$10.52\pm0.04$ \nl
\ion{Mg}{2} & $15.14\pm0.02$ &
	$15.08\pm0.02$\tablenotemark{e} & 
	$13.95\pm0.12$ &  
	$12.50\pm0.02$ & 
	$12.82\pm0.02$ \nl
\ion{Al}{2} & \nodata &
	\nodata & 
	$12.64\pm0.05$ & 
	$11.47\pm0.05$ & 
	$11.27\pm0.06$ \nl
\ion{Al}{3} & $12.01\pm 0.05$ & 
	$12.01\pm 0.05$ & 
	\nodata & \nodata & \nodata \nl
\ion{Si}{2} & $15.14\pm0.02$ &
	$15.10\pm0.02$\tablenotemark{f}  & 
	$13.98\pm0.05$ &
	$12.78\pm0.02$ &
	$12.85\pm0.02$ \nl
\ion{Si}{2}$^*$ & $11.47^{+0.12}_{-0.17}$\phn \phd &
	$11.47^{+0.12}_{-0.17}$\phn \phd & 
	\nodata & \nodata & \nodata \nl
\ion{Si}{3} & \nodata &
	\nodata & 
	$12.65\pm0.08$ & 
	$11.83\pm0.10$ & 
	$10.84^{+0.16}_{-0.21}$\phn \phd \nl
\ion{P}{2} & $13.48\pm0.02$ & 
	$13.45\pm0.02$ & 
	$11.98\pm0.13$ &
	\nodata &
	\nodata \nl
\ion{S}{2} & $15.21\pm0.02$ &  
	$15.19\pm0.02$\tablenotemark{g} & 
	$13.82\pm0.02$ & 
	$<12.85$ & 
	$<12.85$ \nl
\ion{S}{3} & $13.82\pm0.02$ & 
	$13.82\pm0.02$ & 
	$<12.59$ & $<12.53$ & \nodata \nl
\ion{Ca}{2}\tablenotemark{h} & $12.35\pm0.06$ &
	$12.19\pm0.08$ & 
	$11.85\pm0.08$ & 
	\nodata & 
	$\leq11.0$ \nl 
\ion{Ti}{2}\tablenotemark{i} & $11.84\pm0.08$ & 
	$11.78\pm0.09$ & 
	$10.95\pm0.09$ & 
	\nodata & 
	\nodata \nl
\ion{Cr}{2} & $12.62\pm0.03$ &
	$12.47\pm0.04$ & 
	$12.03\pm 0.05$ & 
	\nodata & 
	\nodata \nl
\ion{Mn}{2} & $12.55\pm0.02$ &
	$12.48\pm0.02$ & 
	$11.61\pm0.05$ & 
	$10.82\pm0.09$ &
	$10.55\pm0.14$ \nl
\ion{Fe}{2}\tablenotemark{j} & 
	$14.31\pm0.01$\tablenotemark{k} &
	$14.13\pm0.02$\tablenotemark{k} & 
	$13.78\pm0.02$ & 
	$12.63\pm0.02$ & 
	$12.61\pm0.02$ \nl
\ion{Fe}{3} &  13.37$^{+0.09}_{-0.11}$\phn \phd &
	13.37$^{+0.09}_{-0.11}$\phn \phd &
	\nodata & \nodata & \nodata \nl  
\ion{Ni}{2} & $12.98\pm0.06$\tablenotemark{k} &
	$12.86\pm0.08$\tablenotemark{k} &
	$12.35\pm0.04$ & 
	\nodata & 
	\nodata \nl
\tablebreak
\ion{Cu}{2} & $<11.5$ & \nodata & \nodata & \nodata \nl
\ion{Zn}{2} & $12.57\pm0.08$\tablenotemark{k} &
	$12.55\pm0.08$\tablenotemark{k} & 
	$11.16\pm0.08$ & 
	\nodata & 
	\nodata \nl
\enddata
\tablenotetext{a}{Total sightline column densities derived from
	integrations of $N_a(v)$ profiles, unless otherwise noted.}
\tablenotetext{b}{Adopted column densities for the components
	considered here with $\pm 1 \sigma$ errors, given in atoms
	cm$^{-2}$.  Column densities for components 2--5 are derived
	from our component fitting analysis.  Values for component 1
	were derived through integrations of $N_a(v)$ profiles, unless
	otherwise noted.}
\tablenotetext{c}{Derived using the continuum reconstruction method
	(see Appendix A).}
\tablenotetext{d}{A $2\sigma$ upper limit derived from the
	\wave{1355.6} line.} 
\tablenotetext{e}{Based upon the \ion{Mg}{2} \twowave{1239, 1240} lines.}
\tablenotetext{f}{Based upon the \ion{Si}{2} \wave{1808} line.}
\tablenotetext{g}{Based upon the \ion{S}{2} \wave{1250} line.}
\tablenotetext{h}{The \ion{Ca}{2} data given here are from the
	profile-fitting results of Shull~{\em et~al.} (1977; see their
	Table 1).} 
\tablenotetext{i}{The \ion{Ti}{2} data given here are from the
	profile-fitting results of Welsh~{\em et~al.} (1997; see their
	Table 3).  They have estimated the errors in the column
	density values to be $\sim 10\% - 20\%$.  The errors we have
	adopted in this table are slightly higher than this estimate.} 
\tablenotetext{j}{We have excluded the \ion{Fe}{2} \wave{2344} line
	and those lines with $\lambda < 1200$ \AA\ in deriving this
	value due to possible oscillator strength uncertainties.}  
\tablenotetext{k}{The adopted column density given is derived from our
	component fitting measurements (see text).}
\end{deluxetable}

\clearpage


\begin{landscape}
\begin{deluxetable}{ccccccccccccc}
\tablecolumns{13} 
\tablewidth{0pc} 
\tablecaption{CLOUDY Models of the $\mu$ Columbae \ion{H}{2} 
	Region\label{table:cloudy}} 
\tablehead
{ 
\colhead{$n_{\rm H}$\tablenotemark{a}} 
	& \colhead{$f$\tablenotemark{b}} 
	& \colhead{$R_S$\tablenotemark{c}} 
	& \multicolumn{9}{c}{Log $N(X^i)$ [cm$^{-2}$]\tablenotemark{d}} \\
	\cline{4-13} 
\colhead{[cm$^{-3}$]} 
	& \colhead{} & \colhead{[pc]} 
	& \colhead{S$^{+}$}   & \colhead{S$^{+2}$} 
	& \colhead{P$^{+}$}   & \colhead{P$^{+2}$} 
	& \colhead{N$^{+}$}  & \colhead{Si$^{+}$} 
	& \colhead{Si$^{+3}$} & \colhead{Mg$^{+}$} 
	& \colhead{Mn$^{+}$} 
	& \colhead{Fe$^{+}$} 
}
\startdata 
0.02 & 0.7 & 275 
	& 14.21 & 13.79 & 12.52 & 12.06 & 14.96 & 14.53 & 
	11.90 & 14.47 & 12.39 & 14.18 \nl 
0.05 & 0.5 & 167 
	& 14.23 & 13.84 & 12.53 & 12.14 & 14.99 & 14.55 
	& 11.95 & 14.49 & 12.48 & 14.09 \nl 
0.2 & 0.25 & 84 
	& 14.23 & 13.84 & 12.53 & 12.14 & 14.99 & 14.55 
	& 11.95 & 14.49 & 12.41 & 14.20 \nl 
0.5 & 0.15 & 54
	& 14.22 & 13.82 & 12.52 & 12.13 & 14.98 & 14.55   
	& 11.80 & 14.48 & 12.40 & 14.19 \nl 
\cline{1-3}
\multicolumn{3}{c}{Observed:} & 15.18 & 13.82 & 13.45 & $<12.6$ 
	& 15.7:\tablenotemark{e} & 15.10 & 12.15\tablenotemark{f}
	& 15.11 & 12.48 & 14.09 \nl 
\enddata 
\tablenotetext{a}{The assumed ambient hydrogen density (H atoms
	cm$^{-3}$) for the \ion{H}{2} region model.} 
\tablenotetext{b}{The volume filling factor of the gas within the
	model \ion{H}{2} region.}   
\tablenotetext{c}{The outer radius of the model \ion{H}{2} region.}  
\tablenotetext{d}{Column densities of the indicated ions integrated
	through the model \ion{H}{2} region, integrated from the star
	to $R_S$.  Model calculations are not listed for Zn because of
	the large uncertainties in the atomic parameters for this
	element.}
\tablenotetext{e}{The \ion{N}{2} result quoted here is from SY.  The
	quoted uncertainties of this measurement are $\pm0.3$, though
	given the large measured $W_\lambda$, this value should be
	viewed as highly uncertain.}
\tablenotetext{f}{The \ion{Si}{4} result quoted here is from Brandt
	{\em et al.} (1998).  The measurement is the weighted average
	of the lines $\lambda \lambda$~1394, 1403~\AA, $\log[N({\rm
	Si}^{+3})] = 12.16\pm0.05$.}
\end{deluxetable}

\end{landscape}


\begin{deluxetable}{cccccccc}
\tablecolumns{8}
\tablewidth{0pc}
\tablecaption{Normalized Gas-Phase Abundances\label{table:abundances}}
\tablehead{
\colhead{ } & \colhead{ } & \colhead{ } &
\multicolumn{2}{c}{$[X/{\rm S}]$\tablenotemark{c}} & 
\colhead{ } &
\multicolumn{2}{c}{$[X/{\rm Si}]$\tablenotemark{d}} \nl
\cline{4-5} \cline{7-8}
\colhead{Element} & 
\colhead{$\log (X/{\rm H})$\tablenotemark{a}} &
\colhead{\phs $[X/{\rm H}]$\tablenotemark{b}} &
\colhead{Comp 1} & \colhead{Comp 2} & \colhead{ } &
\colhead{Comp 3} & \colhead{Comp 4} \nl
\colhead{$X$} & \colhead{$\! \! +12.00$} & \colhead{ } &
\colhead{(3 km s$^{-1}$)} & \colhead{(20 km s$^{-1}$)} & 
\colhead{ } &
\colhead{(31 km s$^{-1}$)} & \colhead{(41 km s$^{-1}$)} 
}
\startdata
C  & 8.55 & \phs \nodata 
	& \phs \nodata & \phs \nodata  
	& & \phs \nodata &  $-0.10\pm0.06$ \nl
N  & 7.97 & \phs \nodata 
	& \phs \nodata & \phs \nodata  
	& & \phs \nodata & $-0.10\pm0.03$ \nl
O  & 8.87 & \phs \nodata 
	& \phs \nodata & \phs \nodata  
	& & $-0.52\pm0.04$ & $-0.34\pm0.03$ \nl
Mg & 7.58 & $-0.30\pm0.02$
	& $-0.42\pm0.03$ & $-0.18\pm0.12$
	& & $-0.31\pm0.03$ & $-0.06\pm0.03$ \nl
Al & 6.48 & \phs \nodata 
	& \phs \nodata & $-0.39\pm0.05$
	& & $-0.24\pm0.05$ & $-0.51\pm0.06$ \nl
Si & 7.55 & $-0.26 \pm 0.02$
	& $-0.37\pm0.03$ & $-0.12\pm0.06$ 
	& & \phs \nodata & \phs \nodata \nl
P  & 5.57 & $+0.05 \pm 0.02$
	& $-0.04\pm0.03$ &  $-0.14\pm0.13$ 
	& & \phs \nodata & \phs \nodata \nl
S  & 7.27 & $+0.08 \pm 0.02$
	& \phs \nodata & \phs \nodata  
	& & \phs \nodata & \phs \nodata \nl
Ti & 4.93 & $-0.95\pm0.09$
	& $-1.07\pm0.05$ & $-0.53\pm0.09$ 
	& & \phs \nodata & \phs \nodata \nl
Cr & 5.68 & $-0.92 \pm 0.03$
	& $-1.13\pm0.04$ & $-0.20\pm0.05$ 
	& & \phs \nodata & \phs \nodata \nl
Mn & 5.53 & $-0.84 \pm 0.02$
	& $-0.97\pm0.03$ &  $-0.47\pm0.05$
	& &  $+0.06\pm0.09$ & $-0.28\pm0.14$ \nl
Fe & 7.51 & $-1.06 \pm 0.02$
	& $-1.30\pm0.03$ & $-0.28\pm0.02$
	& & $-0.11\pm0.03$ & $-0.20\pm0.03$ \nl
Ni & 6.25 & $-1.13 \pm 0.06$
	& $-1.31\pm0.08$ & $-0.45\pm0.04$ 
	& & \phs \nodata & \phs \nodata \nl
Zn & 4.65 & $+0.06 \pm 0.08$
	& $-0.02\pm0.08$ & $-0.04\pm0.08$  & & 	
	\phs \nodata & \phs \nodata \nl
\enddata
\tablenotetext{a}{The logarithmic ``solar'' abundances of the
elements, $\log(X/{\rm H})_\odot$, are used in deriving the gas
normalized gas phase abundances.  We have adopted the solar system
meteoritic abundances listed are from Anders \& Grevesse (1989) except
for C, N and O, which are photospheric values from Grevesse \& Noels
(1993).}
\tablenotetext{b}{We present the sightline integrated values of
$[X/{\rm H}]$ in this column.  Thus $[X/{\rm H}] \equiv \log \{N_X
/N_{\rm H} \} - \log \{X/{\rm H}\}_\odot$.  Here we have used the
value of $\log N_{\rm H I} = 19.86\pm0.015$ ($1\sigma$ systematic)
derived in Appendix A as $N_{\rm H}$. This treatment has neglected
contributions from H$^+$.  However, our photoionization modelling (see
\S 4) implies the corrections are relatively small ($\approx -0.04$ to
$-0.05$ dex to the listed values).}
\tablenotetext{c}{For the components 1 and 2 we have referenced the
gas phase abundances to solar by comparing the column densities to
that of sulfur.  Thus $[X/{\rm S}] \equiv \log\{N_X /N_{\rm S}\} -
\log\{X/{\rm S}\}_\odot$. }
\tablenotetext{d}{For components 3 and 4 we have referenced the gas
phase abundances to solar by comparing the column densities to that of
silicon.  Thus $[X/{\rm Si}] \equiv \log\{N_X /N_{\rm Si}\} -
\log\{X/{\rm Si}\}_\odot$.}
\end{deluxetable}

\clearpage


\begin{deluxetable}{lcccccc}
\tablecolumns{7}
\tablewidth{0pc}
\tablecaption{Dust-Phase Abundances for $\mu$ Columbae 
	Clouds\label{table:dustcontent}}
\tablehead{
\colhead{} & \colhead{$(X/{\rm H})_c$\tablenotemark{a}} 
	& \multicolumn{2}{c}{$10^6 (X/{\rm H})_d$\tablenotemark{b}}  & 
	& \multicolumn{2}{c}{$(X/{\rm Si})_d$\tablenotemark{c}} \\
\cline{3-4} \cline{6-7} 
\colhead{Element}  & \colhead{}
& \colhead{Comp. 1} & \colhead{Comp 2}   & 
& \colhead{Comp. 1} & \colhead{Comp 2}   
}
\startdata
& \multicolumn{6}{l}{Solar Abundances (Anders \& Grevesse 1989)}\nl 
\cline{2-7}
Mg  &  $3.8 \times 10^{-5}$ 
	& $23.2 \pm 1.0$
	& $13   \pm 8  $ &
	& $1.16\pm 0.08$ & $1.5 \pm 1.2$ \nl 
Si  &  $3.5 \times 10^{-5}$ 
	& $20.0\pm1.0  $
	& $ 9 \pm 4 $ & 
	& 1.00 & 1.00 \nl 
Fe  &  $3.2 \times 10^{-5}$ 
	& $30.80 \pm 0.08$
	& $15.4 \pm 1.1$ &
	& $1.54 \pm 0.08$ & $1.8 \pm 0.8$ 
	\vspace{0.05in}
	\nl 
Ti  &  $8.5 \times 10^{-8}$
	& $0.0777 \pm 0.0017$
	& $0.060 \pm 0.006$ &
	& $0.0039\pm0.0002 $ 
	& $0.007 \pm0.003 $\nl
Cr  &  $4.8 \times 10^{-7}$ 
	& $0.442 \pm 0.004$
	& $0.18 \pm 0.04$ &
	& $0.0221 \pm 0.0012$
	& $0.021 \pm 0.011$\nl 
Mn  &  $3.4 \times 10^{-7}$
	& $0.302 \pm 0.002$
	& $0.224 \pm 0.015$ &
	& $0.0151 \pm 0.0008$ 
	& $0.026 \pm 0.013$ \nl 
Ni  &  $1.8 \times 10^{-6}$ 
	& $1.690 \pm 0.019 $
	& $1.15 \pm 0.07$ &
	& $0.085 \pm 0.004$
	& $0.13  \pm 0.07 $\nl
\cline{2-7}
& \multicolumn{6}{l}{B-star Abundances 
	(Gies \& Lambert 1992)\tablenotemark{d}}\nl
\cline{2-7}
Mg  &  $2.4 \times 10^{-5}$ 
	& $12.0 \pm 0.8$
	& $ 2   \pm 7  $ &
	& $0.58 \pm 0.05$ & $0.3  \pm 0.7 $ \nl 
Si  &  $3.3 \times 10^{-5}$ 
	& $20.5 \pm 0.8$
	& $10  \pm 3$ &
	& 1.00 & 1.00 \nl 
Fe  &  $5.2 \times 10^{-5}$ 
	& $51.3 \pm 0.1$
	& $38.0 \pm 1.2$ &
	& $2.5  \pm 0.10$ & $3.73\pm 1.19$ \nl 
\cline{2-7}
& \multicolumn{6}{l}{B-star Abundances (Kilian-Montenbruck {\em et al.}
	1994)\tablenotemark{e}}\nl 
\cline{2-7}
Mg  &  $2.4 \times 10^{-5}$ 
	& $16.6 \pm 0.5$
	& $11   \pm 4  $ &
	& $2.05 \pm 0.15$ & $6   \pm   8$ \nl 
Si  &  $1.6 \times 10^{-5}$ 
	& $8.09 \pm 0.5$
	& $1.7 \pm 2.0$ &
	& $1.00$ & $1.00$ \nl 
Fe  &  $2.5 \times 10^{-5}$ 
	& $24.40 \pm 0.004$
	& $16.2 \pm 0.8$ &
	& $3.01 \pm 0.19$ & $9   \pm 11 $ \nl 
\enddata
\tablenotetext{a}{Value for the ``cosmic'' abundance of each element
	using the reference abundances given in the table.}
\tablenotetext{b}{Dust-phase abundance of each listed element in parts
	per million.  This value was derived by assuming S is
	undepleted and accurately traces the distribution of
	\ion{H}{1}.  In each case we have used the (S/H)$_c$ value
	appropriate for the reference system.  These are (S/H)$_c =
	1.9\times10^{-5}$ (Anders \& Grevesse 1989); (S/H)$_c
	=1.6\times10^{-5}$ (Gies \& Lambert 1992); and (S/H)$_c = 9.3
	\times 10^{-6}$ (Kilian-Montenbruck {\em et al.} 1994).}
\tablenotetext{c}{Dust-phase abundance ratio of each listed element to
	Si, using the relative abundances given in the table.}
\tablenotetext{d}{Gies \& Lambert (1992) do not give abundances for Mg,
	Ti, Cr, Mn \& Ni.  We have adopted the Mg abundance derived by
	Kilian-Montenbruck {\em et al.} (1994) here and have not
	derived dust-phase abundances for these other elements.}
\tablenotetext{e}{Kilian-Montenbruck {\em et al.} (1994) do not give
	abundances for Ti, Cr, Mn \& Ni.  We have not derived the
	dust-phase abundances for these elements.}
\end{deluxetable}

\clearpage


\begin{deluxetable}{lccccccc}
\tablecolumns{8}
\tablewidth{0pc}
\tablecaption{Mg, Fe, and Si Dust-Phase Abundances for Diffuse 
	Clouds\label{table:silicates}}
\tablehead{
\colhead{Cloud} & \colhead{10$^6$(Mg/H)$_d$\tablenotemark{a}} 
	& \colhead{10$^6$(Si/H)$_d$\tablenotemark{a}} 
 	& \colhead{10$^6$(Fe/H)$_d$\tablenotemark{a}} 
	& \colhead{(Mg/Si)$_d$\tablenotemark{b}} 
	& \colhead{(Fe/Si)$_d$\tablenotemark{b}} 
	& \colhead{[(Mg+Fe)/Si]$_d$\tablenotemark{c}}
	& \colhead{Ref.\tablenotemark{d}}
}
\startdata
&\multicolumn{6}{c}{Warm Disk Clouds}\nl
\cline{2-7}
$\mu$ Col Comp. 1   
	 & 23.2$\pm$1.0     & 20.0$\pm$1.0   & 30.80$\pm$0.08 
	 & 1.16$\pm$0.08    & 1.54$\pm$0.08  & 2.70$\pm$0.12 
	 & 1 \nl

HD 93521{\em ws} \tablenotemark{e}    
	 & 28$\pm$5       & 22.3$\pm$1.8   & 30.32$\pm$0.14 
	 & 1.24$\pm$0.23  & 1.36$\pm$0.11  & 2.6$\pm$0.3 
	 & 2 \nl
$\zeta$ Oph Comp. A     
	 & 28.7$\pm$1.1   & 24.5$\pm$1.0   & 30.27$\pm$0.19 
	 & 1.17$\pm$0.07  & 1.23$\pm$0.05  & 2.40$\pm$0.08 
	 & 3 \nl
HD 18100
	& 26.4$\pm$1.3    & 20.0$\pm$1.7   & 17.0$\pm$0.5
	& 1.32$\pm$0.13   & 0.85$\pm$0.08  & 2.2$\pm$0.5
	& 4 \nl
HD 215733\tablenotemark{f}
	 & 27.3$\pm$2.0   & 24.8$\pm$1.5   & 30.12$\pm$ 0.21 
         & 1.10$\pm$0.10  & 1.22$\pm$0.07  & 2.32$\pm$0.13 
	 & 5\nl
	\nl
Weighted Mean:
	 & 26.0$\pm$0.6   & 22.4$\pm$0.6   & 30.48$\pm$0.06
        & 1.17$\pm$0.04  & 1.22$\pm$0.03  & 2.46$\pm$0.06
	\nl
\cline{2-7}
&\multicolumn{6}{c}{Warm Halo Clouds}\nl
\cline{2-7}
$\mu$ Col Comp. 2  
	 & 13$\pm$8       &  9$\pm$4       & 15.4$\pm$1.1 
	 & 1.5$\pm$1.2    & 1.8$\pm$0.8    & 3.3$\pm$1.4 
	 & 1 \nl
HD 93521{\em wf} \tablenotemark{e}        
	 & 18$\pm$4         & 15.5$\pm$1.8   & 24.6$\pm$2.0 
	 & 1.1$\pm$0.3      & 1.58$\pm$0.23  & 2.7$\pm$0.4 
	 & 2 \nl
HD 149881\tablenotemark{g}         
	 & 22.5$\pm$1.4    & 15.5$\pm$1.8   & 26.5$\pm$0.5 
	 &  1.45$\pm$0.19  & 1.70$\pm$0.20  & 3.2$\pm$0.3 
	 & 6\nl
HD 215733 Comp. 6\tablenotemark{h}     
	 & 26$\pm$7        & 15$\pm$5       & 27.23$\pm$1.06 
         & 1.8$\pm$0.7     & 1.9$\pm$0.6    & 3.6$\pm$1.0
	 & 5 \nl
	\nl
Weighted Mean:
	 & 21.9$\pm$1.3    & 15.3 $\pm$1.2  & 25.0$\pm$0.4
	 & 1.37$\pm$0.16   & 1.67$\pm$0.14  & 3.11$\pm$0.20
	\nl
\enddata
\tablenotetext{a}{Dust-phase abundance of the elements Mg, Si, and Fe
	 in parts per million.  These values were derived by assuming a
	 species such as S or Zn is undepleted and accurately traces
	 the distribution of \ion{H}{1}. } 
\tablenotetext{b}{Dust-phase abundance ratio of the elements Fe and Mg
	 to Si.}
\tablenotetext{c}{The sum of the dust-phase abundances of Fe and Mg
	 relative to Si.}
\tablenotetext{d}{Reference code for interstellar abundance data:(1)
	 This work; (2) Spitzer \& Fitzpatrick (1993); (3) 
	 Savage {\em et al.} (1992) and Sofia {\em et al.} (1994); (4)
	 Savage \& Sembach (1996); (5) Spitzer \& Fitzpatrick (1995)
	 (6) Fitzpatrick \& Spitzer (1997).}
\tablenotetext{e}{The data for HD 93521 were taken directly from Table
	 3 of Spitzer \& Fitzpatrick (1993) and adjusted for
	 differences in the \ion{Mg}{2} oscillator strengths.  The
	 labels {\em ws} and {\em wf} refer to ``warm slow'' and
	 ``warm fast'' clouds in their naming system.}
\tablenotetext{f}{The gas-phase abundance values used for deriving the
	 dust-phase abundances for HD 215733 are based upon data from
	 Table 4 of Fitzpatrick \& Spitzer (1997).  We have summed
	 their components 4, 15, 17, and 19 for Mg, Si, Fe, and Zn.
	 The dust-phase abundances above are based upon the resulting
	 normalized gas-phase abundances [Mg/Zn]$=-0.42\pm0.08$,
	 [Si/Zn]$=-0.39\pm0.06$, and [Fe/Zn]$=-1.03\pm0.04$, using the
	 suggested \ion{Mg}{2} $f$-value of Fitzpatrick (1997).
	 Following Fitzpatrick \& Spitzer the dust-phase abundances
	 quoted here were obtained by including a gas-phase abundance
	 correction of $-0.13$ dex to account for a measured sub-solar
	 abundance of Zn relative to S along this sightline.}
\tablenotetext{g}{The dust-phase abundances for HD 149881 are based
	 upon data from Spitzer \& Fitzpatrick (1995).  These authors
	 derive the gas-phase abundance of Mg for the integrated
	 sightline ([Mg/Zn]$= -0.39\pm0.04$ after scaling for the
	 oscillator strengths used here).  The Si and Fe values
	 represent the sum of the column densities for their
	 components 1, 2, 4, 6, and 8-10 and are normalized to
	 \ion{Zn}{2} in the same components using data from their
	 Table 4.}
\tablenotetext{h}{The data for this component of the HD 215733
         come directly from Table 7 of Fitzpatrick \& Spitzer (1997),
	 with an adjustment for the [Mg/S] value to account for our adopted
	 oscillator strength.}
\end{deluxetable}


\begin{deluxetable}{lcccccc}
\tablecolumns{7}
\tablewidth{0pc}
\tablecaption{GHRS Echelle Line Spread Functions at 
	$\lambda$1900 \AA \label{table:lsf}}
\tablehead{
\colhead{} &  \colhead{} &  
\multicolumn{2}{c}{Core} & \colhead{} &  
\multicolumn{2}{c}{Halo} \nl
\cline{3-4} \cline{6-7}
\colhead{Source} & \colhead{Apert.} & 
\colhead{FWHM\tablenotemark{a}} &  
\colhead{Frac.\tablenotemark{b}} &  
\colhead{} & 
\colhead{FWHM\tablenotemark{a}} &  
\colhead{Frac.\tablenotemark{b}} }
\startdata
Robinson {\em et al.}
                & LSA & 1.09 & 72.4\% & & 4.21 & 27.6\% \nl
This work
                & LSA & 1.09 & 84.9\% & & 4.21 & 15.1\% \nl
Fitzpatrick \& Spitzer\tablenotemark{c}
		& SSA & 0.95 & 95.9\% & & 5.00 & \phn4.1\% \nl
\enddata
\tablenotetext{a}{The FWHM in diodes of each component Gaussian.}
\tablenotetext{b}{The fraction of the total function in each component
                of the spread function.}
\tablenotetext{c}{For the Fitzpatrick \& Spitzer results we have
		exluded the very broad power-law wings in this
		treatment.  Thus the fractions presented here are
		compared with the ``core'' plus near ``halo''
		components. }
\end{deluxetable}

\pagebreak
\clearpage


\clearpage

\clearpage

\clearpage

\clearpage

\clearpage

\end{document}